\documentclass[onecolumn,showpacs,showkeys,preprintnumbers,amsmath,amssymb,floatfix]{revtex4}

\usepackage{hyperref} % Alessandro

\usepackage[utf8x]{inputenc} 
\usepackage{graphicx}
\usepackage{epsfig}
\usepackage[normalem]{ulem}
\usepackage{color}
\usepackage{ulem}    % AJOUTE PAR SOFIANE POUR BARRER DU TEXTE
\usepackage{enumerate}
\usepackage{lscape}   % AFFICHAGE EN LANDSCAPE POU CERTAINES PAGES
\usepackage[nodisplayskipstretch]{setspace} % INTERLIGNES
\usepackage{type1cm}  % POUR LES LETTRINES
\usepackage{lettrine} % POUR LES LETTRINES
\usepackage{dcolumn}% Align table columns on decimal point
\usepackage{bm}% bold math
\makeatletter
\newcommand*{\defeq}{\mathrel{\rlap{%
                     \raisebox{0.42ex}{$\m@th\cdot$}}%
                     \raisebox{-0.4ex}{$\m@th\cdot$}}%
                     =}                                     % SYMBOLE DE DEFINITION D'UNE EXPRESSION
\makeatother 
\newcommand{\jump}[1]{\big[\!\!\big[{#1}\big]\!\!\big]}     % DOUBLE CROCHET POUR LA NOTATION DES SAUTS
\renewcommand{\cal}[1]{\mathcal{#1}}                        % LETTRES CURSIVES
\newcommand{\calo}[1]{\mathcal{O}\!\left(#1\right)}       % ODRE DE GRANDEUR - GRAND O
\newcommand{\ab}{{\alpha\beta}}                             % COMPOSANTES TENSORIELLES
\newcommand{\lm}{{\ell m}}                                  % MODES
\newcommand{\mE}{\mathcal{E}}                               % ENERGIE PAR UNITE DE MASSE m_0
\newcommand{\mL}{\mathcal{L}}                               % MOMENT ANGULAIRE PAR UNITE DE MASSE m_0
\renewcommand{\tilde}[1]{\overset{\sim}{#1}}
\renewcommand{\tilde}[1]{\widetilde{#1}}                    % QUANTITE TILDEE
\renewcommand{\bar}[1]{\overline{#1}}
\newcommand{\scirc}{{\scalebox{0.5}{$\circ$}}}
\newcommand{\scircc}{{\scalebox{0.5}{$\!\!\circ\circ$}}}
\newcommand{\dott}[1]{\overset{\scirc}{#1}\hspace{0.0pt}}       % DERIVEE PREMIERE TEMPORELLE
\newcommand{\ddott}[1]{\overset{\scircc}{#1}\hspace{0.0pt}}     % DERIVEE SECONDE TEMPORELLE
\newcommand{\ds}{\displaystyle}
\newcommand{\rescale}[2][4]{\scalebox{#1}{$#2$}}

\newcommand{\beq}{\begin{equation}}
\newcommand{\eeq}{\end{equation}}
\newcommand{\ba}{\begin{aligned}}
\newcommand{\ea}{\end{aligned}}
\newcommand{\bea}{\begin{equation}\begin{aligned}}
\newcommand{\eea}{\end{aligned}\end{equation}}

\definecolor{grey1}{rgb}{0.153,0.157,0.133}          % COULEUR_1 DE FOND DU DOCUMENT
\definecolor{grey2}{rgb}{0.074,0.074,0.066}          % COULEUR_2 DE FOND DU DOCUMENT

\definecolor{airforceblue}{rgb}{0.36, 0.54, 0.66} %  Define a new color
\newcommand{\air}{\color{airforceblue}}
\makeatletter
\def\l@subsubsection#1#2{}
\makeatother

\begin{document}
\preprint{Preprint}

\title[Wave-forms by EMRIs]{Indirect (source-free) integration method. I. Wave-forms from geodesic generic orbits of EMRIs}

\author
{
Patxi Ritter\textsuperscript{1,2,3}%\footnote{patxi.ritter@cnrs-orleans.fr,\\ http://lpc2e.cnrs-orleans.fr/$\sim$pritter/}
, 
Sofiane Aoudia\textsuperscript{4,5}%\footnote{aoudia@mpi.aei.de,\\  http://www.researchgate.net/profile/Sofiane\_Aoudia/}
, 
Alessandro D.A.M. Spallicci\textsuperscript{1,6}\footnote{Corresponding author: spallicci@cnrs-orleans.fr, http://lpc2e.cnrs-orleans.fr/$\sim$spallicci/},  
St\'ephane Cordier\textsuperscript{2,7}%\footnote {stephane.cordier@math.cnrs.fr, \\ http://www.univ-orleans.fr/mapmo/membres/cordier/}
}

\affiliation{
\textsuperscript{1}Universit\'e d'Orl\'eans\\ 
Observatoire des Sciences de l'Univers en r\'egion Centre, UMS 3116 \\
Centre Nationale de la Recherche Scientifique\\
Laboratoire de Physique et Chimie de l'Environnement et de l'Espace, UMR 7328\\
\mbox {3A Av. de la Recherche Scientifique, 45071 Orl\'eans, France}\\
\textsuperscript{2}Universit\'e d'Orl\'eans\\
Centre Nationale de la Recherche Scientifique\\
\mbox {Math\'ematiques - Analyse, Probabilit\'es, Mod\`elisation - Orl\'eans, UMR 7349}\\
Rue de Chartres, 45067 Orl\'eans, France\\
\textsuperscript{3}Univerzita Karlova\\
Matematicko-fyzik\'{a}ln\'{i} fakulta, \'{U}stav teoretick\'{e} fyziky\\
V Hole\u{s}ovi\u{c}k\'{a}ch 2, 180 00 Praha 8, \u{C}esk\'{a} Republika\\
\textsuperscript{4}Laboratoire de Physique Th\'{e}orique - Facult\'{e} des Sciences Exactes \\ 
Universit\'{e} de Bejaia, 06000 Bejaia, Algeria\\
\textsuperscript{5}Max Planck Institut f\"{u}r Gravitationphysik, A. Einstein\\
1 Am M\"uhlenberg, 14476 Golm, Deutschland\\
\textsuperscript{6}Chaire Fran\c{c}aise, Universidade do Estado do Rio de Janeiro\\
\mbox{Instituto de F\'isica, Departamento de F\'isica Te{\'o}rica}\\
\mbox{Rua S\~{a}o Francisco Xavier 524, Maracan\~{a}, 20550-900 Rio de Janeiro, Brasil}\\
%{\small\rm spallicci@cnrs-orleans.fr~http://lpc2e.cnrs-orleans.fr/$\sim$spallicci/}\\
\textsuperscript{7}Universit\'{e} Joseph Fourier\\
Agence pour les Math\'{e}matiques en Interaction\\ 
Centre Nationale de la Recherche Scientifique\\
Laboratoire Jean Kuntzmann, UMR 5224\\ 
\mbox{Campus de Saint Martin d'H\'{e}res, Tour IRMA, 51 rue des Math\'{e}matiques, 38041 Grenoble, France} 
} 

\date{14 September 2015}

\begin{abstract}
The Regge-Wheeler-Zerilli (RWZ) wave-equation describes Schwarzschild-Droste black hole perturbations. The source term contains a Dirac distribution and its derivative. We have previously designed a method of integration in time domain. It consists of a finite difference scheme where analytic expressions, dealing with the wave-function discontinuity through the jump conditions, replace the direct integration of the source and the potential. Herein, we successfully apply the same method to the geodesic generic orbits of EMRI (Extreme Mass Ratio Inspiral) sources, at second order. An EMRI is a Compact Star (CS) captured by a Super Massive Black Hole (SMBH). These are considered the best probes for testing gravitation in strong regime. The gravitational wave-forms, the radiated energy and angular momentum at infinity are computed and extensively compared with other methods, for different orbits (circular, elliptic, parabolic, including zoom-whirl). 
\end{abstract}



\maketitle
\vspace{-0.65 cm}
\hspace{1.55 cm}{\footnotesize Mathematics Subject Classification 2010: 35Q75, 35L05, 65M70, 70F05, 83C10, 83C35, 83C57}

%\WhiteOnGrey

%{\blo
%\tableofcontents
%}
%
%\clearpage

%%%%%%%%%%%%%%%%%%%%%%%%%%%%%%%%%%%%%%%%%%%%%%%%%%%%%%%%%%%%%%%%%%%%%%%%%%%%%%%%%%%%%%%%%%%%%%%%%%%%%%%%%%%%%%%%%%%%%%%%%%%%%%%%%%%%%%%%%%%%%%%%%%%%%%%%%%%%%%%%%%%%%%%%%%%%%%%%%%%%%%%%%%%%%%%%%%%%%%%%%%%%%%%%%%%%%%%%%%%%%%%%%%%%%%%%%%%%%%%%%%%%%%%%%%%%%%%%%%%%%%%%%%%%%%%%%%%%%%%%%%%%%%%%%%%%%%%%%%%%%%%%%%%%%%%%%%%%%%%%%%%%%%%%%%%%%%%%%%%%%%%%%%%%%%%%%%%%%%%%%%%%%%%%%%%%%%%%%%%%
\section{Introduction and motivations}
 
It is a current assumption that most galaxies harbour at their centre a supermassive black hole (SMBH), on which stars and compact objects in the neighbourhood inspiral and finally plunge. The process of the capture by a SMBH of a compact star (CS), ideally a stellar-size black hole, is labelled EMRI (Extreme Mass Ratio Inspiral).  It features as one of the most scientifically rewarding sources of gravitational waves for space laser interferometry (SLI), when considering the richness of investigations and tests in gravitation, black hole physics and the astrophysics of galactic centres. An introduction to EMRIs in the context of general relativity is to be found in 
\cite{blspwh11}.  

EMRIs generate complex wave-forms (WFs) which demand an analysis in a multiple parameter space. Further, the WFs are dominated by the back-action effects which easily accumulate during the long approach of the CS towards the SMBH. These sources are studied by means of the perturbative approach, that assumes that the CS is {represented by} a { point} mass perturbing the background spacetime curvature and by that generating gravitational radiation. 

In this work, we deal with Schwarzschild-Droste (SD) black holes \cite{sc16, dro16a, dro16b}, after the names of the discoverers of the first solution of the Einstein field equations \cite{ro02}. 
Much research has been devoted to the study of SD black hole perturbations following the seminal work of Regge and Wheeler \cite{rewh57}. In the gauge named after them (RW), they showed that the evolution of the odd parity vacuum perturbations can be reduced to a wave-equation. This formalism was then complemented by the works of Zerilli for the even and odd parities with a source term \cite{ze70a, ze70b, ze70c}, thereby providing the theoretical framework for treating an SD SMBH of mass $M$ perturbed by a CS of mass ${m_0}\ll M$.
The wave-equations, one for each parity, are named RWZ, henceforth.  Moncrief showed the gauge invariance of the wave-function \cite{mo74}. Among the early contributions, we recall Peters \cite{pe66}, Stachel \cite{st68}, Vishveshwara \cite{vi70}, Edelstein and Vishveshwara \cite{edvi70}.  

For the radial fall case, the WFs in frequency domain were produced first by Zerilli \cite{ze70c}, followed
by Davis {\it et al.} \cite{daruprpr71} and several others, see the references in \cite{sp11}; in time domain by Lousto and Price \cite{lopr97b}, Martel and Poisson \cite{mapo05}, and by the authors \cite{aosp11}.  
For circular and generic orbits, the first WFs in the time domain were produced by Martel \cite{ma04}. They were later followed by Sopuerta and Laguna \cite{sola06}, Field {\it et al.} \cite{fihela09}, Barack and Lousto \cite{balo05} in the harmonic (H) gauge \footnote { The H gauge was proposed by de Donder \cite{dd21} and Lanczos \cite{la22}, but was Einstein himself who wrote it first. This gauge is often named after the Danish Ludvig Lorenz \cite{lo67b}, though he died 24 years before general relativity was published. It is also attributed erroneously to the Dutch Hendrik Lorentz. Instead, the Lorenz gauge of electrodynamics might be also attributed to FitzGerald \cite{hu91}.}. Other contributions include the one by Poisson \cite{po95} in the frequency domain;  Hopper and Evans in frequency and time domains, and RW and H gauges \cite{hoev10,hoev13}. 
Nagar, Damour and Tartaglia {\it et al.} \cite{nadata07} match the effective one body (EOB) computations to the perturbation framework.

The wave-function of the RWZ equation is discontinuous as it belongs to the $C^{-1}$ continuity class \footnote{A function $f$ is said to be of class $C^k$ if the derivatives $f'$, $f''$, ..., $f^{(k)}$ exist and are continuous, where $k\geq 0$. In the literature, the extension to $k<0$ is sometimes adopted for the distributions of Heaviside, Dirac and its derivative.}. 
We construct our algorithm on the jump conditions that the wave-function and its derivatives have to satisfy. The discontinuities  have been discussed in the literature, but the jump conditions were previously used quite differently. In \cite{ha07}, Haas replaces a $C^0$  scalar field with a polynomial, whose coefficients are determined by the jump conditions. Haas uses this technique for integration of the potential term of the wave-equation, turned into $(u,v)$ coordinates, and relies on the traditional splitting into four areas for each of the cells crossed by the particle \cite {lopr97b}. Similar use of the jump conditions is adopted by Barack and Sago  for the gravitational case in the H gauge \cite{basa10}. They use a predictor-corrector scheme that defines the coefficients of the polynomial interpolation quantity $V\psi$. 
Sopuerta and Laguna ensure that the particle always lies at the interface between domain intervals, each remaining homogeneous (Particle-without-Particle), limiting the  jump conditions to act as boundary conditions  \cite{sola06}. 
Canizares and collaborators impose the jump conditions via characteristic fields out of their pseudo-spectral numerical method \cite{caso09,casoja10}.
Field, Hesthaven and Lau follow a similar track in the context of the Galerkin method \cite{fihela09}, or like Hopper and Evans use the jump conditions for checking the correctness of the transformation from frequency to time domain (extended homogeneous solution) \cite{hoev10,hoev13}.

We have shown how our method differs in \cite{aosp11}: the jump conditions are not limited to an auxiliary role, but they are pivotal ingredients in our integration scheme.  We tested our approach with geodesic radial fall at first \cite{aosp11}, and fourth order \cite{rispaoco11}.  

We choose to develop a method in the time domain, because we discern it to be more suited to the computation of the back-action. It is also easily transferable to different physical problems that require anyway solving an hyperbolic equation with a singular source term. The proposed method does not require the use of approximations of $\delta$ distributions. Thus we will treat the particle as a true point particle.
Notwithstanding the { discontinuous behaviour of the metric perturbations and of the wave-function, it is possible to revert this weakness - the lack of smoothness - into a building feature of a new integration method. 
In our  algorithm, largely analytical, the jump conditions implicitly provide the information on the source and the potential. However, for cells not crossed by the particle, we retain the classic approach for integrating the homogeneous wave-equations \cite{lopr97b}. 

For testing our method, we consider the CS moving on the geodesic of the background field, and being unaffected by its own mass and the emitted radiation. 
Herein, we show that { a second} order { algorithm} suffices to acquire both well-behaved WFs infinity, and accurate levels of radiated energy { and angular momentum}.  The WFs, the radiated energy and angular momentum at infinity are extensively compared to previous results, for different orbits (circular, elliptic, parabolic, including zoom-whirl).

The rest of the paper is structured as follows. In Sect. II, after a brief reminder of the formalism on perturbations for { an} SD black hole, we present our own integration method that we have developed to solve the RWZ equation throughout the paper. In Sec. III, the WFs, radiated energy and angular momentum values are displayed and compared with the results from other groups. 
The appendixes display the multipolar expansion and linearised field equations (A); properties of the distributions (B), the explicit form of the jump conditions (C).

Geometric units ($G = c = 1$) are used, unless stated otherwise. The metric signature is $(-, +, +, +)$.

%%%%%%%%%%%%%%%%%%%%%%%%%%%%%%%%%%%%%%%%%%%%%%%%%%%%%%%%%%%%%%%%%%%%%%%%%%%%%%%%%%%%%%%%%%%%%%%%%%%%%%%%%%%%%%%%%%%%%%%%%%%%%%%%%%%%%%%%%%%%%%%%%%%%%%%%%%%%%%%%%%%%%%%%%%%%%%%%%%%%%%%%%%%%%%%%%%%%%%%%%%%%%%%%%%%%%%%%%%%%%%%%%%%%%%%%%%%%%%%%%%%%%%%%%%%%%%%%%%%%%%%%%%%%%%%%%%%%%%%%%%%%%%%%%%%%%%%%%%%%%%%%%%%%%%%%%%%%%%%%%%%%%%%%%%%%%%%%%%%%%%%%%%%%%%%%%%%%%%%%%%%%%%%%%%%%%%%%%%%%

\section{The indirect method}\label{section:equation:onde}
\subsection{Introduction to the RWZ formalism}
At lowest order, the perturbation analysis provides the framework for evaluating the effects - namely the emitted radiation - of a particle of mass $m_0$ moving in a given background spacetime $g_\ab$ determined by the large mass $M$ and solution of the vacuum Einstein equations $\text{G}_\ab[g]=0$. The particle perturbs the geometry through the tensor field $h_\ab$. The quantities involved are such that $\varepsilon= m_0/M = m_0/\mathcal{R}\ll1$
where $\mathcal{R}$ is the characteristic length scale of the background geometry. The amplitude of the fluctuations is small and considered of the order of the mass ratio $h_\ab\sim\mathcal{O}(\epsilon)$. In the total metric $\overline{g}_\ab=g_\ab+h_\ab$, the particle, described by a stress-energy tensor $T_\ab[g+h;\gamma]$ moves 
on the world line $\gamma$, such that

\beq
\text{G}_\ab[g+h]=8\pi T_\ab[g+h;\gamma]~.
\eeq

The particle motion is constrained by the Ricci Curbastro - Bianchi identity \cite{riccicurbastro1888, bianchi1902} \footnote{Toscano argues on the contributions by Ricci Curbastro and Bianchi to the identity equation \cite{toscano2000}.}, for which the first antisymmetric covariant derivative of the Riemann tensor vanishes. It implies $\overline{\nabla}_\beta \text{G}^\ab[\overline{g}]=0$ and thus the conservation of the local stress-energy tensor $\overline{\nabla}_\beta T^\ab[\overline{g}]=0$ where $\overline{\nabla}_\alpha$ is the covariant derivative in the total metric. Since $u^\beta\overline{\nabla}_\beta u^\alpha=0$, the world line is a geodesic of the $\overline{g}_\ab$ metric. 

In absence of an exact solution, we approximate $\overline{g}_\ab$ and $\gamma$. In this limit, we expand the Einstein tensor $\text{G}_\ab[g+h]$ around the vacuum solution for $h_\ab\sim\mathcal{O}(\varepsilon)$
\beq
\text{G}_\ab[g+h]=\text{G}_\ab[g]+\text{G}^{(1)}_\ab[g,h]+\text{G}^{(2)}_\ab[g,h]+\hdots~,
\label{einstein:tensor:expansion}
\eeq
where $\text{G}^{(n)}_\ab[g_\ab,h_\ab]\sim\mathcal{O}(\varepsilon^n)$. For the term of order 0, $h_\ab$ is zero and $g_\ab$ is the vacuum SD solution. The first order term is linear in $h_\ab$, while higher orders are not. Thus, at first perturbation order, $T_\ab[g;\gamma]\sim\mathcal{O}(\varepsilon)$ and $h_\ab$ is governed by the perturbed Einstein equations
\beq
\text{G}^{(1)}_\ab[g,h]=8\pi T_\ab[g;\gamma]+\mathcal{O}(\varepsilon^2)~,
\label{perturbed:EE}
\eeq
where $T_\ab[g;\gamma]$ is the stress-energy of the moving particle in the background metric
\beq
T_\ab[g;\gamma]=m_0\int_\gamma (-g)^{-1/2}u_\alpha u_\beta\delta^{(4)}\left(x-x_p(\tau)\right)d\tau~.
\label{Tab:gback}
\eeq
The particle is located on the world line $\gamma$ through the coordinates $x_p^\alpha(\tau)$; $u^\alpha$ is the four-velocity and $\tau$ the proper time associated to the background metric $g_\ab$; $\delta^{(4)}\left(x-x_p(\tau)\right)=\delta(x^0-x_p^0)\delta(x^1-x_p^1)\delta(x^2-x_p^2)\delta(x^3-x_p^3)$ is the four-dimensional Dirac distribution, and $g$ the determinant of the metric $g_\ab$.
The linear operator of the Einstein perturbed tensor is 
\beq
\begin{aligned}
\text{G}^{(1)}_\ab[g,h]=
&-\frac{1}2\nabla^\gamma\nabla_\gamma h_\ab 
+ \nabla_\beta\nabla^\gamma h_{\alpha\gamma}+\nabla_\alpha\nabla^\gamma h_{\beta\gamma}
-R_{\gamma\alpha\delta\beta}h^{\gamma\delta}
-\frac{1}2\nabla_\beta\nabla_\alpha h +\\
&R^\gamma_{\ \alpha}h_{\beta\gamma}+R^\gamma_{\ \beta}h_{\alpha\gamma}
-\frac{1}2g_\ab\left(\nabla^\delta\nabla^\gamma h_{\delta\gamma}-\nabla^\gamma\nabla_\gamma h\right)
-\frac{1}2h_\ab R
+\frac{1}2g_\ab h_{\gamma\delta}R^{\gamma\delta},
\end{aligned}
\label{G:linear}
\eeq
where $R^\alpha_{\ \beta\gamma\delta}$ is the Riemann tensor, $R_\ab=R^\gamma_{\ \alpha\gamma\beta}$ and $R=R^{\ \alpha}_\alpha$ the Ricci Curbastro \cite{riccicurbastro1888}, respectively tensor and scalar computed in the background metric $g_\ab$. Similarly, for the covariant derivative $\nabla_\alpha$ expressed in metric $g_\ab$. The term (\ref{G:linear}) is simplified to the extent that $g_\ab$ is a solution of the vacuum ($R_\ab=0$) 

\beq
\text{G}^{(1)}_\ab[g,h]=-\frac{1}2\nabla^\gamma\nabla_\gamma h_\ab 
+ \nabla_\beta\nabla^\gamma h_{\alpha\gamma}+\nabla_\alpha\nabla^\gamma h_{\beta\gamma}
-R_{\gamma\alpha\delta\beta}h^{\gamma\delta}
-\frac{1}2\nabla_\beta\nabla_\alpha h
-\frac{1}2g_\ab\left(\nabla^\delta\nabla^\gamma h_{\delta\gamma}-\nabla^\gamma\nabla_\gamma h\right)~.
\label{G:linear:vacuum}
\eeq

Equation (\ref{G:linear:vacuum}) may be further simplified by a gauge transformation; adopting the H gauge for which 
$\nabla_\beta\bar{h}^\ab=0$, Eq. (\ref{perturbed:EE}) becomes

\beq
g^{\gamma\delta}\nabla_\gamma\nabla_\delta\bar{h}^{\ab}+2R^{\ \alpha\ \beta}_{\gamma\ \delta\ }\bar{h}^{\gamma\delta}=-16\pi \delta T^{\ab}+\mathcal{O}(\varepsilon^2)~,
\label{field:eq:linear:lorenz}
\eeq
where $\bar{h}_\ab = {h}_\ab - 1/2\, {g}_\ab\, {h}$ is the perturbed trace-reversed field relative to the background metric $g_\ab$.
For the Ricci Curbastro - Bianchi identity satisfied at first order $\text{G}^{(1)}_\ab$, that is from Detweiler \cite{de05} 

\beq
\nabla^\alpha \text{G}^{(1)}_\ab[g,h]=0~, 
\eeq
the stress-energy tensor $T_\ab$ must be conserved in the background metric

\beq
\nabla^\alpha T_\ab[g;\gamma]\sim\mathcal{O}(\varepsilon^2)~, 
\label{T:conservation}
\eeq 
and $\gamma$ must be approximately a geodesic of the background spacetime. Thus the equation of motion remains

\beq
u^\alpha\nabla_\alpha u^\beta\sim\mathcal{O}(\varepsilon)~.
\eeq

For the perturbative analysis at second order, the tensor $T_\ab$ is to be conserved not in the background metric but in the first order perturbed metric. So for solving the field equations at second order, the stress-energy tensor must depend on the first order perturbed metric: the source of $n$-th order equation implies a $T_\ab$ at $(n-1)$-th order. 

%Equation (\ref{G:linear:vacuum}) can be written as 
%
%\beq
%h_{\ab;\sigma}^{~~~~;\sigma} - 2 \mathcal J_{(\alpha;\beta)} +2 R_{\sigma\alpha\rho\beta} h^{\sigma\rho} + h_{;\ab} 
%+g_{\ab} \left( \mathcal J_{\sigma}^{~;\sigma} -h_{;\sigma}^{~~;\sigma}\right) = -16 \pi T_{\ab}.
%\label{perturbed:EE2} 
%\eeq
%where $  \mathcal J_\alpha = h_{\ab}^{~~~;\beta}$.

Due to the spherical symmetry of the SD field, it is possible to split the perturbations in two different parity families, even and odd, \cite{rewh57,ze70c}. In a suitable spherical tensorial harmonics basis and in SD coordinates $x^\alpha=\{t,r,\theta,\phi\}$}, the even and odd perturbations are expressed by the following matrices

\beq
h^{e~\lm}_{\ab} = 
\left(
\begin{matrix}
f H_{0}^\lm Y^\lm
&  H_{1}^\lm Y^\lm
&  h_0^{(e)\lm}  Y_{,\theta}^\lm
& h_0^{(e)\lm} Y_{,\phi}^\lm \cr
sym 
&  f^{-1}H_{2}^\lm Y^\lm
&  h_1^{(e)\lm} Y_{,\theta}^\lm
& h_1^{(e)\lm}  Y_{,\phi}^\lm
\cr                        
sym
&  sym
&  
r^2 \left [ K^\lm  Y^\lm  + G^\lm  Y_{,\theta\theta}^\lm  
\right ] 
& r^2 G^\lm 
\left (Y_{,\theta\phi}^\lm  
- \cot\theta Y_{,\phi}^\lm  
\right ) 
\cr
sym 
&  sym 
&  sym 
&  r^2\sin^2\theta\left[ 
K^\lm +G^\lm \left(\frac{\ds Y_{,\phi\phi}^\lm }{\ds\sin^2\theta} 
+ \cot \theta Y_{,\theta}^\lm  
\right ) \right ] 
\end{matrix}
\right ), 
\label{eq:rwzeven}
\eeq

\beq
h^{o~\lm}_{\ab} = 
\left(
\begin{matrix}
0  
&  0 
&  - h_0^\lm \sec \theta Y_{,\phi}^\lm  
& h_0^\lm \sin \theta Y_{,\theta}^\lm  
\cr
sym 
&  0 
&  - h_1^\lm \sec\theta Y_{,\phi}^\lm  
& h_1^\lm \sin\theta Y_{,\theta}^\lm 
\cr                  
sym
&  sym
&  h_2^\lm  \sec\theta\left (Y_{,\theta\phi}^\lm  - \cot \theta Y_{,\phi}^\lm \right ) 
& {\ds \frac{1}{2}}h_2^\lm  \left (\sec \theta Y_{,\phi\phi}^\lm  + \cos\theta \cot\theta Y_{,\theta}^\lm  -  
\sin \theta Y_{,\theta\theta}^\lm \right ) 
\cr
sym 
&  sym 
&  sym 
& - h_2^\lm  \left (\sin \theta Y_{,\theta\phi}^\lm  - \cos\theta Y_{,\phi}^\lm \right )
\end{matrix}
\right ), 
\label{eq:rwzodd}
\eeq
{ where $f = (1-2M/r^2)$, $H_0$, $H_1$, $H_2$, $h^{(e)}_0$, $h^{(e)}_1$, $K$, $G$ and $h_0$, $h_1$, $h_2$ are functions of $(t,r)$, and $Y(\theta,\phi )$ represent scalar spherical harmonics. The RW gauge poses $h^{(e)}_0=h^{(e)}_1=G=0$ and $h_0=h_1=0$.}
The odd and even parity wave-functions are built through a combination of the perturbation functions (three odd and seven even parity perturbations). 
Incidentally, we deal in this paper only with modes $\ell \geq 2$ which are radiative. In general relativity, there is no contribution to gravitational radiation by monopole and dipole components.  

Different algebraic combinations have been used by different authors over time: Zerilli \cite{ze70c}, Moncrief \cite{mo74}, Cunningham, Price and Moncrief (CPM) \cite{cuprmo78}, Nagar and Rezzolla \cite{nare05}. We choose to work with the convention of Lousto and Price \cite{lopr97b}, equivalent to the CPM convention, up to a normalisation factor. Thus{, in the RW gauge}

\beq
\psi^\lm_{e}(t,r)=\frac{r}{\lambda+1}\left[K^\lm+\frac{rf}{\lambda r+3M}\left(H_2^\lm-r\frac{\partial K^\lm}{\partial r}\right)\right]
+ \frac{rf}{\lambda r+3M}\left[r^2\frac{\partial G^\lm}{\partial r}-2h_1^\lm\right]~,
\label{psi:e} 
\eeq

\beq
\psi^\lm_{o}(t,r)=\frac{r}{\lambda}\left[r^2\frac{\partial}{\partial r}\left(\frac{h_0^\lm}{r^2}\right)-\frac{\partial h_1^\lm}{\partial t}\right]~,
\label{psi:o}
\eeq
where {$\lambda=(\ell-1)(\ell+2)/2$}. Casting the energy-momentum tensor in spherical harmonics, App. (A), we get the RWZ wave-equation 

\beq
\left[-\partial^2_t+\partial^2_{r^*} - V^{\ell}_{e/o}(r)\right]\psi^{\ell m}_{e/o} (t,r) = S^{\ell m}_{e/o}(t,r)~,  
\label{eq:rwz}
\eeq
where $r^* = r + 2M\ln (r/2M - 1)$ is the tortoise coordinate. 
The two potentials $V^{\ell}_{e}(r)$ and $V^{\ell}_{o}(r)$ are given by
\beq
V^{\ell}_e(r) = 2 f \frac{\lambda^2(\lambda+1)r^3+3\lambda^2Mr^2+9\lambda M^2r+9M^3}{r^3(\lambda r+3M)^2 }~,\\
\label{veven}
\eeq
\beq
V^{\ell}_o(r) = 2 f \left(\frac{\lambda+1}{r^2}-\frac{3M}{r^3}\right)~.
\label{vodd}
\eeq

The source term (\ref{eq:rwz}) involves a Dirac delta $\delta$ and its distributional spatial derivative $\delta'\defeq d\delta/dr$. It is weighted by two functions of time $\mathcal{F}^{\ell m}_{e/o}$ and $\mathcal{G}^{\ell m}_{e/o}$ for even and odd modes 
\beq
S^{\ell m}_{e/o}(t,r) =  \mathcal{G}^{\ell m}_{e/o}(t)\delta\left(r - r_p(t)\right) + \mathcal{F}^{\ell m}_{e/o}(t)\delta'\left(r - r_p(t)\right)~;
\label{eq:rwz:source} 
\eeq
for the even modes

\beq
\mathcal{F}^{\ell m}_{e}(t)=-\frac{8\pi m_0}{\lambda+1}u^tr_pf(r_p)\frac{\dott{r}_p^2-f(r_p)^2}{\lambda r_p+3M}Y^{\lm\star}~,
\label{F:e}
\eeq
\beq
\begin{aligned}
\mathcal{G}^{\ell m}_{e}(t)=&\frac{8\pi m_0}{\lambda+1}u^t\left[
2\frac{r_p\dott{r}_pf(r_p)}{\lambda r_p+3M}\frac{d}{dt}Y^{\lm\star}
-\frac{r_pf(r_p)}{\lambda}\dott{\theta}_p\dott{\phi}_pX^{\lm\star}
-\frac{r_pf(r_p)}{2\lambda}\left(\dott{\theta}_p^2-\sin^2\theta_p\dott{\phi}_p^2\right)W^{\lm\star} + \right.\\
&\frac{r_p^2f(r_p)^2}{\lambda r_p+3M}\left(\dott{\theta}_p^2+\sin^2\theta_p\dott{\phi}_p^2\right)Y^{\lm\star}
+\frac{\dott{r}_p^2\left[(\lambda+1)(6r_pM+\lambda r_p^2)+3M^2\right]}{r_p(\lambda r_p+3M)^2}Y^{\lm\star} - \\
&\left.\frac{f(r_p)^2\left[r_p^2\lambda(\lambda+1)+6\lambda r_pM+15M^2\right]}{r_p(\lambda r_p+3M)^2}Y^{\lm\star}\right]~;\\
\end{aligned}
\label{G:e}
\eeq
while for the odd modes

\beq
\mathcal{F}^{\ell m}_{o}(t)=\frac{8\pi m_0}{\lambda(\lambda+1)}u^tr_p\left[\dott{r}_p^2-f(r_p)^2\right]\mathcal{A}^{\lm\star}~,
\label{F:o}
\eeq
\bea
\mathcal{G}^{\ell m}_{o}(t)=&-\frac{8\pi m_0}{\lambda(\lambda+1)}u^tr_p\Bigg\{\Bigg[r_p\frac{d}{dt}(u^t\dott{r}_p)+
&u^t\left(\dott{r}_p^2-f(r_p)\right)\Bigg]\mathcal{A}^{\lm\star}+\dott{r}_p\frac{d}{dt}\mathcal{A}^{\lm\star}\Bigg\}~,
\eea
where the asterisk notes complex conjugation, $u^t$ is the time component of the particle's four-velocity, { $r_p=r_p(t)$ is the particle's position at time $t$}, and "$\rescale[0.5]{\circ}$" means the total time derivative. The angular parts of the harmonics are evaluated at  $(\theta_p(t),\phi_p(t))$, and

\begin{align}
    &X^\lm(\theta,\phi)=2\left(\partial_\theta\partial_\phi-\cot\theta\partial_\phi\right)Y^\lm(\theta,\phi)~,\label{X}\\
    &W^\lm(\theta,\phi)=\left(\partial^2_\theta-\cot\theta\partial_\theta-\sin^{-2}\theta\partial^2_\phi\right)Y^\lm(\theta,\phi)
~,\label{W}\\
    &\mathcal{A}^{\lm}(\Theta,\Phi) = \left(\dot\Theta\sin^{-1}\Theta\partial_\phi-\sin\Theta\dot{\Phi}\partial_\theta\right)Y^\lm(\Theta,\Phi)~.\label{Acal}
\end{align}

The  highly singular nature of the source term in  Eq. (\ref{eq:rwz:source}) is responsible for the discontinuity of the wave-function at $r=r_p(t)$. The wave-function $\psi^\lm_{(e/o)}$ is otherwise piecewise continuous, and we distinguish its behaviour on the right and left hand side of the particle world line.

For the even parity, from Eq. (\ref{psi:e}) and the Hamiltonian constraint { Eq. (\ref{EE:1:e})} in App. (A), we express $K^\lm(t,r)$ and $H_2^\lm(t,r)$in terms of $\psi^\lm_{(e)}$ and the source { term} $T^{(i)\ell m}$, Eq. (\ref{coef:T})

\beq
K^\lm=\frac{6M^2+3M\lambda r+\lambda(\lambda+1)r^2}{r^2(\lambda r+3M)}\psi^\lm_{e} +f\frac{\partial}{\partial r}\psi^\lm_{e}-\frac{8\pi r^3}{(\lambda+1)(\lambda r+3M)}T^{(1)\lm}~,
\label{K:psi}
\eeq

\beq
\begin{aligned}
H_2^\lm&=-\frac{9M^3+9\lambda M^2r+3\lambda^2Mr^2+\lambda^2(\lambda+1)r^3}{r^2(\lambda r+3M)^2}\psi^\lm_{e}
+\frac{3M^2-\lambda Mr+\lambda r^2}{r(\lambda r+3M)}\frac{\partial}{\partial r}\psi^\lm_{e}+rf\frac{\partial^2}{\partial r^2}\psi^\lm_{e} +\\
&\frac{8\pi r^3(\lambda r^2(\lambda-2)+10\lambda rM-9rM+27M^2)}{(\lambda+1)rf(\lambda r+3M)^2)}T^{(1)\lm}
-\frac{8\pi r^4}{(\lambda+1)(\lambda r+3M)}\frac{\partial}{\partial r}T^{(1)\lm}~.
\end{aligned}
\label{H2:psi}
\eeq
Similarly, from { Eq. (\ref{EE:2:e})} and the partial time derivative of Eqs. (\ref{K:psi}, \ref{H2:psi}), we obtain $H_1^\lm(t,r)$
\beq
H_1^\lm=r\frac{\partial^2}{\partial r\partial t}\psi^\lm_{e}+\frac{\lambda r^2-3M\lambda r-3M^2}{rf(\lambda r+3M)}\frac{\partial}{\partial t}\psi^\lm_{e}
-\frac{8\pi r^5}{(\lambda+1)rf(\lambda r+3M)}\frac{\partial}{\partial t}T^{(1)\lm}\\
+\frac{4\sqrt{2}i\pi r^2}{(\lambda+1)}T^{(2)\lm}~, 
\label{H1:psi}
\eeq
and from { Eq. (\ref{EE:7:e})} $H_0^\lm(t,r)$
{
\beq
H_0^\lm(t,r)=H_2^\lm(t,r)+\frac{16\pi r^2}{\sqrt{2\lambda(\lambda+1)}}T^{(9)\lm}~.
\label{H0:psi}
\eeq
}

For the odd parity, Eqs. (\ref{EE:1:o}-\ref{EE:3:o}) provide $h_0^\lm(t,r)$ and $h_1^\lm(t,r)$ 

\beq
h_0^\lm=\frac{1}2f\frac{\partial}{\partial r}\left(r\psi^\lm_{o}\right)+\frac{4\pi r^3}{\lambda\sqrt{\lambda+1}}T^{(6)\lm}~,
\label{h0:psi}
\eeq\beq
h_1^\lm=\frac{1}2rf^{-1}\frac{\partial}{\partial t}\psi^\lm_{o}+\frac{4\pi i r^3}{\lambda\sqrt{\lambda+1}}T^{(7)\lm}~.
\label{h1:psi}
\eeq

{ Through the wave-function $\psi^{\ell m}_{e/o}$, we have access to the total energy $\dott{E}^{\infty}$ and angular momentum $\dott{L}^{\infty}$ radiated to infinity at the observer position ($r^*_{\text{obs}}\rightarrow + \infty$), and across the event horizon ($r^*_{\text{eh}}\rightarrow - \infty$), namely $\dott{E}^\text{eh}$ and $\dott{L}^\text{eh}$. The relations are} 
\beq
\dott{E}^{\text{eh},\infty}=\frac{1}{64}\sum_{\ell m}\frac{(\ell+2)!}{(\ell-2)!}\left[|\dott{\psi}^{\ell m}_e|^2+|\dott{\psi}^{\ell m}_o|^2\right]~,
\label{Edot}
\eeq

\beq
\dott{L}^{\text{eh},\infty}=\frac{im}{64}\sum_{\ell m}\frac{(\ell+2)!}{(\ell-2)!}\left[\bar{\psi}^{\ell m}_e\dott{\psi}^{\ell m}_e+\bar{\psi}^{\ell m}_o\dott{\psi}^{\ell m}_o\right]~,
\label{Ldot}
\eeq
{ again the dot indicating the total time  derivative}.

%%%%%%%%%%%%%%%%%%%%%%%%%%%%%%%%%%%%%%%%%%%%%%%%%%%%%%%%%%%%%%%%%%%%%%%%%%%%%%%%%%%%%%%%%%%%%%%%%%%%%%%%%%%%%%%%%%%%%%%%%%%%%%%%%%%%%%%%%%%%%%%%%%%%%%%%%%%%%%%%%%%%%%%%%%%%%%%%%%%%%%%%%%%%%%%%%%%%%%%%%%%%%%%%%%%%%%%%%%%%%%%%%%%%%%%%%%%%%%%%%%%%%%%%%%%%%%%%%%%%%%%%%%%%%%%%%%%%%%%%%%%%%%%%%%%%%%%%%%%%%%%%%%%%%%%%%%%%%%%%%%%%%%%%%%%%%%%%%%%%%%%%%%%%%%%%%%%%%%%%%%%%%%%%%%%%%%%%%%%%
\subsection{Equations of motion}
The 4-velocity vector $u^\mu=dz^\mu/d\tau$ of a point particle { moving in } an SD black hole geometry obeys the geodesic equation

\beq
u^\mu\nabla_\mu u^\nu=0~.
\label{geodeq}
\eeq

For motion { in the orbital plane} $\theta=\pi/2$, the following system must be satisfied, see Hagihara \cite{ha31}, 
Darwin \cite{da59,da61}, Chandrasekhar \cite{ch83}, and Cutler {\it et al.} \cite{cukepo94}.

\beq
\begin{aligned}
&u^t = \mE/f~,\\
&(u^r)^2 = \mE^2-f\left(1-\mL^2/r_p^2\right)~,\\
&u^\theta=0~,\\
&u^\phi=\mL/r_p^2~,
\end{aligned}
\eeq
where {$\mathcal E$} and {$\mathcal L$} are the two integrals of motion corresponding to the particle specific energy and angular momentum, respectively. Instead of searching for a solution $r(t)$, it is customary to refer to the single-valued parameter $\chi$, such that 
{
\beq
r_p(\chi)=\frac{pM}{1+e\cos(\chi)}~,
\label{partpos1}
\eeq
}
where $p$ and $e$ correspond to the { semi-latus} rectum and the eccentricity, respectively. For circular $e = 0$, and elliptic orbits $0<e<1$, the angle range is $0 < \chi < 2 \pi$; further, $e=1$ and $-\pi < \chi < \pi$ for parabolic orbits, $e<1$ and $-\arccos (-e^{-1})< \chi < \arccos (e^{-1})$ for hyperbolic orbits. 
The parameters $p, e$  are related to the constants of motion
{
\beq
\mE^2=\frac{(p-2)^2-4e^2}{p(p-3-e^2)},\ \ \mL^2=\frac{M^2p^2}{p-3-e^2}
\eeq }
and to the periastron and apoastron
{
\beq
r_\text{peri}=pM/(1+e),\ \ { r_\text{apo}} = pM/(1-e)\ .
\eeq
}
The particle position is now described by Eq. (\ref{partpos1}) and by 

\beq
\frac{d\chi}{dt}= \frac{(p-2-2e\cos\chi)(1+e\cos\chi)^2}{Mp^2} 
 \sqrt{\frac{p-6-2e\cos\chi}{(p-2)^2-4e^2}}~, 
\label{partpos2}
\eeq

\beq
\frac{d\phi_p}{dt}=\frac{(p-2-2e\cos\chi)(1+e\cos\chi)^2}{p^{3/2}M\sqrt{(p-2)^2-4e^2}}
\label{partpos3}~.
\eeq

The system { of Eqs. (\ref{partpos1},\ref{partpos2},\ref{partpos3}}) is not well defined for $p<6+2e$ that characterises the regime of unstable orbits. Only orbits { lying} outside the separatrix ($e\ge0, p\ge6+2e$) will be considered in this paper.

%%%%%%%%%%%%%%%%%%%%%%%%%%%%%%%%%%%%%%%%%%%%%%%%%%%%%%%%%%%%%%%%%%%%%%%%%%%%%%%%%%%%%%%%%%%%%%%%%%%%%%%%%%%%%%%%%%%%%%%%%%%%%%%%%%%%%%%%%%%%%%%%%%%%%%%%%%%%%%%%%%%%%%%%%%%%%%%%%%%%%%%%%%%%%%%%%%%%%%%%%%%%%%%%%%%%%%%%%%%%%%%%%%%%%%%%%%%%%%%%%%%%%%%%%%%%%%%%%%%%%%%%%%%%%%%%%%%%%%%%%%%%%%%%%%%%%%%%%%%%%%%%%%%%%%%%%%%%%%%%%%%%%%%%%%%%%%%%%%%%%%%%%%%%%%%%%%%%%%%%%%%%%%%%%%%%%%%%%%%%
\subsection{Jump conditions}\label{section:conditions:de:saut}

In App. (B), we recall some properties of the distributions, that shall be used herein. 
The discontinuity in $\psi(t,r)$ along the path at $r=r_p$  
is labelled of the \textit{first kind} because the values of $\psi$ on both sides of $r_p$ at constant $t$
\beq
\lim_{\epsilon\to0}\psi(t,r_p-\epsilon)\quad\text{and}\quad\lim_{\epsilon\to0}\psi(t,r_p+\epsilon)~,
\label{limit:left:right}
\eeq
are different but finite. The wave-function $\psi(t,r)$ belongs to the $C^\infty$ class in $\mathbb{R}^2\!\setminus\!\{(t,r)\ |\ r\!=\!r_p(t)\}$, while the functions $\psi^+(t,r)$ and $\psi^-(t,r)$ belong to the $C^\infty$ class and satisfy 

\begin{align}
&\psi^-(t,r)=\psi(t,r)\ \forall\ r\in]-\infty,r_p[~,\\
&\psi^-(t,r_p)=\lim_{\epsilon\to0}\psi(t,{ r_p}-\epsilon)~,
\label{psi:moins:rp}
\end{align}
\begin{align}
&\psi^+(t,r)=\psi(t,r)\ \forall\ r\in]r_p,+\infty[~,\\
&\psi^+(t,r_p)=\lim_{\epsilon\to0}\psi(t,{ r_p}+\epsilon)~.
\label{psi:plus:rp}
\end{align}

We thus can write $\psi$ as a formal distribution given by
\beq
\psi(t,r) = \psi^+ \mathcal{H}_+ + \psi^-\mathcal{H}_-~,  
\label{distrib:psi}
\eeq
where $\mathcal{H}_+\defeq\mathcal{H}(r-r_p(t))$ and $\mathcal{H}_-\defeq\mathcal{H}(r_p(t)-r)$, are Heaviside or step distributions. By using the properties described by Eqs. (\ref{t:derivative:delta},\ref{r:derivative:heaviside},\ref{t:derivative:heaviside}), the derivatives of Eq. (\ref{distrib:psi}) are

\beq
\partial_r\psi(t,r) = \partial_r\psi^+{\cal H}_+ + \partial_r\psi^-{\cal H}_- + \left(\psi^+ -\psi^-\right) \delta~,
\label{distrib:drpsi}
\eeq
{
\beq
\partial_t\psi(t,r) = \partial_t\psi^+{\cal H}_+ + \partial_t\psi^- {\cal H}_-- \dott{r}_p\left(\psi^+ -\psi^-\right)\delta~,
\label{distrib:dtpsi}
\eeq
}

\beq
\begin{aligned}
\partial^2_{r}\psi(t,r) = \partial^2_{r}\psi^+ {\cal H}_+  +  \partial^2_{r}\psi^- {\cal H}_- + 2\left( \partial_{r}\psi^+ - \partial_{r}\psi^- \right) \delta + \left(\psi^+ - \psi^-\right) \delta'~, 
\end{aligned}
\label{distrib:drrpsi}
\eeq
\beq
\begin{aligned}
\partial^2_{t}\psi(t,r) = \partial^2_{t}\psi^{+}{\cal H}_+  + { \partial_{t}^2 \psi^{-} {\cal H}_-} - 2\dott{r}_p\left(\partial_{t}\psi^+ -\partial_{t}\psi^-\right)\delta- \ddott{r}_p\left(\psi^+ -\psi^-\right)\delta +\dott{r}_p^2\left(\psi^+ -\psi^-\right)\delta'~,
\end{aligned}
\label{distrib:dttpsi}
\eeq
where $\delta \defeq \delta_{r_p} \defeq \delta(r-r_p(t))$ and $\delta' \defeq \delta'_{r_p}\defeq \partial_r\delta(r-r_p(t))$ are Dirac or delta distribution. Given the nature of the discontinuity (first kind), the jump at $\psi$ is simply the difference of the limit values of $\psi$ on the two sides of the path $r_p(t)$

\beq
\begin{aligned}
\jump{\psi}=\lim_{\epsilon\to0}\Big[\psi(t,r_p+\epsilon)-\psi(t,r_p-\epsilon)\Big]=\psi^+(t,r_p)-\psi^-(t,r_p)~, 
\label{jump:psi}
\end{aligned}
\eeq
which is ultimately a function of $t$ only. According to Eq. (\ref{distrib:drpsi}) for which { $\lim_{\epsilon\to0}\partial_r\psi(t,r_p\pm\epsilon)=\partial_r\psi^\pm(t,r_p)$}, the jumps of the partial derivatives of $\psi$ are written as 

\beq
\jump{\partial_r^n\partial_t^m\psi}=\partial_r^n\partial_t^m\psi^+(t,r_p)-\partial_r^n\partial_t^m\psi^-(t,r_p)~.
\label{jump:drdtpsi}
\eeq

The quantities $\jump{\partial_r^n\partial_t^m\psi}\ \forall\ n,m\in\mathbb{N}$ are the \textit {jump conditions} on the wave-function. Thus, by exploiting the properties of Eqs. (\ref{prop:delta}, \ref{prop:delta:prime}) and the definition of Eq. (\ref{jump:drdtpsi}), Eqs. (\ref{distrib:drpsi}-\ref{distrib:dttpsi}) are rewritten as

\beq
\partial_r\psi(t,r) = \partial_r\psi^+{\cal H}_+ + \partial_r\psi^-{\cal H}_- + \jump{\psi}\delta~,
\label{distrib:drpsi:reduced}
\eeq
\beq
\partial_t\psi(t,r) = \partial_t\psi^+{\cal H}_+ + \partial_t\psi^- {\cal H}_-- \dott{r}_p\jump{\psi}\delta~,
\label{distrib:dtpsi:reduced}
\eeq
\beq
\partial^2_{r}\psi(t,r) = \partial^2_{r}\psi^+ {\cal H}_+  +  \partial^2_{r}\psi^- {\cal H}_- + \jump{\partial_r\psi}\delta + \jump{\psi}\delta'~, 
\label{distrib:drrpsi:reduced}
\eeq
\beq
\begin{aligned}
\partial^2_{t}\psi(t,r) = \partial^2_{t}\psi^{+}{\cal H}_+  + \partial_{t}^2 \psi^{-} {\cal H}_- +
\left(\dott{r}_p^2\jump{\partial_r\psi}-\ddott{r}_p\jump{\psi}-2\dott{r}_p\frac{d}{dt}\jump{\psi}\right)\delta + \dott{r}_p^2\jump{\psi}\delta'~.
\end{aligned}
\label{distrib:dttpsi:reduced}
\eeq

Since $\partial^2_{r^*}=f^2\partial^2_r+ff'\partial_r$ and inserting Eqs. (\ref{distrib:psi}, \ref{distrib:drpsi:reduced}, \ref{distrib:drrpsi:reduced}, \ref{distrib:dttpsi:reduced}) into 
Eq. (\ref{eq:rwz}), the jump conditions are thus related to $\mathcal{F}$ and $\mathcal{G}$ of the source term

\beq
\begin{aligned}
\left[\Big(f(r_p)^2-\dott{r}_p^2\Big)\!\jump{\partial_r\psi}+\!\Big(\ddott{r}_p-f(r_p)f'(r_p)\Big)\!\jump{\psi}- 2\dott{r}_p\frac{d}{dt}\jump{\psi}\right]\delta+\Big(f(r_p)^2-\dott{r}_p^2\Big)\delta'\!=\!\mathcal{G}(t)\delta+\mathcal{F}(t)\delta'~.
\end{aligned}
\label{distrib:rwz2}
\eeq
We have recurred to $\psi^\pm$, solutions of the homogeneous equation $\mathcal{Z}[\psi^\pm]=0$, where $\mathcal{Z}\defeq-\partial^2_t+\partial^2_{r^*}-V(r)$, and to the identity $f^2\delta'=f(r_p)^2\delta'-2f(r_p)f'(r_p)\delta$ { used in }$f^2\partial^2_r\psi$.

%%%%%%%%%%%%%%%%%%%%%%%%%%%%%%%%%%%%%%%%%%%%%%%%%%%%%%%%%%%%%%%%%%%%%%%%%%%%%%%%%%%%%%%%%%%%%%%%%%%%%%%%%%%%%%%%%%%%%%%%%%%%%%%%%%%%%%%%%%%%%%%%%%%%%%%%%%%%%%%%%%%%%%%%%%%%%%%%%%%%%%%%%%%%%%%%%%%%%%%%%%%%%%%%%%%%%%%%%%%%%%%%%%%%%%%%%%%%%%%%%%%%%%%%%%%%%%%%%%%%%%%%%%%%%%%%%%%%%%%%%%%%%
%%%%%%%%%%%%%%%%%%%%%%%%%%%%%%%%%%%%%%%%%%%%%%%%%%%%%%%%%%%%%%%%%%%%%%%%%%%%%%%%%%%%%%%%%%%%%%%
\subparagraph*{0\textsuperscript{th} order jump.}
The jump condition $\jump{\psi}$ is identified by equating the coefficients of $\delta'$ on both sides of Eq. 
(\ref{distrib:rwz2})

\beq
\jump{\psi}=\frac{\mathcal{F}(t)}{f(r_p)^2-\dott{r}_p^2}~.
\label{jump}
\eeq

%%%%%%%%%%%%%%%%%%%%%%%%%%%%%%%%%%%%%%%%%%%%%%%%%%%%%%%%%%%%%%%%%%%%%%%%%%%%%%%%%%%%%%%%%%%%%%%%%%%%%%%%%%%%%%%%%%%%%%%%%%%%%%%%%%%%%%%%%%%%%%%%%%%%%%%%%%%%%%%%%%%%%%%%%%%%%%%%%%%%%%%%%%%%%%%%%%%%%%%%%%%%%%%%%%%%%%%%%%%%%%%%%%%%%%%%%%%%%%%%%%%%%%%%%%%%%%%%%%%%%%%%%%%%%%%%%%%%%%%%%%%%%
%%%%%%%%%%%%%%%%%%%%%%%%%%%%%%%%%%%%%%%%%%%%%%%%%%%%%%%%%%%%%%%%%%%%%%%%%%%%%%%%%%%%%%%%%%%%%%%
\subparagraph*{1\textsuperscript{st} order jumps.}
Similarly, $\jump{\partial_r\psi}$ is identified through the coefficients of $\delta$ 

\beq
\begin{aligned}
\jump{\partial_r\psi}=&\frac{1}{f(r_p)^2-\dott{r}_p^2}\left[\mathcal{G}(t)+\Big(f(r_p)f'(r_p)-\ddott{r}_p\Big)\jump{\psi}-
2\dott{r}_p\frac{d}{dt}\jump{\psi}\right]~.
\end{aligned}
\label{jumpdr}
\eeq

By using the relation on the total derivative, Eq. (\ref{tot:derivative}), we get

\beq
\jump{\partial_t\psi}=\frac{d}{dt}\jump{\psi}-\dott{r}_p\jump{\partial_r\psi}~.
\label{jumpdt}
\eeq

%%%%%%%%%%%%%%%%%%%%%%%%%%%%%%%%%%%%%%%%%%%%%%%%%%%%%%%%%%%%%%%%%%%%%%%%%%%%%%%%%%%%%%%%%%%%%%%%%%%%%%%%%%%%%%%%%%%%%%%%%%%%%%%%%%%%%%%%%%%%%%%%%%%%%%%%%%%%%%%%%%%%%%%%%%%%%%%%%%%%%%%%%%%%%%%%%%%%%%%%%%%%%%%%%%%%%%%%%%%%%%%%%%%%%%%%%%%%%%%%%%%%%%%%%%%%%%%%%%%%%%%%%%%%%%%%%%%%%%%%%%%%%
%%%%%%%%%%%%%%%%%%%%%%%%%%%%%%%%%%%%%%%%%%%%%%%%%%%%%%%%%%%%%%%%%%%%%%%%%%%%%%%%%%%%%%%%%%%%%%%
\subparagraph*{2\textsuperscript{nd} order jumps.}
Taking into account that $\mathcal{Z}[\psi^\pm]=0$, and $\mathcal{Z}[\psi^+-\psi^-]=0$ in the limit $r\to r_p$, the homogeneous RWZ equation becomes

\beq
-\jump{\partial_t^2\psi}+f(r_p)^2\jump{\partial_r^2\psi}+f(r_p)f'(r_p)\jump{\partial_r\psi}-V(r_p)\jump{\psi}=0~.
\label{jump:rwz}
\eeq

Injecting 

\beq
\begin{aligned}
\jump{\partial_t^2\psi}&=\frac{d}{dt}\jump{\psi}-\dott{r}_p\jump{\partial_r\partial_t\psi}=\frac{d}{dt}\jump{\psi}-\dott{r}_p\jump{\partial_r\psi}+\dott{r}_p^2\jump{\partial_r^2\psi}~ ,
\end{aligned}
\label{compo:derivatives:tt}
\eeq

into Eq. (\ref{jump:rwz}), we get the second spatial derivative
\beq
\begin{aligned}
\jump{\partial_r^2\psi}=&\frac{1}{f(r_p)^2-\dott{r}_p^2} \left[\frac{d}{dt}\jump{\partial_t\psi} - f(r_p)f'(r_p)\jump{\partial_r\psi} -
\dott{r}_p\frac{d}{dt}\jump{\partial_r\psi} + V(r_p)\jump{\psi}\right]~.
\end{aligned}
\label{jumpdrr}
\eeq

The jumps on the second time derivative and mixed derivative are expressed respectively by

\begin{align}
&\jump{\partial_t^2\psi}=\jump{\partial_{r^*}^2\psi}-V(r_p)\jump{\psi}~,
\label{jumpdtt}\\
&\jump{\partial_t\partial_r\psi}=\jump{\partial_r\partial_t\psi}=\frac{d}{dt}\jump{\partial_r\psi}-\dott{r}_p\jump{\partial_r^2\psi}~,
\label{jumpdrt}
\end{align}
where $\jump{\partial_{r^*}^2\psi}=f(r_p)^2\jump{\partial_r^2\psi}+f(r_p)f'(r_p)\jump{\partial_r\psi}$.

%%%%%%%%%%%%%%%%%%%%%%%%%%%%%%%%%%%%%%%%%%%%%%%%%%%%%%%%%%%%%%%%%%%%%%%%%%%%%%%%%%%%%%%%%%%%%%%%%%%%%%%%%%%%%%%%%%%%%%%%%%%%%%%%%%%%%%%%%%%%%%%%%%%%%%%%%%%%%%%%%%%%%%%%%%%%%%%%%%%%%%%%%%%%%%%%%%%%%%%%%%%%%%%%%%%%%%%%%%%%%%%%%%%%%%%%%%%%%%%%%%%%%%%%%%%%%%%%%%%%%%%%%%%%%%%%%%%%%%%%%%%%%
%%%%%%%%%%%%%%%%%%%%%%%%%%%%%%%%%%%%%%%%%%%%%%%%%%%%%%%%%%%%%%%%%%%%%%%%%%%%%%%%%%%%%%%%%%%%%%%
\subparagraph*{3\textsuperscript{rd} order jumps.}
{ The first space partial derivative of $\mathcal{Z}[\psi^+-\psi^-]=0$ for the limit $r\rightarrow r_p$ leads to} 

\beq
\begin{aligned}
-\jump{\partial_r\partial_t^2\psi}+f^2\jump{\partial_r^3\psi}+3ff'\jump{\partial_r^2\psi}+
\left(f'^2+ff''-V\right)\jump{\partial_r\psi}-V'\jump{\psi}=0~.
\end{aligned}
\label{jumpdr:rwz}
\eeq

Since 

\beq
\jump{\partial_r\partial_t^2\psi}=\frac{d}{dt}\jump{\partial_r\partial_t\psi}-\dott{r}_p\frac{d}{dt}\jump{\partial_r^2\psi}+\dott{r}_p^2\jump{\partial_r^3\psi}~.
\label{compo:derivatives:rtt}
\eeq
by inserting Eq. (\ref{compo:derivatives:rtt}) into Eq. (\ref{jumpdr:rwz}), we get

\beq
\begin{aligned}
\jump{\partial_r^3\psi}=&\frac{1}{f^2-\dott{r}_p^2}
\left[
\frac{d}{dt}\jump{\partial_r\partial_t\psi} 
-\dott{r}_p\frac{d}{dt}\jump{\partial_r^2\psi}
- \left(f'^2+ff''-V\right)\jump{\partial_r\psi}
-3ff'\jump{\partial_r^2\psi} 
-V'\jump{\psi}
\right]~.
\end{aligned}
\label{jumpdrrr}
\eeq

The other mixed derivative is obtained through a proper combination

\beq
\jump{\partial_t\partial_r^2\psi}=\frac{d}{dt}\jump{\partial_r^2\psi}-\dott{r}_p\jump{\partial_r^3\psi}~, 
\label{jumpdtrr}
\eeq

while { $\jump{\partial_r\partial_t^2\psi}$ is determined by inverting Eq. (\ref{jumpdr:rwz}). In general, if we { get} an explicit solution for $\jump{\partial_r^{m+2n}}$, any jump of the type $\jump{\partial_r^m\partial_t^{2n}\psi}$ is connected to the wave-equation through the homogeneous operator $\cal{Z}$.}

%%%%%%%%%%%%%%%%%%%%%%%%%%%%%%%%%%%%%%%%%%%%%%%%%%%%%%%%%%%%%%%%%%%%%%%%%%%%%%%%%%%%%%%%%%%%%%%%%%%%%%%%%%%%%%%%%%%%%%%%%%%%%%%%%%%%%%%%%%%%%%%%%%%%%%%%%%%%%%%%%%%%%%%%%%%%%%%%%%%%%%%%%%%%%%%%%%%%%%%%%%%%%%%%%%%%%%%%%%%%%%%%%%%%%%%%%%%%%%%%%%%%%%%%%%%%%%%%%%%%%%%%%%%%%%%%%%%%%%%%%%%%%
%%%%%%%%%%%%%%%%%%%%%%%%%%%%%%%%%%%%%%%%%%%%%%%%%%%%%%%%%%%%%%%%%%%%%%%%%%%%%%%%%%%%%%%%%%%%%%%
\subparagraph*{4\textsuperscript{th} order jumps.}

The double space derivation of $\mathcal{Z}[\psi^+-\psi^-]=0$ with the limit 
for $r\rightarrow r_p$ leads to 

\beq
\begin{aligned}
-\jump{\partial^2_r\partial_t^2\psi}+f^2\jump{\partial^4_r\psi}+2ff'\jump{\partial^3_r\psi}+
\Big[4\left(f'^2+ff''\right)-V\Big]\jump{\partial^2_r\psi}+
\Big(3f'f''+ff'''-2V'\Big)\jump{\partial_r\psi}-V'\jump{\psi}=0~.
\end{aligned} 
\label{jumpdrr:rwz}
\eeq

Since

\beq
\jump{\partial^2_r\partial_t^2\psi}=\frac{d}{dt}\jump{\partial_r^2\partial_t^{\ }\psi}-\dott{r}_p\frac{d}{dt}\jump{\partial_r^3\psi}+\dott{r}_p^2\jump{\partial_r^4\psi}~,
\label{compo:derivatives:rrtt}
\eeq
{ by inserting Eq. (\ref{compo:derivatives:rrtt}) into Eq. (\ref{jumpdrr:rwz}), we get}

\beq
\begin{aligned}
\jump{\partial_r^4\psi}=&\frac{1}{f^2-\dott{r}_p^2}
\left\{
\frac{d}{dt}\jump{\partial_r^2\partial_t\psi} 
-\dott{r}_p\frac{d}{dt}\jump{\partial_r^3\psi}
2ff'\jump{\partial_r^3\psi}
\Big[4\left(f'^2+ff''\right)-V\Big]\jump{\partial_r^2\psi}
-\Big(3f'f''+ff''-2V'\Big)\jump{\partial_r\psi}
-V'\jump{\psi}
\right\}~,
\end{aligned}
\label{jumpdrrrr}
\eeq

then

\beq
\jump{\partial_t\partial_r^3\psi}=\frac{d}{dt}\jump{\partial_r^3\psi}-\dott{r}_p\jump{\partial_r^4\psi}~.
\label{jumpdtrrr}
\eeq

The jump $\jump{\partial_t^2\partial_r^2\psi}$ is computed when solving Eq. (\ref{jumpdrr:rwz}). The jump $\jump{\partial_t^3\partial_r\psi}$ is obtained by differentiating $\mathcal{Z}[\psi^+-\psi^-]=0$ with respect to $r$ first and to $t$ later, and finally taking the limit 
$r\rightarrow r_p$. 

%%%%%%%%%%%%%%%%%%%%%%%%%%%%%%%%%%%%%%%%%%%%%%%%%%%%%%%%%%%%%%%%%%%%%%%%%%%%%%%%%%%%%%%%%%%%%%%%%%%%%%%%%%%%%%%%%%%%%%%%%%%%%%%%%%%%%%%%%%%%%%%%%%%%%%%%%%%%%%%%%%%%%%%%%%%%%%%%%%%%%%%%%%%%%%%%%%%%%%%%%%%%%%%%%%%%%%%%%%%%%%%%%%%%%%%%%%%%%%%%%%%%%%%%%%%%%%%%%%%%%%%%%%%%%%%%%%%%%%%%%%%%%
%%%%%%%%%%%%%%%%%%%%%%%%%%%%%%%%%%%%%%%%%%%%%%%%%%%%%%%%%%%%%%%%%%%%%%%%%%%%%%%%%%%%%%%%%%%%%%%
\subparagraph*{$N > 2$\textsuperscript{nd} order jumps.  }

{ In general, by} knowing the jumps at $N-1$ order, those at order $N$ are computed with the following recipe:

\begin{itemize}
\item Take the limit for $r\rightarrow r_p$ of the $N\!-\!2$ spatial derivative of $\mathcal{Z}[\psi^+-\psi^-]=0$
\beq
\jump{\partial_r^{N-2}\partial_t^2\psi} = \sum_{k=0}^ND_{(N,k)}(r_p)\jump{\partial_r^k\psi}~.
\label{jumpdrN:rwz}
\eeq
\item{Express $\jump{\partial_r^{N-2}\partial_t^2\psi}$ as a function of the previously computed jumps
\beq
\jump{\partial^{N-2}_r\partial_t^2\psi}=\frac{d}{dt}\jump{\partial_r^{N-2}\partial_t^{2}\psi}-\dott{r}_p\frac{d}{dt}\jump{\partial_r^{N-1}\psi}+\dott{r}_p^2\jump{\partial_r^{N}\psi}~. 
\label{compo:derivatives:rNtt}
\eeq
}
\item Inject Eq. (\ref{compo:derivatives:rNtt}) into Eq. (\ref{jumpdrN:rwz})
\beq
\begin{aligned}
\jump{\partial_r^{N}\psi}=&\frac{1}{D_{(N,N)}-\dott{r}_p}\left[
\frac{d}{dt}\jump{\partial_r^{N-2}\partial_t^{\ }\psi}
-\dott{r}_p\frac{d}{dt}\jump{\partial_r^{N-1}\psi}
-\sum_{k=0}^{N-1}D_{(N,k)}(r_p)\jump{\partial_r^k\psi}
\right]~.
\end{aligned}
\eeq
\item The explicit form of $\jump{\partial_r^{N-2}\partial_t^2\psi}$ is obtained via Eq. (\ref{compo:derivatives:rNtt}) or by reversing Eq. (\ref{jumpdrN:rwz}).
\item Mixed derivatives of the type {$\jump{\partial_r^m\partial_t^{2n}\psi}$} are directly related to the jumps $\jump{\partial_r^k\psi}$ for $n,m\in\mathbb{N}$ such that $2n+m=N$ { and for $k\leq N$ because} in the homogeneous case, the operator $\partial_t^{2n}$ can be seen as a combination involving spatial derivatives only 

\beq
\partial_t^{2n}=\underbrace{(\partial_{r^*}^2-V)\circ(\partial_{r^*}^2-V)\circ\cdots\circ(\partial_{r^*}^2-V)}_{\text{n times}}
\eeq
{ where $\circ$ is the composition operator.} 
\item Mixed derivatives of the type {$\jump{\partial_r^m\partial_t^{2n+1}\psi}$}, for $n,m\in\mathbb{N}$ such that $2n+m+1=N$, are directly related to the jumps $\jump{\partial_r^k\partial_t^{\ }\psi}$ for $k\leq N-1$ and $\jump{\partial_r^k\psi}$ for $k\leq N$
\end{itemize}

In App. (C), the jump conditions appear in explicit form. 

%%%%%%%%%%%%%%%%%%%%%%%%%%%%%%%%%%%%%%%%%%%%%%%%%%%%%%%%%%%%%%%%%%%%%%%%%%%%%%%%%%%%%%%%%%%%%%%%%%%%%%%%%%%%%%%%%%%%%%%%%%%%%%%%%%%%%%%%%%%%%%%%%%%%%%%%%%%%%%%%%%%%%%%%%%%%%%%%%%%%%%%%%%%%%%%%%%%%%%%%%%%%%%%%%%%%%%%%%%%%%%%%%%%%%%%%%%%%%%%%%%%%%%%%%%%%%%%%%%%%%%%%%%%%%%%%%%%%%%%%%%%%%
%%%%%%%%%%%%%%%%%%%%%%%%%%%%%%%%%%%%%%%%%%%%%%%%%%%%%%%%%%%%%%%%%%%%%%%%%%%%%%%%%%%%%%%%%%%%%%%
\subsection{Numerical implementation}
\label{section:numerical:implementation}
\subsubsection{Domain, boundary and initial conditions}

\begin{figure}[h!]
\centering
\includegraphics[width=0.6\linewidth]{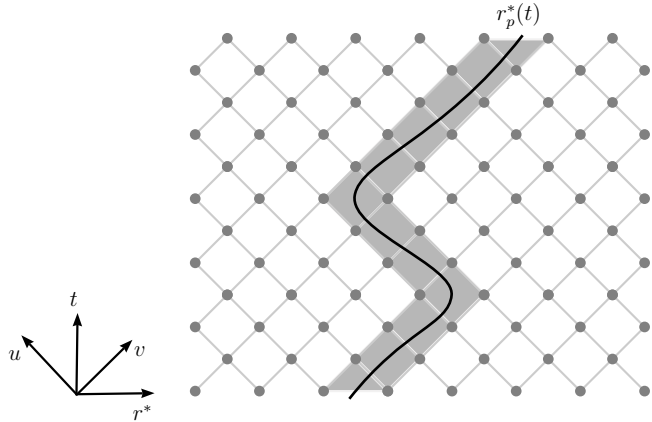}
\caption{Discretised numerical spacetime}
\label{fig01}
\end{figure}

\begin{figure}[h!]
\centering
\includegraphics[width=0.3\linewidth]{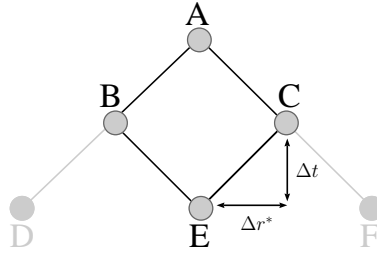}
\caption{Geometry of an empty cell, {\it i.e.} never occupied by the particle during its motion.}
\label{fig02}
\end{figure}

Following previous work \cite{lopr97b,mapo02}, we consider a staggered numerical grid  where two nodes are separated by { $\Delta r^*$ }and $\Delta t$, in space and time respectively, Figs. (\ref{fig01},\ref{fig02}). The null line for the retarded time is $u=t-r^*$, while $v=t+r^*$ for the advanced time.

The support of the source term is localised solely on the world line. Two kinds of numerical cells appear, Fig. (\ref{fig01}): (i) the cells, not crossed by the world line, and dealt with the homogeneous equation \cite{lopr97b,mapo02}; (ii) the cells crossed by the particle path where the solution is discontinuous due to the source term, and dealt with the jump conditions.

%%%%%%%%%%%%%%%%%%%%%%%%%%%%%%%%%%%%%%%%%%%%%%%%%%%%%%%%%%%%%%%%%%%%%%%%%%%%%%%%%%%%%%%%%%%%%%%%%%%%%%%%%%%%%%%%%%%%%%%%%%%%%%%%%%%%%%%%%%%%%%%%%%%%%%%%%%%%%%%%%%%%%%%%%%%%%%%%%%%%%%%%%%%%%%%%%%%%%%%%%%%%%%%%%%%%%%%%%%%%%%%%%%%%%%%%%%%%%%%%%%%%%%%%%%%%%%%%%%%%%%%%%%%%%%%%%%%%%%%%%%%%%
%%%%%%%%%%%%%%%%%%%%%%%%%%%%%%%%%%%%%%%%%%%%%%%%%%%%%%%%%%%%%%%%%%%%%%%%%%%%%%%%%%%%%%%%%%%%%%%

\begin{figure}[h!]
\centering
\includegraphics[width=0.60\linewidth]{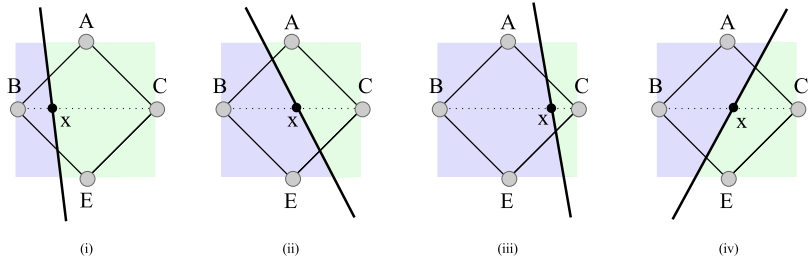}
\caption{First order mesh for the cells crossed by the particle. The regions $\Omega_+$ and $\Omega_-$ refer to the points "on the right" and "on the left" of the trajectory, respectively.}
\label{fig03}
\end{figure}

\begin{figure}[h!]
\centering
\includegraphics[width=0.6\linewidth]{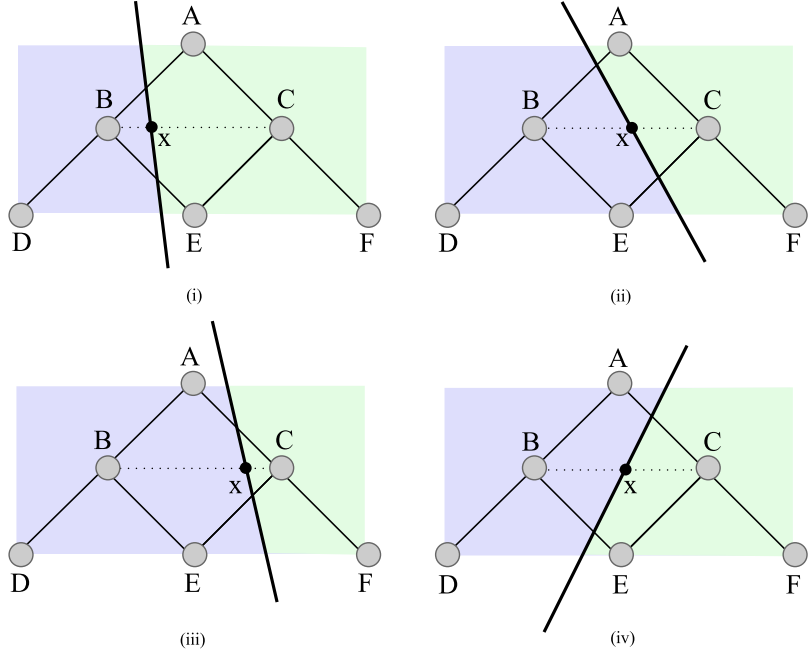}
\caption{Second order mesh for the cells crossed by the particle. The number of grid nodes raises to six, though the cases remain four.}
\label{fig04}
\end{figure}

\begin{figure}[h!]
\centering
\includegraphics[width=0.4\linewidth]{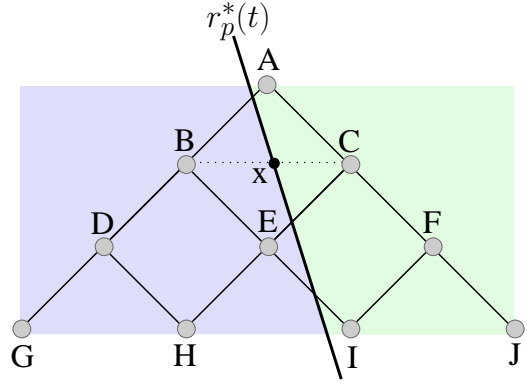}
\caption{Fourth order mesh for the cells crossed by the particle. The number of grid nodes raises to ten. For generic orbits the world line crosses the stencil in eight possible ways, among which the one shown here.}
\label{fig05}
\end{figure}

In our finite element integration scheme,  the field value at the top of the filled cell, point A, Figs. (\ref{fig03}-\ref{fig05}), is drawn from known and causally related quantities { and from the information on junctions provided by the jump conditions.} 
{ We use an $r^*$-grid ranging between $-2500$ M and $2500 $ M for an observer located at $r^*\!=\!r^*_\text{obs} = 1500$ M. 
The grid is bound by a spacelike hypersurface $t=0$ which contains the initial gravitational content of our system; on the left by the excision hypersurface taken at $r_\text{min}/2M=1+\epsilon$ (we typically take $r^*_\text{min}/2M=-1500$), and on the right at $r^*_\text{max}/2M=1500>r^*_\text{obs}$, consistent with a far observer approximation.} We solve the problem of boundary conditions by using a large grid, thereby avoiding any spurious reflections in the time window explored by the signal. The boundaries fulfill the conditions: $r^*_\text{min}\leq r^* \leq r^*_\text{max}$,  and $0 \leq t \leq t_{end}$. 
Alternatively, it would be conceivable to use hyperboloidal surfaces and a compactification of the radial coordinaten as proposed by Zengino\u{g}lu \cite{ze10}. 

Conversely to the more difficult problem of radial fall, initial conditions are not relevant for periodic orbits. Indeed, non-physical initial data are smoothed out, whenever sufficient time for their propagation is assigned. For an observer at infinity and various types of 
initial data, the time convergence towards the physical solution is fully satisfactory. In { closed} orbits, { a} period is often sufficient to avoid the propagation of the spurious initial radiation. We pose the following null initial { conditions} 
{
\beq
\psi(r,0)=d\psi/dt(r,0)=0~.
\eeq
}

%%%%%%%%%%%%%%%%%%%%%%%%%%%%%%%%%%%%%%%%%%%%%%%%%%%%%%%%%%%%%%%%%%%%%%%%%%%%%%%%%%%%%%%%%%%%%%%%%%%%%%%%%%%%%%%%%%%%%%%%%%%%%%%%%%%%%%%%%%%%%%%%%%%%%%%%%%%%%%%%%%%%%%%%%%%%%%%%%%%%%%%%%%%%%%%%%%%%%%%%%%%%%%%%%%%%%%%%%%%%%%%%%%%%%%%%%%%%%%%%%%%%%%%%%%%%%%%%%%%%%%%%%%%%%%%%%%%%%%%%%%%%%
%%%%%%%%%%%%%%%%%%%%%%%%%%%%%%%%%%%%%%%%%%%%%%%%%%%%%%%%%%%%%%%%%%%%%%%%%%%%%%%%%%%%%%%%%%%%%%%
\subsubsection{Notation}

\begin{figure}[h!]
\centering
\includegraphics[width=0.6\linewidth]{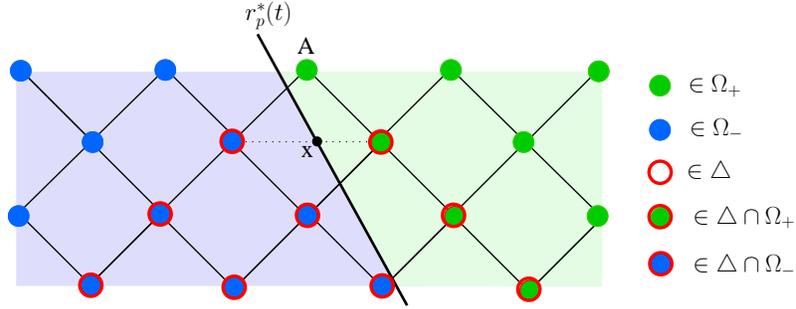}
\caption{{ Few cells near the world-line (see the text on notation for point labelling).}}
\label{fig06}
\end{figure}

We define, Fig. (\ref{fig06}),

\begin{itemize}
\item $\psi_i$ the value of $\psi$ taken at the grid point $i$ of coordinates $(t_i, r^*_i)$,
\item $V_i$ the value of the potential $V$ taken at the grid point $i$ of coordinates $(t_i, r^*_i)$,
\item $\jump{\partial_{r^*}^n\partial_t^m\psi}_\text{x}$ the jump conditions taken at a trajectory point $\text{x}$ of coordinates $(t_\text{x}, r^*_\text{x}\!=\!r_p^*(t_x))$,
\item $\partial^p_{r*}\partial^q_t\psi^\pm_\text{x}=\lim_{\epsilon\to0}\partial^p_{r*}\partial^q_t\psi^\pm\left(r_p(t_\text{x})\pm\epsilon,t_\text{x}\right)$,
\item $\Omega_+\!=\!\left\{(t_i,r^*_i)\ |\ r^*_i>r^*_p(t_i)\right\}$, all grid points "on the right" of the world line,
\item $\Omega_-\!=\!\left\{(t_i,r^*_i)\ |\ r^*_i<r^*_p(t_i)\right\}$, all grid points "on the left" of the world line,
\item $\psi(\Omega_\pm)\!=\!\left\{\psi_i\ |\ i\in\Omega_\pm\right\}$; thus, by definition, $\psi^+_i\in\psi(\Omega_+)$ and $\psi^-_i\in\psi(\Omega_-)$,
\item $\triangle\!=\!\left\{(t_i, r^*_i)\ |\ t_i<t_A,|r^*_i/t_i|<1\right\}$ the set of points belonging to the past light cone of the upper cell node $A$, {\it i.e.}, at first order, $\triangle\!=\!\{B,C,E\}$, Fig. (\ref{fig03}); at second order, $\triangle\!=\!\{B,C,D,E,F\}$, Fig. (\ref{fig04}); at fourth order, $\triangle\!=\!\{B,C,D,E,F,G,H,I,J\}$, Fig. (\ref{fig05}), 
\item { $h=\Delta r^*=\Delta t$},  
\item { $\Psi_i=V_i\psi_i$}.
\end{itemize}

%%%%%%%%%%%%%%%%%%%%%%%%%%%%%%%%%%%%%%%%%%%%%%%%%%%%%%%%%%%%%%%%%%%%%%%%%%%%%%%%%%%%%%%%%%%%%%%%%%%%%%%%%%%%%%%%%%%%%%%%%%%%%%%%%%%%%%%%%%%%%%%%%%%%%%%%%%%%%%%%%%%%%%%%%%%%%%%%%%%%%%%%%%%%%%%%%%%%%%%%%%%%%%%%%%%%%%%%%%%%%%%%%%%%%%%%%%%%%%%%%%%%%%%%%%%%%%%%%%%%%%%%%%%%%%%%%%%%%%%%%%%%%
%%%%%%%%%%%%%%%%%%%%%%%%%%%%%%%%%%%%%%%%%%%%%%%%%%%%%%%%%%%%%%%%%%%%%%%%%%%%%%%%%%%%%%%%%%%%%%%
\subsubsection{Algorithm}

%%%%%%%%%%%%%%%%%%%%%%%%%%%%%%%%%%%%%%%%%%%%%%%%%%%%%%%%%%%%%%%%%%%%%%%%%%%%%%%%%%%%%%%%%%%%%%%%%%%%%%%%%%%%%%%%%%%%%%%%%%%%%%%%%%%%%%%%%%%%%%%%%%%%%%%%%%%%%%%%%%%%%%%%%%%%%%%%%%%%%%%%%%%%%%%%%%%%%%%%%%%%%%%%%%%%%%%%%%%%%%%%%%%%%%%%%%%%%%%%%%%%%%%%%%%%%%%%%%%%%%%%%%%%%%%%%%%%%%%%%%%%%
%%%%%%%%%%%%%%%%%%%%%%%%%%%%%%%%%%%%%%%%%%%%%%%%%%%%%%%%%%%%%%%%%%%%%%%%%%%%%%%%%%%%%%%%%%%%%%%
\paragraph{Empty cells.}
The evolution of $\psi$ is mostly governed by the solution of the homogeneous RWZ equation as the source occupies only one cell per time step.

%%%%%%%%%%%%%%%%%%%%%%%%%%%%%%%%%%%%%%%%%%%%%%%%%%%%%%%%%%%%%%%%%%%%%%%%%%%%%%%%%%%%%%%%%%%%%%%%%%%%%%%%%%%%%%%%%%%%%%%%%%%%%%%%%%%%%%%%%%%%%%%%%%%%%%%%%%%%%%%%%%%%%%%%%%%%%%%%%%%%%%%%%%%%%%%%%%%%%%%%%%%%%%%%%%%%%%%%%%%%%%%%%%%%%%%%%%%%%%%%%%%%%%%%%%%%%%%%%%%%%%%%%%%%%%%%%%%%%%%%%%%%%
%%%%%%%%%%%%%%%%%%%%%%%%%%%%%%%%%%%%%%%%%%%%%%%%%%%%%%%%%%%%%%%%%%%%%%%%%%%%%%%%%%%%%%%%%%%%%%%
\subparagraph{2\textsuperscript{nd} order empty cells.}

\begin{figure}[h!]
\centering
\includegraphics[width=0.4\linewidth]{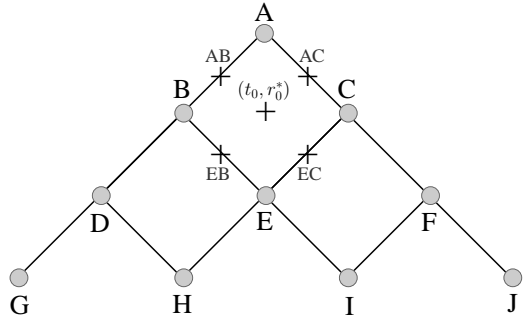}
\caption{The stencil used for solving the homogeneous RWZ equation. The points { BE, CE, AB, AC} indicated by the crosses are not grid points but extra points for the 4\textsuperscript{th} order scheme, Eq. (\ref{scheme:empty:4th}).}
\label{fig07}
\end{figure}

We use the classical finite difference scheme,  Figs. (\ref{fig02},\ref{fig07}), \cite{lopr97b, mapo02}. 
It is useful to introduce the Eddington-Finkelstein coordinates $(u,v)$ \cite{ed24,fi58} connected to $(t,r^*)$ coordinates via a bi-continuous application $\mu$ such that

\beq
\mu:\binom{t}{r^*}\mapsto\binom{u=t-r^*}{v=t+r^*}~,
\eeq
\beq
\mu^{-1}:\binom{u}{v}\mapsto\binom{t=(v+u)/2}{r^*=(v-u)/2}~.
\eeq

{ The integration of the homogeneous RWZ equation} on the whole cell leads to 
\beq
\begin{aligned}
\int_\text{cell}[\partial^2_{r^*}-\partial^2_t]\psi dr^*dt=-\int_\text{cell}4\partial_u\partial_v\psi dudv
=-4[\psi_A-\psi_B-\psi_C+\psi_E]~,
\end{aligned}
\eeq

while the potential is approximated by
\beq
\int_\text{cell}V\psi dr^*dt\approx h^2V_A[\psi_A+\psi_B+\psi_C+\psi_E]~.
\eeq

The value at the upper cell node is given by
\beq
\psi_A =\psi_E+(\psi_B+\psi_C)\left(1-\frac{1}2h^2V_A\right)+\mathcal{O}(h^4)~.
\label{scheme:empty:2nd}
\eeq

%%%%%%%%%%%%%%%%%%%%%%%%%%%%%%%%%%%%%%%%%%%%%%%%%%%%%%%%%%%%%%%%%%%%%%%%%%%%%%%%%%%%%%%%%%%%%%%%%%%%%%%%%%%%%%%%%%%%%%%%%%%%%%%%%%%%%%%%%%%%%%%%%%%%%%%%%%%%%%%%%%%%%%%%%%%%%%%%%%%%%%%%%%%%%%%%%%%%%%%%%%%%%%%%%%%%%%%%%%%%%%%%%%%%%%%%%%%%%%%%%%%%%%%%%%%%%%%%%%%%%%%%%%%%%%%%%%%%%%%%%%%%%
%%%%%%%%%%%%%%%%%%%%%%%%%%%%%%%%%%%%%%%%%%%%%%%%%%%%%%%%%%%%%%%%%%%%%%%%%%%%%%%%%%%%%%%%%%%%%%%
\subparagraph{4\textsuperscript{th} order empty cells.}

We use the scheme designed in \cite{ha07}, derived from Lousto \cite{lo05}. Like for Eq. (\ref{scheme:empty:2nd}), the integrated wave-operator over the cell is drawn from the { four nodes of the cell}. 
The integration of { the potential} is achieved by applying the Simpson's rule for the double integral

\beq
\iint_\text{cell} \Psi dr^*dt=\left(\frac{h}{3}\right)^2\!\!\Big[\Psi_A+\Psi_B+\Psi_E+\Psi_C+16\Psi_0+\left(\Psi_{BE}+\Psi_{AB}+\Psi_{CE}+\Psi_{AC}\right)\Big]+{\cal O}(h^6)~,
\label{gda}
\eeq
where $\Psi_{BE}, \Psi_{AB}, \Psi_{CE}, \Psi_{AC}$ refer to the points { outside} the mesh. Their contribution, without resampling the grid,  is obtained by applying Eq. (\ref{scheme:empty:2nd}) on $\psi_{AB}$ and $\psi_{AC}$ and developing the potential $V(r)$ around $r_0$, spatial coordinate of the cell central point

\[
 \Psi_{BE}+\Psi_{AB}+\Psi_{CE}+\Psi_{AC}= 2V_0\psi_0\Big[1-\frac{1}2\left(\frac{h}2\right)^2V_0\Big] + 
\]
\beq
 V_{BE}\psi_B\Big[1-\frac{1}2\left(\frac{h}2\right)^2V_{BE}\Big] + V_{CE}\psi_C\Big[1-\frac{1}2\left(\frac{h}2\right)^2\!\!V_{CE}\Big] +
\frac{1}2\Big[V_{BE}-2V_0+V_{CE}\Big]\!\!\left(\psi_B+\psi_C\right)+{\cal O}(h^4)~~.
\label{eq:sumPsi}
\eeq

The quantity $\Psi (r_0) = \Psi_0$ is approximated by a linear combination of the values of $\Psi_i$ for $i\in\triangle=\{B,C,D,E,F,G,H,$ $I,J\}$

\begin{align}
%\nonumber
\Psi_0=&\frac{1}{16}\Big[8\Psi_B+8\Psi_E+8\Psi_C-4\Psi_{D}-4\Psi_{F}+\Psi_{G}-\Psi_{H}-\Psi_{I}+\Psi_{J}\Big]+{\cal O}(h^4)~~.
\label{eq.Psi0}
\end{align}

Finally, we get 

\begin{widetext}
\beq
\begin{aligned}
\psi_A\approx-&\psi_E + \psi_B\Big[1 - \frac{1}4\left(\frac{h}3\right)^2 + \frac{1}{16}\left(\frac{h}3\right)^4V_0\Big] \left(V_0+V_B\right)+ 
\psi_C\Big[1 - \frac{1}4\left(\frac{h}3\right)^2 + \frac{1}{16}\left(\frac{h}3\right)^4V_0\Big] \left(V_0+V_C\right) - \\
& \left(\frac{h}3\right)^2\Big[1-\frac{1}4\left(\frac{h}3\right)^2V_0\Big]\Big[\Psi_{BE}+\Psi_{AB}+\Psi_{CE}+\Psi_{AC}+4\Psi_0\Big]~,
\end{aligned}
\label{scheme:empty:4th}
\eeq
\end{widetext}

%%%%%%%%%%%%%%%%%%%%%%%%%%%%%%%%%%%%%%%%%%%%%%%%%%%%%%%%%%%%%%%%%%%%%%%%%%%%%%%%%%%%%%%%%%%%%%%%%%%%%%%%%%%%%%%%%%%%%%%%%%%%%%%%%%%%%%%%%%%%%%%%%%%%%%%%%%%%%%%%%%%%%%%%%%%%%%%%%%%%%%%%%%%%%%%%%%%%%%%%%%%%%%%%%%%%%%%%%%%%%%%%%%%%%%%%%%%%%%%%%%%%%%%%%%%%%%%%%%%%%%%%%%%%%%%%%%%%%%%%%%%%%
%%%%%%%%%%%%%%%%%%%%%%%%%%%%%%%%%%%%%%%%%%%%%%%%%%%%%%%%%%%%%%%%%%%%%%%%%%%%%%%%%%%%%%%%%%%%%%%
\paragraph{Full cells.}

Equations (\ref{scheme:empty:2nd}, \ref{scheme:empty:4th}) cannot be used any longer being $\psi$ not regular.  Cells crossed by the trajectory imply discontinuity of $\psi$, and an {\it ad hoc} numerical scheme has been conceived. 
For an easy introduction, the reader may refer to \cite{aosp11, rispaoco11}. 
The strategy is the following: 

\begin{itemize}
\item For $\alpha_i$ being the weighting coefficients, establish a relation of the type
\beq
\psi^\pm_A=\sum_{i\in\triangle}\alpha_i\psi^\pm_i~.
\label{extrapolation}
\eeq
\item Expand the $\psi$ function in Taylor series at $A$ and at each point of $\triangle$, around the x point; { mark the intersection between the world line and the segment $[BC]$ of the cell}. Then

\beq
\psi^\pm_i=\sum_{n+m\leq N}{T^{(n,m)}_{i}\partial^n_{r*}\partial^m_t\psi^\pm_\text{x}}+\calo{h^{N+1}}~,
\label{taylor}
\eeq
where $T^{(n,m)}_{i}=(r^*_i-r_\text{x})^n(t_i-t_\text{x})^m/n!m!$ are the Taylor coefficients. By injecting Eq. (\ref{taylor}) into Eq.  (\ref{extrapolation}), we get by identification the $\alpha_i$ coefficients. Normally, at order $N$, $(N+1)(N+2)/2$ points are needed to satisfy the $(N+1)(N+2)/2$ relationships leading to the $\alpha_i$ { of}  Eq. (\ref{extrapolation}). However, since $\partial^2_t\psi^\pm_i=[\partial^2_{r^*}-V_i]\psi^\pm_i$,  $2N+1$ points are sufficient.

\item Since $i\in\Omega_+$ (or $i\in\Omega_-$), $\psi^-_i$ (or $\psi^+_i$) { is not well defined}. Therefore, the jump conditions $\partial^n_{r*}\partial^m_t\psi^\pm_\text{x}=\partial^n_{r*}\partial^m_t\psi^\mp_\text{x}\pm\jump{\partial_{r^*}^n\partial_t^m\psi}_\text{x}$ render meaningful $\psi^-_i$ and $\psi^+_i$ for $i\in\Omega_+$ and $i\in\Omega_-$, respectively. The junction term $\xi_i$ is such that

\begin{align}
\nonumber
\psi^\pm_i = \sum_{n+m\leq N}{T^{(n,m)}_{i}\Big(\partial^n_{r*}\partial^m_t\psi^\mp_\text{x}}\pm\jump{\partial_{r^*}^n\partial_t^m\psi}_\text{x}\Big)+ 
\calo{h^{N+1}} \approx\psi^\mp_i\pm\xi_i\quad\text{for }i\in\Omega_\mp~,
\label{taylor:jump}
\end{align}
where
\beq
\psi^\mp_i=\sum_{n+m\leq N}{T^{(n,m)}_{i}\partial^n_{r*}\partial^m_t\psi^\mp_\text{x}}~,
\eeq
and
\beq
\xi_i=\sum_{n+m\leq N}{T^{(n,m)}_{i}\jump{\partial_{r^*}^n\partial_t^m\psi}_\text{x}}~.
\label{xii}
\eeq
\item If $A\in\Omega_-$, then Eq. (\ref{extrapolation}) is rewritten as 
\begin{align}
\nonumber
\psi^-_A=\sum_{i\in\triangle}\alpha_i\psi^-_i~ =\sum_{i\in\triangle\cap\Omega_-}\alpha_i\psi^-_i+\sum_{i\in\triangle\cap\Omega_+}\alpha_i\psi^-_i = \sum_{i\in\triangle\cap\Omega_-}\alpha_i\psi^-_i+\sum_{i\in\triangle\cap\Omega_+}\alpha_i\Big(\psi^+_i-\xi_i\Big)~,
\end{align}
The first { sum} refers to the directly accessible grid points for which $\psi^-_i$ is well defined, since $i\in\Omega_-$; instead, the second sum to grid points for which $\psi^-_i$ is not defined, since $i\in\Omega_+$, and thus associated to $\psi^+_i$. { Accordingly}, we have 

\beq
\psi^-_A=\sum_{i\in\triangle}\alpha_i\psi^\pm_i-\!\!\sum_{i\in\triangle\cap\Omega_+}\!\!\alpha_i\xi_i~.
\label{formule_moins}
\eeq

Similarly, if $A\in\Omega_+$, then

\beq
\psi^+_A=\sum_{i\in\triangle}\alpha_i\psi^\pm_i-\!\!\sum_{i\in\triangle\cap\Omega_+}\!\!\alpha_i\xi_i+\xi_A~.
\label{formule_plus}
\eeq

\item The general formula can be finally summarised by

\beq
\psi_A=\sum_{i\in\triangle}\alpha_i\psi_i+\!\!\!\!\sum_{n+m\leq N}{\!\!\beta^{(n,m)}\jump{\partial_{r^*}^n\partial_t^m\psi}_\text{x}}+\calo{h^{N+1}}~,
\label{formule_generale}
\eeq
%\beq
%\nonumber\text{where }\beta^{(n,m)}\defeq-\sum_{i\in\triangle_A}\alpha_iT^{(n,m)}_i\text{ with }\alpha_A=-1\text{ and }\triangle_A=\left(\triangle\cup\{A\}\right)\cap\Omega_+~.
%\eeq
where $\beta^{(n,m)}\defeq-\sum_{i\in\triangle_A}\alpha_iT^{(n,m)}_i$ with $\alpha_A=-1\text{ and }\triangle_A=\left(\triangle\cup\{A\}\right)\cap\Omega_+$. The second term of Eq. (\ref{formule_generale}) takes into account the jump conditions only from the points contained in $\Delta_A$, {\it i.e.} the points belonging to the past light cone of $A$, including the latter (that is to say $\triangle\cup\{A\}$), and to the right hand side of the path (that is to say $\left(\triangle\cup\{A\}\right)\cap\Omega_+$).
\end{itemize}

We now apply this strategy at increasing order. The first order serves the purpose of an introduction to the application. We then develop, the second order and the fourth order, which will be used in this work (Part I) and in the companion work (Part II), respectively.   

%%%%%%%%%%%%%%%%%%%%%%%%%%%%%%%%%%%%%%%%%%%%%%%%%%%%%%%%%%%%%%%%%%%%%%%%%%%%%%%%%%%%%%%%%%%%%%%%%%%%%%%%%%%%%%%%%%%%%%%%%%%%%%%%%%%%%%%%%%%%%%%%%%%%%%%%%%%%%%%%%%%%%%%%%%%%%%%%%%%%%%%%%%%%%%%%%%%%%%%%%%%%%%%%%%%%%%%%%%%%%%%%%%%%%%%%%%%%%%%%%%%%%%%%%%%%%%%%%%%%%%%%%%%%%%%%%%%%%%%%%%%%%
%%%%%%%%%%%%%%%%%%%%%%%%%%%%%%%%%%%%%%%%%%%%%%%%%%%%%%%%%%%%%%%%%%%%%%%%%%%%%%%%%%%%%%%%%%%%%%%
\subparagraph{1\textsuperscript{st} order full cells.}

For $N=1$, {\it i.e.} $\calo{\Delta r^{2*},\Delta t^2}$, Eq. (\ref{taylor}) gives 

\begin{align}
&\psi^-_A=\psi^-_\text{x}+(\epsilon-\Delta r^*)\partial_{r^*}\psi^-_\text{x}+\Delta t\partial_t\psi^-_\text{x}~,\\
&\psi^-_B=\psi^-_\text{x}-(2\Delta r^*-\epsilon)\partial_{r^*}\psi^-_\text{x}~,\\
&\psi^-_C=\psi^-_\text{x}+\epsilon\partial_{r^*}\psi^-_\text{x}~,\\
&\psi^-_D=\psi^-_\text{x}+(\epsilon-\Delta r^*)\partial_{r^*}\psi^-_\text{x}-\Delta t\partial_t\psi^-_\text{x}~,
\end{align}
where $\epsilon\defeq r^*_C-r^*_\text{x}$. Solving for $\psi^-_A=\sum_{i\in\triangle}\alpha_i\psi^-_i$ { leads to reverse the following linear system}

\beq
\left(\begin{matrix}1&\epsilon\!-\!\Delta r^*&0\\1&\epsilon&0\\1&\epsilon\!-\!\Delta r^*&-1\end{matrix}\right)\cdot
\left(\begin{matrix}\alpha_B\\\alpha_C\\\alpha_E\end{matrix}\right)=\left(\begin{matrix}1\\\epsilon\!-\!\Delta r^*\\\Delta t\end{matrix}\right)~.
\eeq
The solution is $\alpha_B=\alpha_C=-\alpha_E=1$. The trajectory crosses the cell into four possible ways, Fig. (\ref{fig03}). For the case (i), $\triangle=\{B,C,E\}$; $\triangle\cap\Omega_-=\{B\}$; $\triangle\cap\Omega_+=\{C,E\}$, and according to Eq. (\ref{xii})
\begin{align}
&\xi_A=\jump{\psi}_\text{x}+(\epsilon-\Delta r^*)\jump{\partial_{r^*}\psi}_\text{x}+\Delta t\jump{\partial_t\psi}_\text{x}~,\label{xiA_1st}\\
&\xi_B=\jump{\psi}_\text{x}-(2\Delta r^*-\epsilon)\jump{\partial_{r^*}\psi}_\text{x}~,\label{xiB_1st}\\
&\xi_C=\jump{\psi}_\text{x}+\epsilon\jump{\partial_{r^*}\psi}_\text{x}~,\label{xiC_1st}\\
&\xi_E=\jump{\psi}_\text{x}+(\epsilon-\Delta r^*)\jump{\partial_{r^*}\psi}_\text{x}-\Delta t\jump{\partial_t\psi}_\text{x}~.\label{xiE_1st}
\end{align}

For the case (i), $\psi^+_A$ is computed from $\psi^-_B$, $\psi^+_C$, $\psi^+_E$. Equation (\ref{formule_plus}) then gives

\beq
\begin{aligned}
\text{(i)}\ \psi^+_A \approx\psi^-_B+\psi^+_C-\psi^+_E+\xi_A-\xi_C+\xi_E 
\approx\psi^-_B+\psi^+_C-\psi^+_E + \jump{\psi}_\text{x} - (2\Delta r^*-\epsilon)\jump{\partial_{r*}\psi}_\text{x}~.
\label{first1}
\end{aligned}
\eeq

Applying Eqs. (\ref{formule_moins}, \ref{formule_plus}, \ref{xiA_1st}-\ref{xiE_1st}),  and changing the points 
in $\triangle\cap\Omega_+$ and $\triangle\cap\Omega_-$ according to the (ii), (iii) or (iv) cases, Fig. (\ref{fig03}), we get

\begin{align}
%\text{(i)}\ &\psi^+_A \approx \psi^-_B+\psi^+_C-\psi^+_E + \jump{\psi}_\text{x} - (2\Delta r^*-\epsilon)\jump{\partial_{r*}\psi}_\text{x}~,\\
\text{(ii)}\ &\psi^+_A \approx \psi^-_B+\psi^+_C-\psi^-_E + \Delta t\jump{\partial_t\psi}_\text{x} - \Delta r^*\jump{\partial_{r*}\psi}_\text{x}~, \label{first2}\\
\text{(iii)}\ &\psi^-_A \approx \psi^-_B+\psi^+_C-\psi^-_E - \jump{\psi}_\text{x} - \epsilon\jump{\partial_{r*}\psi}_\text{x}~, \label{first3}\\
\text{(iv)}\ &\psi^-_A \approx \psi^-_B+\psi^+_C-\psi^+_E - \Delta t\jump{\partial_t\psi}_\text{x} - \Delta r^*\jump{\partial_{r*}\psi}_\text{x}~\label{first4}.
\end{align}

At first order, the method provides the value of the cell upper node as function of the other three nodes and of analytic expressions.

%%%%%%%%%%%%%%%%%%%%%%%%%%%%%%%%%%%%%%%%%%%%%%%%%%%%%%%%%%%%%%%%%%%%%%%%%%%%%%%%%%%%%%%%%%%%%%%%%%%%%%%%%%%%%%%%%%%%%%%%%%%%%%%%%%%%%%%%%%%%%%%%%%%%%%%%%%%%%%%%%%%%%%%%%%%%%%%%%%%%%%%%%%%%%%%%%%%%%%%%%%%%%%%%%%%%%%%%%%%%%%%%%%%%%%%%%%%%%%%%%%%%%%%%%%%%%%%%%%%%%%%%%%%%%%%%%%%%%%%%%%%%%
%%%%%%%%%%%%%%%%%%%%%%%%%%%%%%%%%%%%%%%%%%%%%%%%%%%%%%%%%%%%%%%%%%%%%%%%%%%%%%%%%%%%%%%%%%%%%%%
\subparagraph{2\textsuperscript{nd} order full cells.}

We set $N=2$ in Eq. (\ref{taylor}), and use $\partial^2_t\psi^\pm_i=[\partial^2_{r^*}-V_i]\psi^\pm_i$ to limit to five the number of grid points.  The latter are chosen in the past light cone of $A$, $\triangle=\{B,C,D,E,F\}$, Fig. (\ref{fig04}). We have 

\begin{align}
&\alpha_B=1-{1\over2}\epsilon hV_\text{x}~,\\
&\alpha_C=1-{1\over2}h(2h-\epsilon)V_\text{x}~,\\
&\alpha_D=\alpha_F={1\over8}\epsilon (2h-\epsilon)V_\text{x}~,\\
&\alpha_E=-1-{1\over4}\epsilon (2h-\epsilon)V_\text{x}~,
\end{align}
where $\Delta r^*=\Delta t=h$. Again,  the trajectory intersects the stencil in four possible configurations, Fig. (\ref{fig04}). The junction functions $\xi_i$ take the following form

\begin{widetext}
\begin{subequations}
\begin{align}
&
\xi_A=\left(1\!-\!\frac{1}2\Delta t^2V_\text{x}\right)\jump{\psi}_\text{x}
\!+\!(\epsilon\!-\!\Delta r^*)\jump{\partial_{r^*}\psi}_\text{x}
\!+\!\Delta t\jump{\partial_t\psi}_\text{x}
\!+\!\frac{1}2(\epsilon\!-\!\Delta r^*)^2\jump{\partial^2_{r^*}\psi}_\text{x}
\!+\!\Delta t(\epsilon\!-\!\Delta r^*)\jump{\partial_{r^*}\partial_t\psi}_\text{x}~,
\\
&\xi_B=\jump{\psi}_\text{x}
\!-\!(2\Delta r^*\!-\!\epsilon)\jump{\partial_{r^*}\psi}_\text{x}
\!+\!\frac{1}2(2\Delta r^*\!-\!\epsilon)^2\jump{\partial^2_{r^*}\psi}_\text{x}~,\\
&\xi_C=\jump{\psi}_\text{x}
\!+\!\epsilon\jump{\partial_{r^*}\psi}_\text{x}
\!+\!\frac{1}2\epsilon^2\jump{\partial^2_{r^*}\psi}_\text{x}~,\\
&\xi_D=\left(1\!-\!\frac{1}2\Delta t^2V_\text{x}\right)\jump{\psi}_\text{x}
\!-\!(3\Delta r^*\!-\!\epsilon)\jump{\partial_{r^*}\psi}_\text{x}
\!-\!\Delta t\jump{\partial_t\psi}_\text{x}
\!+\!\frac{1}2(3\Delta r^*\!-\!\epsilon)^2\jump{\partial^2_{r^*}\psi}_\text{x}
\!+\!\Delta t(3\Delta r^*\!-\!\epsilon)\jump{\partial_{r^*}\partial_t\psi}_\text{x}~,
\\
&\xi_E=\left(1\!-\!\frac{1}2\Delta t^2V_\text{x}\right)\jump{\psi}_\text{x}
\!+\!(\epsilon\!-\!\Delta r^*)\jump{\partial_{r^*}\psi}_\text{x}
\!-\!\Delta t\jump{\partial_t\psi}_\text{x}
\!+\!\frac{1}2(\epsilon\!-\!\Delta r^*)^2\jump{\partial^2_{r^*}\psi}_\text{x}
\!-\!\Delta t(\epsilon\!-\!\Delta r^*)\jump{\partial_{r^*}\partial_t\psi}_\text{x}~,
\\
&\xi_F=\left(1\!-\!\frac{1}2\Delta t^2V_\text{x}\right)\jump{\psi}_\text{x}
\!+\!(\epsilon\!+\!\Delta r^*)\jump{\partial_{r^*}\psi}_\text{x}
\!-\!\Delta t\jump{\partial_t\psi}_\text{x}
\!+\!\frac{1}2(\epsilon\!+\!\Delta r^*)^2\jump{\partial^2_{r^*}\psi}_\text{x}
\!-\!\Delta t(\epsilon\!+\!\Delta r^*)\jump{\partial_{r^*}\partial_t\psi}_\text{x}~.
\label{xi2}
\end{align}
\end{subequations}
\end{widetext}

Applying Eqs. (\ref{formule_plus}, or \ref{formule_moins}), together with Eqs. (\ref{xi2}), and properly referring to the points associated to either $\triangle\cap\Omega_+$ or $\triangle\cap\Omega_-$ for the (i), (ii), (iii) and (iv) cases, Fig. (\ref{fig04}),  the following relationships are obtained

\begin{align}
\text{(i)}\ &\psi^+_A \approx \sum_{i\in\triangle}\alpha_i\psi^\pm_i  - \left(\alpha_C\xi_C+\alpha_E\xi_E+\alpha_F\xi_F-\xi_A\right)~,\\
\text{(ii)}\ &\psi^+_A \approx \sum_{i\in\triangle}\alpha_i\psi^\pm_i  - \left(\alpha_C\xi_C+\alpha_F\xi_F-\xi_A\right)~,\\
\text{(iii)}\ &\psi^-_A \approx \sum_{i\in\triangle}\alpha_i\psi^\pm_i - \left(\alpha_C\xi_C+\alpha_F\xi_F\right)~,\\
\text{(iv)}\ &\psi^-_A \approx \sum_{i\in\triangle}\alpha_i\psi^\pm_i  - \left(\alpha_C\xi_C+\alpha_E\xi_E+\alpha_F\xi_F\right)~,
\end{align}
where $\sum_{i\in\triangle}\alpha_i\psi^\pm_i=\alpha_B\psi^-_B+\alpha_C\psi^+_C+\alpha_D\psi^-_D+\alpha_E\psi^+_E+\alpha_F\psi^+_F$ .

%%%%%%%%%%%%%%%%%%%%%%%%%%%%%%%%%%%%%%%%%%%%%%%%%%%%%%%%%%%%%%%%%%%%%%%%%%%%%%%%%%%%%%%%%%%%%%%%%%%%%%%%%%%%%%%%%%%%%%%%%%%%%%%%%%%%%%%%%%%%%%%%%%%%%%%%%%%%%%%%%%%%%%%%%%%%%%%%%%%%%%%%%%%%%%%%%%%%%%%%%%%%%%%%%%%%%%%%%%%%%%%%%%%%%%%%%%%%%%%%%%%%%%%%%%%%%%%%%%%%%%%%%%%%%%%%%%%%%%%%%%%%%
%%%%%%%%%%%%%%%%%%%%%%%%%%%%%%%%%%%%%%%%%%%%%%%%%%%%%%%%%%%%%%%%%%%%%%%%%%%%%%%%%%%%%%%%%%%%%%%
\subparagraph{4\textsuperscript{th} order full cells.}

The extension to fourth order necessitates ten points in the stencil, Fig. (\ref{fig05}), $\triangle=\{B,C,D,E,F,G,H,I,J\}$. 
The expressions of the jump conditions and $\alpha_i$ coefficients achieve a { considerable length}. 
The fourth order accuracy is useful when we will discuss the case of radial orbits under SF. Therein, the third derivative of the wave-function $\psi$ at the position of the particle guarantees an adequate accuracy. 

Though the trajectory forcefully cuts the area limited by the $\triangle$ points (else there is no particle), it does not necessarily cut the cell $ABCE$. Therefore, Eq. (\ref{eq.Psi0}) is not applicable, and we take a Taylor expansion {\cite{ha07}} at $t=t_0$ constant. If $r^*_p(t_0)>r^*_C$, then 

\beq
\begin{aligned}
\Psi_0=&\frac{1}{16}\Big[5V(r^*_0-h)\psi(t_0,r^*_0-h)+15V(r^*_0-3h)\psi(t_0,r^*_0-3h)-\\
& 5V(r^*_0-5h)\psi(t_0,r^*_0-5h)+ V(r^*_0-7h)\psi(t_0,r^*_0-7h)\Big]+{\cal O}(h^4)~,
\end{aligned}
\eeq

whereas if $r^*_p(t_0)<r^*_B$

\beq
\begin{aligned}
\Psi_0=&\frac{1}{16}\Big[5V(r^*_0+h)\psi(t_0,r^*_0+h)+
15V(r^*_0+3h)\psi(t_0,r^*_0+3h)- \\
& 5V(r^*_0+5h)\psi(t_0,r^*_0+5h)+
V(r^*_0+7h)\psi(t_0,r^*_0+7h)\Big]+{\cal O}(h^4)~.
\end{aligned}
\eeq

%%%%%%%%%%%%%%%%%%%%%%%%%%%%%%%%%%%%%%%%%%%%%%%%%%%%%%%%%%%%%%%%%%%%%%%%%%%%%%%%%%%%%%%%%%%%%%%
%%%%%%%%%%%%%%%%%%%%%%%%%%%%%%%%%%%%%%%%% SUBSECTION %%%%%%%%%%%%%%%%%%%%%%%%%%%%%%%%%%%%%%%%%%
%%%%%%%%%%%%%%%%%%%%%%%%%%%%%%%%%%%%%%%%%%%%%%%%%%%%%%%%%%%%%%%%%%%%%%%%%%%%%%%%%%%%%%%%%%%%%%%

%%%%%%%%%%%%%%%%%%%%%%%%%%%%%%%%%%%%%%%%%%%%%%%%%%%%%%%%%%%%%%%%%%%%%%%%%%%%%%%%%%%%%%%%%%%%%%%%%%%%%%%%%%%%%%%%%%%%%%%%%%%%%%%%%%%%%%%%%%%%%%%%%%%%%%%%%%%%%%%%%%%%%%%%%%%%%%%%%%%%%%%%%%%%%%%%%%%%%%%%%%%%%%%%%%%%%%%%%%%%%%%%%%%%%%%%%%%%%%%%%%%%%%%%%%%%%%%%%%%%%%%%%%%%%%%%%%%%%%%%%%%%%
%%%%%%%%%%%%%%%%%%%%%%%%%%%%%%%%%%%%%%%%%%%%%%%%%%%%%%%%%%%%%%%%%%%%%%%%%%%%%%%%%%%%%%%%%%%%%%%
\section{WFs from geodesic generic orbits}

We deal here with orbits that a CS pursues in its approach to a SMBH, before plunge (for geodesic radial infall, see \cite{aosp11}). 
The different orbits are obtained by inserting the values of eccentricity $e$ and semi-latus rectum $p$ in Eqs. (\ref{partpos1}-\ref{partpos3}), together with the initial conditions { $\phi_p(t = 0)$ and $\chi(t = 0)$.} 

%%%%%%%%%%%%%%%%%%%%%%%%%%%%%%%%%%%%%%%%%%%%%%%%%%%%%%%%%%%%%%%%%%%%%%%%%%%%%%%%%%%%%%%%%%%%%%%%%%%%%%%%%%%%%%%%%%%%%%%%%%%%%%%%%%%%%%%%%%%%%%%%%%%%%%%%%%%%%%%%%%%%%%%%%%%%%%%%%%%%%%%%%%%%%%%%%%%%%%%%%%%%%%%%%%%%%%%%%%%%%%%%%%%%%%%%%%%%%%%%%%%%%%%%%%%%%%%%%%%%%%%%%%%%%%%%%%%%%%%%%%%%%
%%%%%%%%%%%%%%%%%%%%%%%%%%%%%%%%%%%%%%%%%%%%%%%%%%%%%%%%%%%%%%%%%%%%%%%%%%%%%%%%%%%%%%%%%%%%%%%

\subsection{Circular orbits}

\begin{center}
    \begin{figure}[h!]
        \includegraphics[width=0.6\linewidth]{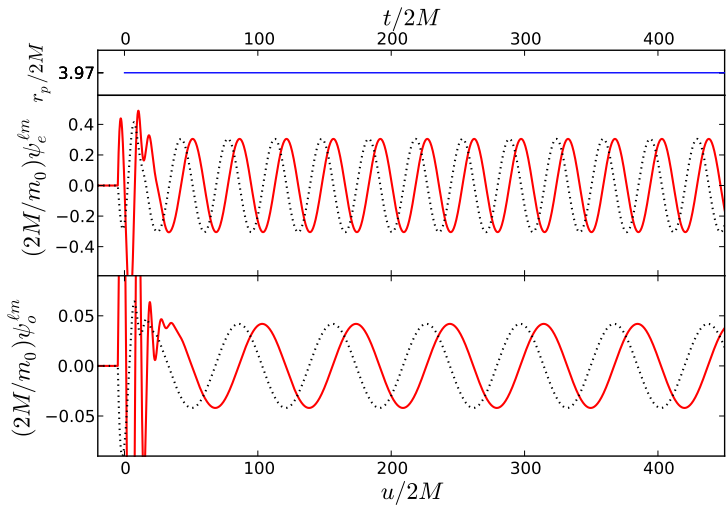}
        \caption{The $\psi$ wave-function in terms of the coordinates $t/2M$ and $u = (t - r*_\text{ obs})/2M$, for the even quadrupolar mode $(\ell,m)$=$(2,2)$ (top) and the odd mode $(\ell,m)$=$(2,1)$ (bottom); orbital parameters $(e,p)=(0, 12)$. We draw the real { (solid line)} and the imaginary { (dotted line)} parts, their dephasing being $\Delta u=\frac{\pi}2m^{-1}\Omega^{-1}=\frac{\pi}2m^{-1}Mp^{3/2}$.
        % The values in SI units for $\psi$ are obtained through multiplication by {\rosso $m/r_{\rm g}$} - in geometrised units - and by $c^2 G^{-0.5}$.
        }
        \label{fig08}
    \end{figure}
\end{center}

In a circular orbit, the CS { turns around} the SMBH at { $r_p=pM$}. 
In Fig. (\ref{fig08}), two examples of WFs at infinity are shown, $(\ell,m)$=$(2,2)$ for { the} even parity, and $(\ell,m)$=$(2,1)$ for { the} odd parity, { each with real and imaginary parts}. The oscillatory shape of the gravitational radiation derives from the circularity of the orbit. The wave-function oscillates with a frequency $m\Omega$ for a given mode $(\ell,m)$, where {$\Omega=d\phi_p/dt=M^{-1}p^{-3/2}$} is the angular velocity. 
{From} Eq. (\ref{Edot}), the total energy flux is (for the $m$-{symmetry}, $\dott{E}_{\ell,-m}=\dott{E}_{\ell,m}$) 

\beq
\dott{E}=\sum_{\ell=2}^{\ell_\text{max}}\sum_{m=-\ell}^\ell\dott{E}_{\ell m}=\sum_{\ell=2}^{\ell_\text{max}}{\left[\dott{E}_{\ell 0}+2\sum_{m=1}^\ell\dott{E}_{\ell m}\right]}~,
\eeq

From Eq. (\ref{Ldot})  the total angular momentum flux is 

\beq
\dott{L}=2\sum_{\ell=2}^{\ell_\text{max}}\sum_{m=1}^\ell\dott{L}_{\ell m}~.
\eeq
 
If the particle follows an equatorial path, some simplifications can be imposed without loss 
of generality. Indeed, if $\ell+m$ is even, { then} $\psi^{\ell m}_o=0$, { whereas} if $\ell+m$ is odd, { then} $\psi^{\ell m}_e=0$. 

For a better accuracy we use a Richardson extrapolation on the flux values. In { practice,} $\dott{E}_{\ell m}$ and $\dott{L}_{\ell m}$ are { computed} for various grid step { sizes} $h/2^n$ with $n=0,1,2,3$ and $h/2M=0.1$, and then extrapolated by the first terms of the Richardson recurrence formula. 

In Tabs. \ref{table_Edot}, \ref{table_Ldot}, the energy $\dott{E}_{\ell m}^\infty$ (units of $M^2/m_0^2$) and the angular momentum fluxes $\dott{L}_{\ell m}$ (units of $M/m^2_0$) at infinity are computed, for different ${\ell m}$ modes and $\ell \leq 5$; the semi-latus rectum is $p=7.9456$. 
For $\dott{E}_{\ell m}^\infty $, our results are compared to those of Poisson \cite{po95,po97}, Martel  \cite{ma04}, Barack and Lousto \cite{balo05}, Sopuerta and Laguna \cite{sola06}. Our total values differ with the first by $0.005\%$, with the second by $0.096\%$, with the third by $0.009\%$, and with the fourth by $0.001\%$.

For $\dott{L}_{\ell m}^\infty$, our results are compared to those of Poisson \cite{po95,po97}, Martel  \cite{ma04}, Sopuerta and Laguna \cite{sola06}. Our total values differ with the first by $0.007\%$, with the second by $0.2\%$, and with the third by $0.005\%$.

The total energy $\dott{E}_{\ell m}^\infty$ from Nagar, { Damour} and Tartaglia \cite{nadata07}, differs from our value by about $0.8 \%$, while the total angular momentum $\dott{L}_{\ell m}^\infty $ by about $0.7 \%$.

% \begin{widetext}
\renewcommand{\arraystretch}{1.4} %donne la distance entre les lignes%
\setlength{\tabcolsep}{0.006\linewidth} %donne la distance entre les colonnes%
\begin{table*}
        \begin{tabular}{*{2}{c}*{5}{c}}
          \hline\hline
          $\ell$ & $m$ & $\dott{E}^\infty_{\ell m}$ 
& $\dott{E}^\infty_{\ell m}$ \cite{po95,po97} 
& $\dott{E}^\infty_{\ell m}$ \cite{ma04} 
& $\dott{E}^\infty_{\ell m}$ \cite{balo05} 
& $\dott{E}^\infty_{\ell m}$ \cite{sola06}\\
          \hline
            2 & 1 & $8.1680.10^{-07}$ 
& $8.1633.10^{-07}\   [0.06\%]\ \ $ 
& $8.1623.10^{-07}\ [0.07\%]$ 
& $8.1654.10^{-07}\    [0.03\%]\ \ $ 
& $8.1662.10^{-07}\  [0.02\%]\ \ $\\
            2 & 2 & $1.7064.10^{-04}$ 
& $1.7063.10^{-04}\   [0.006\%]$ 
& $1.7051.10^{-04}\ [0.07\%]$ 
& $1.7061.10^{-04}\       [0.02\%]\ \ $ 
& $~1.7064.10^{-04}\  [\!<\!0.001\%]$\\
            3 & 1 & $2.1757.10^{-09}$ 
& $2.1731.10^{-09}\   [0.1\%]\ \ \ $ 
& $2.1741.10^{-09}\ [0.07\%]$ 
& $2.1734.10^{-09}\   [0.1\%]\ \ \ $ 
& $2.1732.10^{-09}\ [0.1\%]\ \ \ $\\
            3 & 2 & $2.5203.10^{-07}$ 
& $2.5199.10^{-07}\  [0.02\%]\ \ $ 
& $2.5164.10^{-07}\ [0.2\%]\ $ 
& $2.5207.10^{-07}\    [0.01\%]\ \ $ 
& $2.5204.10^{-07}\  [0.002\%]$\\
            3 & 3 & $2.5471.10^{-05}$ 
& $2.5471.10^{-05}\  [0.001\%]$ 
& $2.5432.10^{-05}\ [0.2\%]\ $ 
& $2.5479.10^{-05}\       [0.03\%]\ \ $ 
& $2.5475.10^{-05}\  [0.02\%]\ \ $\\
            4 & 1 & $8.4124.10^{-13}$ 
& $8.3956.10^{-13}\  [0.2\%]\ \ \ $ 
& $8.3507.10^{-13}\ [0.7\%]\ $ 
& $8.3982.10^{-13}\   [0.2\%]\ \ \ $ 
& $8.4055.10^{-13}\ [0.08\%]\ \ $\\
            4 & 2 & $2.5099.10^{-09}$ 
& $2.5091.10^{-09}\  [0.03\%]\ \ $ 
& $2.4986.10^{-09}\ [0.5\%]\ $ 
& $2.5099.10^{-09}\    [0.002\%]$ 
& $2.5099.10^{-09}\     [0.002\%]$\\
            4 & 3 & $5.7750.10^{-08}$ 
& $5.7751.10^{-08}\  [0.001\%]$ 
& $5.7464.10^{-08}\ [0.5\%]\ $ 
& $5.7759.10^{-08}\       [0.02\%]\ \ $ 
& $5.7765.10^{-08}\  [0.03\%]\ \ $\\
            4 & 4 & $4.7251.10^{-06}$ 
& $4.7256.10^{-06}\  [0.01\%]\ \ $ 
& $4.7080.10^{-06}\ [0.4\%]\ $ 
& $4.7284.10^{-06}\    [0.07\%]\ \ $ 
& $4.7270.10^{-06}\  [0.04\%]\ \ $\\
            5 & 1 & $1.2632.10^{-15}$ 
& $1.2594.10^{-15}\  [0.3\%]\ \ \ $ 
& $1.2544.10^{-15}\ [0.7\%]\ $ 
& $1.2598.10^{-15}\   [0.3\%]\ \ \ $ 
& $1.2607.10^{-15}\ [0.2\%]\ \ \ $\\
            5 & 2 & $2.7910.10^{-12}$ 
& $2.7896.10^{-12}\  [0.05\%]\ \ $ 
& $2.7587.10^{-12}\ [1.2\%]\ $ 
& $2.7877.10^{-12}\    [0.1\%]\ \ \ $ 
& $2.7909.10^{-12}\ [0.003\%]$\\
            5 & 3 & $1.0933.10^{-09}$ 
& $~1.0933.10^{-09}\  [\!<\!0.001\%]$ 
& $1.0830.10^{-09}\ [0.9\%]\ $ 
& $1.0934.10^{-09}\  [0.009\%]$ 
& $1.0936.10^{-09}\     [0.03\%]\ \ $\\
            5 & 4 & $1.2322.10^{-08}$ 
& $1.2324.10^{-08}\  [0.01\%]\ \ $ 
& $1.2193.10^{-08}\ [1.1\%]\ $ 
& $1.2319.10^{-08}\    [0.03\%]\ \ $ 
& $1.2329.10^{-08}\  [0.05\%]\ \ $\\
            5 & 5 & $9.4544.10^{-07}$ 
& $9.4563.10^{-07}\  [0.02\%]\ \ $ 
& $9.3835.10^{-07}\ [0.8\%]\ $ 
& $9.4623.10^{-07}\    [0.08\%]\ \ $ 
& $9.4616.10^{-07}\  [0.08\%]\ \ $\\
            \hline
          \multicolumn{2}{c}{ Total } & $2.0293.10^{-04}$ 
& $2.0292.10^{-04}\ [0.005\%]$ 
& $~2.0273.10^{-04}\ [0.096\%]$ 
& $2.0291.10^{-04}\ [0.009\%]$ 
& $~2.0293.10^{-04}\ [<\!0.001\%]$\\
          \hline\hline
        \end{tabular}
        \caption{ Circular orbit: energy flux $dE_{\ell m}/dt$ values at infinity, for different ${\ell m}$ modes and $\ell < 5$ (units of $M^2/m_0^2$). The semi-latus rectum is $p=7.9456$. The first column lists our results, and the second those of Poisson \cite{po95,po97}, the third of Martel  \cite{ma04}, the fourth of Barack and Lousto \cite{balo05}, the fifth of Sopuerta and Laguna \cite{sola06}.}
        \label{table_Edot}
\end{table*}
%\end{widetext}

\renewcommand{\arraystretch}{1.4} %donne la distance entre les lignes%
\setlength{\tabcolsep}{0.0248\linewidth} %donne la distance entre les colonnes%
\begin{table*}
        \begin{tabular}{*{2}{c}*{4}{c}}
          \hline\hline
          $\ell$ & $m$ & $\dott{L}^\infty_{\ell m}$ 
& $\dott{L}^\infty_{\ell m}$ \cite{po95,po97} 
& $\dott{L}^\infty_{\ell m}$ \cite{ma04} 
& $\dott{L}^\infty_{\ell m}$ \cite{sola06}\\
          \hline
            2 & 1 & $1.8294.10^{-05}$ 
& $1.8283.10^{-05}\ [0.06\%]\ \ $   
& $1.8270.10^{-05}\ [0.1\%]\ \ $ 
& $1.8289.10^{-05}\ [0.03\%]\ \ $ \\
            2 & 2 & $3.8218.10^{-03}$ 
& $3.8215.10^{-03}\ [0.009\%]$      
& $3.8164.10^{-03}\ [0.1\%]\ \ $ 
& $3.8219.10^{-03}\ [0.002\%]$ \\
            3 & 1 & $4.8729.10^{-08}$ 
& $4.8670.10^{-08}\ [0.1\%]\ \ \ $  
& $4.8684.10^{-08}\ [0.09\%]\ $  
& $4.8675.10^{-08}\ [0.1\%]\ \ \ $ \\
            3 & 2 & $5.6448.10^{-06}$ 
& $5.6439.10^{-06}\ [0.02\%]\ \ $    
& $5.6262.10^{-06}\ [0.3\%]\ \ $ 
& $5.6450.10^{-06}\ [0.003\%]$ \\
            3 & 3 & $5.7048.10^{-04}$ 
& $~5.7048.10^{-04}\ [\!<\!0.001\%]$ 
& $5.6878.10^{-04}\ [0.2\%]\ \ $ 
& $5.7057.10^{-04}\ [0.02\%]\ \ $ \\
            4 & 1 & $1.8841.10^{-11}$ 
& $1.8803.10^{-11}\ [0.2\%]\ \ \ $  
& $1.8692.10^{-11}\ [0.8\%]\ \ $ 
& $1.8825.10^{-11}\ [0.09\%]\ \ $ \\
            4 & 2 & $5.6213.10^{-08}$ 
& $5.6195.10^{-08}\ [0.03\%]\ \ $    
& $5.5926.10^{-08}\ [0.5\%]\ \ $ 
& $5.6215.10^{-08}\ [0.003\%]$ \\
            4 & 3 & $1.2934.10^{-06}$ 
& $1.2934.10^{-06}\ [0.003\%]$       
& $1.2933.10^{-06}\ [0.01\%]\ $  
& $1.2937.10^{-06}\ [0.02\%]\ \ $ \\
            4 & 4 & $1.0583.10^{-04}$ 
& $1.0584.10^{-04}\ [0.01\%]\ \ $    
& $1.0518.10^{-04}\ [0.6\%]\ \ $ 
& $1.0586.10^{-04}\ [0.03\%]\ \ $ \\
            5 & 1 & $2.8293.10^{-14}$ 
& $2.8206.10^{-14}\ [0.3\%]\ \ \ $  
& $2.8090.10^{-14}\ [0.7\%]\ \ $ 
& $2.8237.10^{-14}\ [0.2\%]\ \ \ $ \\
            5 & 2 & $6.2509.10^{-11}$ 
& $6.2479.10^{-11}\ [0.05\%]\ \ $   
& $6.1679.10^{-11}\ [1.3\%]\ \ $ 
& $6.2509.10^{-11}\ [0.001\%]$ \\
            5 & 3 & $2.4487.10^{-08}$ 
& $2.4486.10^{-08}\ [0.002\%]$      
& $2.4227.10^{-08}\ [1.1\%]\ \ $ 
& $2.4494.10^{-08}\ [0.03\%]\ \ $ \\
            5 & 4 & $2.7598.10^{-07}$ 
& $2.7603.10^{-07}\ [0.02\%]\ \ $   
& $2.7114.10^{-07}\ [1.8\%]\ \ $ 
& $2.7613.10^{-07}\ [0.05\%]\ \ $ \\
            5 & 5 & $2.1175.10^{-05}$ 
& $2.1179.10^{-05}\ [0.02\%]\ \ $   
& $2.0933.10^{-05}\ [1.2\%]\ \ $ 
& $2.1190.10^{-05}\ [0.07\%]\ \ $ \\
            \hline
          \multicolumn{2}{c}{ Total } & $4.5449.10^{-03}$ 
& $4.5446.10^{-03}\ [0.007\%]$ 
& $\!\!4.5369.10^{-03}\ [0.2\%]$ 
& $4.5452.10^{-03}\ [0.005\%]$\\
          \hline\hline
        \end{tabular}
\caption{ Circular orbit: angular momentum flux $dL_{\ell m}/dt$ values at infinity, for different ${\ell m}$ modes and $\ell < 5$ (units of $M/m^2_0$). The semi-latus rectum is $p=7.9456$. The first column lists our results, and the second those of Poisson \cite{po95,po97}, the third of Martel  \cite{ma04}, the fourth of Sopuerta and Laguna \cite{sola06}.}
        \label{table_Ldot}
\end{table*}

%\begin{widetext}
\renewcommand{\arraystretch}{1.2}
\setlength{\tabcolsep}{10pt} %donne la distance entre les collones%
\begin{table*}
        \begin{tabular}{*{2}{c}*{4}{c}}
          \hline\hline
          $e$ & $p$ & $<\dott{E}^\infty>$ & $<\dott{E}^\infty>$ \cite{cukepo94} & $<\dott{E}^\infty>$ \cite{ma04} & $<\dott{E}^\infty>$ \cite{sola06}\\
          \hline
          $0.188917$ & $7.50478$ & $3.1617.10^{-04}$ & $3.1680.10^{-04}[0.2\%]$ & $3.1770.10^{-04}[0.5\%]$ & $3.1640.10^{-04}[0.07\%]$\\
          $0.764124$ & $8.75455$ & $2.1026.10^{-04}$ & $2.1008.10^{-04}[0.09\%]$ & $2.1484.10^{-04}[2.1\%]$ & $2.1004.10^{-04}[0.1\%]$\\
          \hline\hline
        \end{tabular}
        \caption{ Elliptic orbit: average of the energy flux (units of $M^2/m_0^2$), computed according to Eq. (\ref{eeccaveq}), taken over a few periods in the case of two elliptic orbits $(e,p)$=$(0.188917,7.50478)$ and $(0.764124,8.75455)$. The differences with Cutler {\it et al.} \cite{cukepo94}, Martel \cite{ma04}, and Sopuerta and Laguna \cite{sola06} are shown}
        \label{Eeccave}
\end{table*}
%\end{widetext}

%\begin{widetext}
\begin{table*}
        \begin{tabular}{*{2}{c}*{4}{c}}
          \hline\hline
          $e$ & $p$ & $<\dott{L}^\infty>$ & $<\dott{L}^\infty>$ \cite{cukepo94} & $<\dott{L}^\infty>$ \cite{ma04} & $<\dott{L}^\infty>$ \cite{sola06}\\
          \hline
          $0.188917$ & $7.50478$ & $5.9550.10^{-03}$ & $5.9656.10^{-03}[0.2\%]$ & $5.9329.10^{-03}[0.4\%]$ & $5.9555.10^{-03}[0.008\%]$\\
          $0.764124$ & $8.75455$ & $2.7531.10^{-03}$ & $2.7503.10^{-03}[0.1\%]$ & $2.7932.10^{-03}[1.4\%]$ & $2.7505.10^{-03}[0.09\%]$\\
          \hline\hline
        \end{tabular}
        \caption{ Elliptic orbit: average of the angular momentum flux (units of $M/m_0^2$), computed according to Eq. (\ref{leccaveq}), taken over a few periods in the case of two elliptic orbits $(e,p)$=$(0.188917,7.50478)$ and $(0.764124,8.75455)$. The differences with Cutler {\it et al.} \cite{cukepo94}, Martel \cite{ma04}, and Sopuerta and Laguna \cite{sola06} are shown.}
        \label{Leccave}
\end{table*}
%\end{widetext}

\renewcommand{\arraystretch}{1.2}
\setlength{\tabcolsep}{12pt} %donne la distance entre les colonnes%
\begin{table*}
        \begin{tabular}{*{5}{c}}
          \hline\hline
          $\ell$ & $<\dott{E}_\ell^\infty>$ & $<\dott{E}_\ell^\infty>$ \cite{hoev10} & $<\dott{L}_\ell^\infty>$ & $<\dott{L}_\ell^\infty>$ \cite{hoev10} \\
          \hline
          $2$    & $1.571333.10^{-04}$    & $1.57133846.10^{-04}[0.0004\%]$    & $2.092406.10^{-03}$     & $2.09219582.10^{-03}[0.01\%]\ $     \\
          $3$    & $3.776283.10^{-05}$    & $3.77696202.10^{-05}[0.02\%]\ \ \ $      & $4.745961.10^{-04}$     & $4.74663748.10^{-04}[0.01\%]\ $     \\
          $4$    & $1.149375.10^{-05}$    & $1.14987458.10^{-05}[0.04\%]\ \ \ $      & $1.399210.10^{-04}$     & $1.39978027.10^{-04}[0.04\%]\ $     \\
          $5$    & $3.837470.10^{-06}$    & $3.84046353.10^{-06}[0.08\%]\ \ \ $      & $4.575322.10^{-05}$     & $4.57886526.10^{-05}[0.08\%]\ $     \\
          \hline
          Total  & $2.102273.10^{-04}$    & $2.10242676.10^{-04}[0.007\%]\ $     & $2.752676.10^{-03}$     & $2.75262625.10^{-03}[0.002\%]$     \\
          \hline\hline
        \end{tabular}
        \caption{ Elliptic orbit: average of the $\ell$-mode energy (units of $M^2/m_0^2$) and angular momentum (units of $M/m^2_0$) fluxes radiated to infinity and taken over a few periods in the case of an elliptic orbit  $(e,p)$=$(0.764124,8.75455)$. They are computed according to Eqs. (\ref{eeccaveq}, \ref{leccaveq}); further each $\ell$-mode is obtained by summing the flux over all the azimuthal $m$-modes.  The differences with Hopper and Evans \cite{hoev10} are shown.}
        \label{hopper_ecc}
\end{table*}

%\begin{widetext}
\renewcommand{\arraystretch}{1.2}
\setlength{\tabcolsep}{8pt} %donne la distance entre les collones%

\begin{table*}%[h]
        \begin{tabular}{c*{3}{c}*{3}{c}}
          \hline\hline
          $p$ & $E^\infty$ & $E^\infty$ \cite{ma04} & $E^\infty$ \cite{sola06} & $E^\text{eh}$ & $E^\text{eh}$ \cite{ma04} & $E^\text{eh}$ \cite{sola06}\\
          \hline
          $8.00001$ & $3.5820$ & $3.6703[2.4\%]$ & $3.5603[0.6\%]$ & $1.8900.10^{-1}$ & $1.8876.10^{-1}[0.1\%]$ & $1.8884.10^{-1}[0.008\%]$ \\
          $8.001$   & $2.2350$ & $2.2809[2.0\%]$ & $2.2212[0.6\%]$ & $1.1349.10^{-1}$ & $1.1260.10^{-1}[0.8\%]$ & $1.1339.10^{-1}[0.09\%]$  \\
          \hline\hline
        \end{tabular}
\caption{ Parabolic orbit: energy radiated to infinity $E^\infty$, and to the horizon $E^{\text{eh}}$ (units of $M/m_0^2$), computed according to Eq. (\ref{epar}), for $p\simeq 8$. The differences with the results of Martel \cite{ma04}, and Sopuerta and Laguna \cite{sola06} are shown.}
        \label{Epara}
\end{table*}
%\end{widetext}

\renewcommand{\arraystretch}{1.2}
\setlength{\tabcolsep}{8pt} %donne la distance entre les collones%
\begin{table*}
        \begin{tabular}{c*{3}{c}*{3}{c}}
          \hline\hline
          $p$ & $L^\infty$ & $L^\infty$ \cite{ma04} & $L^\infty$ \cite{sola06} & $L^\text{eh}$ & $L^\text{eh}$ \cite{ma04} & $L^\text{eh}$ \cite{sola06}\\
          \hline
          $8.00001$ & $2.9596.10^{1}$ & $3.0133.10^{1}[1.8\%]$ & $2.9415.10^{1}[0.6\%]$ & $1.5137$         & $1.5208[0.5\%]$         & $1.5112[0.2\%]$ \\
          $8.001$   & $1.8813.10^{1}$ & $1.9088.10^{1}[1.4\%]$ & $1.8704.10^{1}[0.6\%]$ & $9.0964.10^{-1}$ & $9.1166.10^{-1}[0.2\%]$ & $9.0783.10^{-1}[0.2\%]$  \\
          \hline\hline
        \end{tabular}
\caption{ Parabolic orbit: angular momentum radiated to infinity $L^\infty$, and to the horizon $L^{\text{eh}}$ (units of $1/m_0^2$), computed according to Eq. (\ref{lpar}), for $p\simeq 8$. The differences with the result of Martel \cite{ma04}, and Sopuerta and Laguna \cite{sola06} are shown.}
        \label{Lpara}
\end{table*}

%%%%%%%%%%%%%%%%%%%%%%%%%%%%%%%%%%%%%%%%%%%%%%%%%%%%%%%%%%%%%%%%%%%%%%%%%%%%%%%%%%%%%%%%%%%%%%%%%%%%%%%%%%%%%%%%%%%%%%%%%%%%%%%%%%%%%%%%%%%%%%%%%%%%%%%%%%%%%%%%%%%%%%%%%%%%%%%%%%%%%%%%%%%%%%%%%%%%%%%%%%%%%%%%%%%%%%%%%%%%%%%%%%%%%%%%%%%%%%%%%%%%%%%%%%%%%%%%%%%%%%%%%%%%%%%%%%%%%%%%%%%%%
%%%%%%%%%%%%%%%%%%%%%%%%%%%%%%%%%%%%%%%%%%%%%%%%%%%%%%%%%%%%%%%%%%%%%%%%%%%%%%%%%%%%%%%%%%%%%%%
\subsection{Elliptic orbits}

\begin{center}
    \begin{figure}[h!]
        \includegraphics[width=0.6\linewidth]{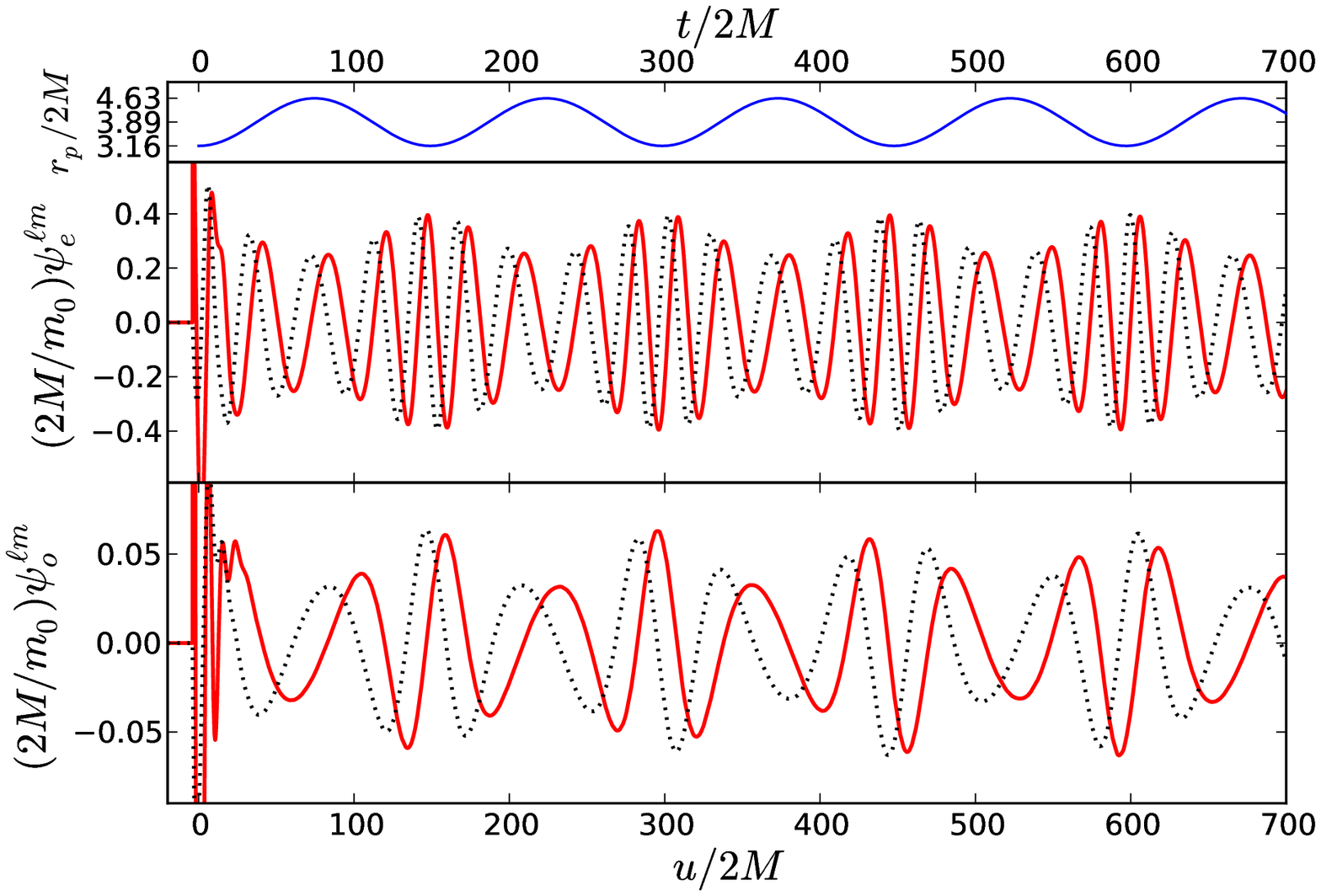}
        \caption{The $\psi$ wave-function in terms of the coordinates $t/2M$ and  $u = (t - r*_\text{\air obs})/2M$, for the even quadrupolar mode $(\ell,m)$=$(2,2)$ (top) and the odd mode $(\ell,m)$=$(2,1)$ (bottom); orbital parameters $(e,p)=(0.188917, 7.50478)$. We draw the real { (solid line)} and the imaginary { (dotted line)} parts.
        % The values in SI units for $\psi$ are obtained through multiplication by {\rosso $m/r_{\rm g}$} - in geometrised units - and by $c^2 G^{-0.5}$
        }
        \label{fig09}
    \end{figure}
\end{center}

In Fig. (\ref{fig09}), an elliptic orbit has been considered, $(e,p)\!=\!(0.188917,7.50478)$, { for $(\ell,m)$=$(2,2)$ and $(\ell,m)$=$(2,1)$ modes. Both real and imaginary parts are shown. 
The average radiated energy and angular momentum fluxes are given by}

\beq
<\dott{E}>=\frac{1}{T_2-T_1}\int_{T_1}^{T_2}\dott{E}dt
\label{eeccaveq}
\eeq

\beq
<\dott{L}>=\frac{1}{T_2-T_1}\int_{T_1}^{T_2}\dott{L}dt
\label{leccaveq}
\eeq
where $T_2-T_1=kT_\text{orb}$, { $k\geqslant5$}. 

In Tabs. \ref{Eeccave}, \ref{Leccave}, the average of the energy $\dott{E}^\infty$ (units of $M^2/m_0^2$) and 
the angular momentum $\dott{L}^\infty$ (units of $M/m_0^2$) fluxes at infinity are computed over few periods; for $(e,p)=(0.188917, 7.50498)$ and $(e,p) = (0.764124, 8.75455)$.  Our results are compared with those of Cutler {\it et al.} \cite{cukepo94}, Martel \cite{ma04}, and Sopuerta and Laguna \cite{sola06}. For $\dott{E}^\infty$, our results differ with the first by up to $0.2\%$, with the second by up to $2.1\%$, and with the third by up to $0.1\%$. 

In Tab. \ref{hopper_ecc}, the average of the $\ell$-mode energy (units of $M^2/m_0^2$) and angular momentum (units of $M/m^2_0$) fluxes radiated to infinity, taken over a few periods for $(e,p)$=$(0.764124,8.75455)$, are computed. Each $\ell$-mode is obtained by summing the flux over all the azimuthal $m$-modes such that $(\dott{E}^\infty_\ell,\dott{L}^\infty_\ell)=\sum_{m=-\ell}^\ell(\dott{E}^\infty_{\ell m},\dott{L}^\infty_{\ell m})$. Our results differ from those of Hopper and Evans \cite{hoev10} \footnote{Hopper and Evans compare their results with those of Fujita {\it et al.} \cite{fuhita09}.} by $0.007\%$ in energy flux, and by $0.002\%$ in angular momentum flux.

%For $\dott{L}^\infty$, our results differ with the first by up to $0.6\%$, with the second by up to $0.8\%$, and with the third by up to $0.09\%$.        

\begin{center}
    \begin{figure}[h!]
        \includegraphics[width=0.6\linewidth]{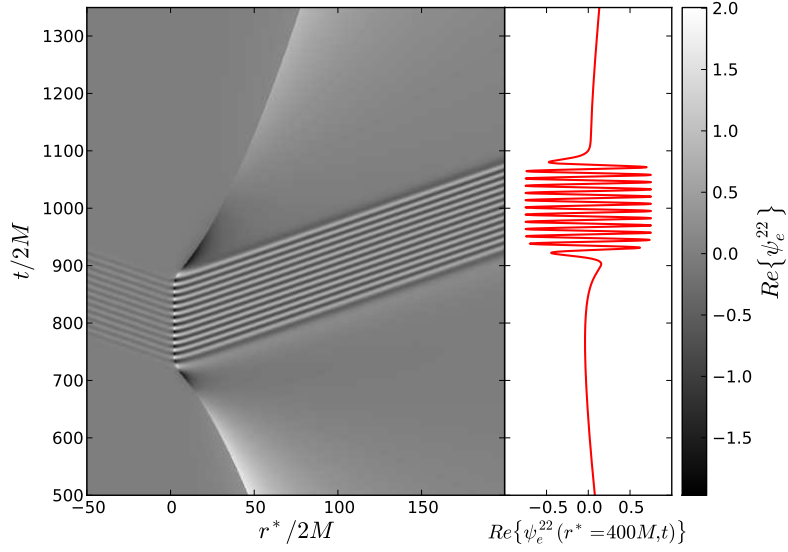}
        \caption{Even $\psi(r^*,t)$ 2D-function for the zoom-whirl case of orbital parameters $(e,p)=(1.0,~8.00001)$, and of mode $(\ell,m) = (2, 2)$ (left panel). The wave propagates on both sides of the world line, and it is recorded by an observer located at $400M$ (right panel).}
        \label{fig10}
    \end{figure}
\end{center}

\begin{center}
    \begin{figure}[h!]
        \includegraphics[width=0.6\linewidth]{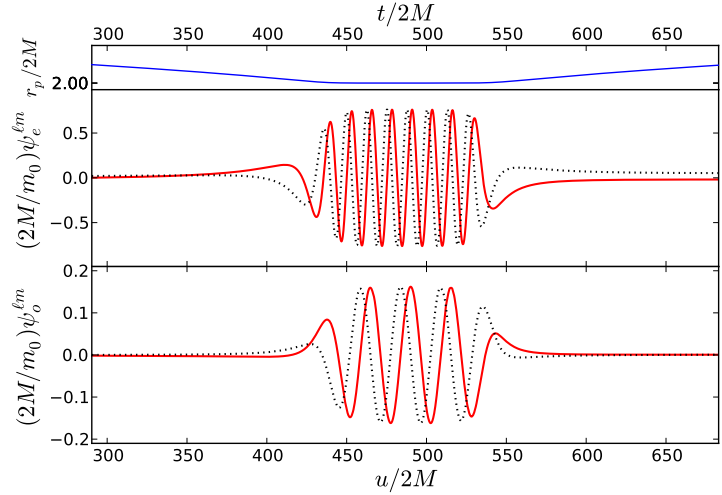}
        \caption{The $\psi$ wave-function in terms of the coordinates $t/2M$ and $u = (t - r*_\text{ obs})/2M$, for the even quadrupolar mode $(\ell,m)$=$(2,2)$ (top) and the odd mode $(\ell,m)$=$(2,1)$ (bottom); orbital parameters $(e, p) = (1, 8.001)$. We draw the real { (solid line)} and the imaginary { (dotted line)} parts.
        % The values in SI units for $\psi$ are obtained through multiplication by {\rosso $m/r_{\rm g}$} - in geometrised units - and by $c^2 G^{-0.5}$.
        }
        \label{fig11}
    \end{figure}
\end{center}

%%%%%%%%%%%%%%%%%%%%%%%%%%%%%%%%%%%%%%%%%%%%%%%%%%%%%%%%%%%%%%%%%%%%%%%%%%%%%%%%%%%%%%%%%%%%%%%%%%%%%%%%%%%%%%%%%%%%%%%%%%%%%%%%%%%%%%%%%%%%%%%%%%%%%%%%%%%%%%%%%%%%%%%%%%%%%%%%%%%%%%%%%%%%%%%%%%%%%%%%%%%%%%%%%%%%%%%%%%%%%%%%%%%%%%%%%%%%%%%%%%%%%%%%%%%%%%%%%%%%%%%%%%%%%%%%%%%%%%%%%%%%%
%%%%%%%%%%%%%%%%%%%%%%%%%%%%%%%%%%%%%%%%%%%%%%%%%%%%%%%%%%%%%%%%%%%%%%%%%%%%%%%%%%%%%%%%%%%%%%%
\subsection{Parabolic orbits}

Parabolic orbits include the zoom-whirl case. Indeed, for $e=1$ and $p\gtrsim8$, the orbit, though close to the separatrix, stays in the stability domain. When the particle reaches the periastron it momentarily circles the SMBH, leading to oscillations in the WFs, Figs. (\ref{fig10},\ref{fig11}). The energy and angular momentum are obtained by integrating the flux over a symmetric lapse of time to the periastron

\beq
E=\int_{T_1}^{T_2}\dott{E}dt~,
\label{epar}
\eeq

\beq
L=\int_{T_1}^{T_2}\dott{L}dt~,
\label{lpar}
\eeq
where $(T_1+T_2)/2=T_\text{peri}$ is the { instant at which} {$r_p=r_\text{peri}=pM/2$}. In practice, we use $T_1=T_\text{peri}-600M$ and $T_2=T_\text{peri}+600M$.

In Tab. \ref{Epara}, the energy radiated to infinity $E^\infty$, and to the horizon $E^{\text{eh}}$ (units of $M/m_0^2$), computed according to Eq. (\ref{epar}), { are shown still for $p\simeq 8$}. 
In Tab. \ref{Lpara}, the angular momentum radiated to infinity $L^\infty$, and to the horizon $L^{\text{eh}}$ (units of $1/m_0^2$), computed according to Eq. (\ref{lpar}), { are shown still for $p\simeq 8$}. Our results are compared to those of Martel \cite{ma04} and Sopuerta and Laguna \cite{sola06}. With the former work, our results differ by up to $2.4\%$ for $E^\infty$ and up to $0.8\%$ for $E^{\text{eh}}$; with the latter, our results differ by up to $0.6\%$ for $E^\infty$, and up to $0.09\%$ for $E^{\text{eh}}$. 
Further, with the former work, our results differ by up to $1.8\%$ for $L^\infty$ and up to $0.5\%$ for $L^{\text{eh}}$; with the latter, our results differ by up to $0.6\%$ for $L^\infty$, and up to $0.2\%$ for $L^{\text{eh}}$.

%%%%%%%%%%%%%%%%%%%%%%%%%%%%%%%%%%%%%%%%%%%%%%%%%%%%%%%%%%%%%%%%%%%%%%%%%%%%%%%%%%%%%%%%%%%%%%%%%%%%%%%%%%%%%%%%%%%%%%%%%%%%%%%%%%%%%%%%%%%%%%%%%%%%%%%%%%%%%%%%%%%%%%%%%%%%%%%%%%%%%%%%%%%%%%%%%%%%%%%%%%%%%%%%%%%%%%%%%%%%%%%%%%%%%%%%%%%%%%%%%%%%%%%%%%%%%%%%%%%%%%%%%%%%%%%%%%%%%%%%%%%%%
%%%%%%%%%%%%%%%%%%%%%%%%%%%%%%%%%%%%%%%%%%%%%%%%%%%%%%%%%%%%%%%%%%%%%%%%%%%%%%%%%%%%%%%%%%%%%%%
\section{Conclusions and perspectives}

Pursuing the work presented for radial fall at first order \cite{aosp11} and fourth order \cite{rispaoco11}, the applicability of the source-free or indirect method has been extended to generic orbits by means of a second order algorithm.   

The indirect method is based on finite differences, and it employs the jump conditions on the wave-function and its derivatives to step forward in the time domain grid to the next cell crossed by the particle; it avoids the regularisation of the Dirac distribution (and derivatives), and deals with the particle as a true singularity. 

Its performance may be summarised as follows: i) in  terms of radiated energy and angular momentum, it confirms the results of previous literature, obtaining a relative error in many cases down to 0.001 \%; ii) it provides sufficiently smooth wave-forms with relatively low computational cost for our future developments;  iii) the algorithm is { quite comprehensible} and easy to implement, thanks to the analytic expressions dealing with the source and the potential.

In the companion paper (Part II), the upgrading of the scheme to fourth allows the computation of the self force in the RW gauge for radial fall. 

Beyond the scenario of EMRIs, it would be worth exploring the indirect method for all wave-equations with a singular source term.

\section*{Acknowledgements}

We thank C. F. Sopuerta (Barcelona), L. Blanchet (Paris) and L. Burko (Huntsville) for the comments received. P. Ritter's thesis, on which this paper is partly based, received an honorable mention by the Gravitational Wave International (GWIC) and Stefano Braccini Thesis Prize Committee in 2013.

\appendix

%%%%%%%%%%%%%%%%%%%%%%%%%%%%%%%%%%%%%%%%%%%%%%%%%%%%%%%%%%%%%%%%%%%%%%%%%%%%%%%%%%%%%%%%%%%%%%%
%%%%%%%%%%%%%%%%%%%%%%%%%%%%%%%%%%%%%%%%%%%%%%%%%%%%%%%%%%%%%%%%%%%%%%%%%%%%%%%%%%%%%%%%%%%%%%%
%%%%%%%%%%%%%%%%%%%%%%%%%%%%%%%%%%%%%%%%%%%%%%%%%%%%%%%%%%%%%%%%%%%%%%%%%%%%%%%%%%%%%%%%%%%%%%%

\section{Multipolar expansion and linearised field equations in the RW gauge}
\label{Multipolar.Expansion}
The stress-energy tensor of a point particle with mass $m_0$ moving along the world-line $\gamma$ is

\beq
  T^\ab =m_0\int_\gamma(-g)^{-1/2}u^\alpha u^\beta\delta^{(4)}\left(x^\alpha-x_p^\alpha(\tau)\right)d\tau
  =m_0\frac{u^\alpha u^\beta}{r^2\sin\theta}{ u^t} \delta\left(r-r_p(t)\right)\delta\left(\theta-\theta_p(t)\right)\delta\left(\phi-\phi_p(t)\right)~,
  \label{T:local}
\eeq
which is a rank 2 symmetric tensor expandable as
\beq
T_\ab=\sum_\lm\sum_{i=1}^{10}T^{(i)\lm}(t,r)Y_\ab^{(i)\lm}(\theta,\phi)~.
\label{dvlp:T}
\eeq
{
For the tensorial spherical harmonics $Y_\ab^{(i)\lm}$, see Tab. \ref{table:Yi:to:YZ}. The functions $T^{(i)\lm}(t,r)$ are given by
}
\beq
T^{(i)\lm}(t,r)=\int_{\mathbb{S}^2}\eta^{\alpha\gamma}\eta^{\beta\delta}T_\ab\left(Y_{\gamma\delta}^{(i)\ell m}\right)^\star d\Omega~.
\eeq
where $T_\ab$ is given by Eq. (\ref{T:local}), and $\eta_\ab=\text{diag}(1,1,r^2,r^2\sin^2\theta)$. The explicit computation of $T^{(i)\lm}$ leads to

\begin{widetext}
\begin{subequations}
  \begin{align}
  &T^{(1)\lm}(t,r)=m_0u^t\left(\frac{dr_p}{dt}\right)^2r^{-2}f(r)^{-2}\delta\left(r-r_p(t)\right)Y^{\lm\star}\left(\theta_p,\phi_p\right)\\
  &T^{(2)\lm}(t,r)=\sqrt{2}im_0u^t\frac{dr_p}{dt}r^{-2}\delta\left(r-r_p(t)\right)Y^{\lm\star}\left(\theta_p,\phi_p\right)\\
  &T^{(3)\lm}(t,r)=m_0u^tr^{-2}f(r)^2\delta\left(r-r_p(t)\right)Y^{\lm\star}\left(\theta_p,\phi_p\right)\\
  &T^{(4)\lm}(t,r)=2\mathfrak{m}(\ell)^{-1}im_0u^tr^{-1}f(r)\delta\left(r-r_p(t)\right)\frac{d}{dt}Y^{\lm\star}\left(\theta_p,\phi_p\right)\\
  &T^{(5)\lm}(t,r)=2\mathfrak{m}(\ell)^{-1}m_0u^t\frac{dr_p}{dt}r^{-1}f(r)^{-1}\delta\left(r-r_p(t)\right)\frac{d}{dt}Y^{\lm\star}\left(\theta_p,\phi_p\right)\\
  &T^{(6)\lm}(t,r)=-2\mathfrak{m}(\ell)^{-1}m_0u^tr^{-1}f(r)\delta\left(r-r_p(t)\right)\left[\frac{1}{\sin\theta_p}\frac{\partial Y^{\lm\star}}{\partial\phi_p}\frac{d\theta_p}{dt}-\sin\theta_p\frac{\partial Y^{\lm\star}}{\partial\theta_p}\frac{d\phi_p}{dt}\right]\\
  &
  \begin{aligned}
  T^{(7)\lm}(t,r)=-2\mathfrak{m}(\ell)^{-1}im_0u^t\frac{dr_p}{dt}r^{-1}f(r)^{-1}\delta\left(r-r_p(t)\right)\left[\frac{1}{\sin\theta_p}\frac{\partial Y^{\lm\star}}{\partial\phi_p}\frac{d\theta_p}{dt}-\sin\theta_p\frac{\partial Y^{\lm\star}}{\partial\theta_p}\frac{d\phi_p}{dt}\right]
  \end{aligned}\\
  &
  \begin{aligned}
  T^{(8)\lm}(t,r)=&-\mathfrak{n}(\ell)^{-1}im_0u^t\delta\left(r-r_p(t)\right)\left\{\frac{1}2\left[\left(\frac{d\theta_p}{dt}\right)^2-\sin^2\theta_p\left(\frac{d\phi_p}{dt}\right)^2\right]\frac{X^{\lm\star}\left(\theta_p,\phi_p\right)}{\sin\theta_p}- \right.\\
  &\left.\sin\theta_p\frac{d\phi_p}{dt}\frac{d\theta_p}{dt}W^{\lm\star}\left(\theta_p,\phi_p\right)\right\}
  \end{aligned}\\
  &\begin{aligned}
  T^{(9)\lm}(t,r)=& \mathfrak{n}(\ell)^{-1}m_0u^t\delta\left(r-r_p(t)\right)\left\{\frac{1}2\left[\left(\frac{d\theta_p}{dt}\right)^2-\sin^2\theta_p\left(\frac{d\phi_p}{dt}\right)^2\right]W^{\lm\star}\left(\theta_p,\phi_p\right)+ \right.\\
  &\left.\frac{d\phi_p}{dt}\frac{d\theta_p}{dt}X^{\lm\star}\left(\theta_p,\phi_p\right)\right\}\\
  \end{aligned}\\
  &T^{(10)\lm}(t,r)=\frac{1}{\sqrt{2}}m_0u^t\delta\left(r-r_p(t)\right)\left[\left(\frac{d\theta_p}{dt}\right)^2+\sin^2\theta_p\left(\frac{d\phi_p}{dt}\right)^2\right]Y^{\lm\star}\left(\theta_p,\phi_p\right)
  \end{align}
  \label{coef:T}
\end{subequations}
\end{widetext}
where $X^\lm(\theta,\phi)$ and $W^\lm(\theta,\phi)$ are given by Eqs. ({\ref{X},\ref{W}), and 
\begin{align}
  &\mathfrak{m}(\ell)=\sqrt{2\ell(\ell+1)}~,\\
  &\mathfrak{n}(\ell)=\sqrt{\frac{1}2\ell(\ell+1)(\ell-1)(\ell+2)}~.
\end{align}

{
We perform a multipolar expansion of the perturbation metric tensor
\beq
h_\ab=\sum_\lm\sum_{i=1}^{10}h^{(i)\lm}(t,r)Y_\ab^{(i)\lm}(\theta,\phi)~,
\label{dvlp:h}
\eeq
where the functions $h^{(i)\lm}$ are linked to the RW metric perturbation functions, see Tab. \ref{table:Yi:to:YZ}. Inserting Eqs. (\ref{dvlp:T}, \ref{dvlp:h}) into Eq. (\ref{perturbed:EE}), we get for each radiative modes ($\ell\geq2$), the linearised field equations in terms of the RW perturbation functions and stress-energy tensor coefficients, Eq. (\ref{coef:T}). For even parity, for which in the RW gauge $h_0^{(e)\lm}=h_1^{(e)\lm}=G^\lm=0$ for $\forall\ \ell\geq2$, we have
}
\begin{widetext}
\begin{subequations}
    \beq
    \begin{aligned}
    f(r)\left[ f(r)\frac{\partial^2K^\lm}{\partial r^2}
    +\frac{1}r\left(3-\frac{5M}r\right)\frac{\partial K^\lm}{\partial r}
    -\frac{1}rf(r)\frac{\partial H_2^\lm}{\partial r}
    -\frac{1}{r^2}\left(H_2^\lm-K^\lm\right)- \right.
    \frac{\ell(\ell+1)}{2r^2}\left(\!\!\right.&\left.\left.H_2^\lm+ K^\lm\right)\right.\Big ]=\\
    &-8\pi T^{(1)\lm}~,
    \label{EE:1:e}
    \end{aligned}
    \eeq
    \\
    \beq
    \frac{\partial}{\partial t}\left[\frac{\partial K^\lm}{\partial r}+\frac{1}r\left(K^\lm-H_2^\lm\right)-\frac{M}{r^2f(r)}K^\lm\right]
    -\frac{\ell(\ell+1)}{2r^2}H_1^\lm=-4\sqrt{2}\pi iT^{(2)\lm}~,
    \label{EE:2:e}
    \eeq
    \\
    \beq
    \begin{aligned}
    \frac{1}{f(r)^{2}}\frac{\partial^2K^\lm}{\partial t^2}
    -\frac{r-M}{r^2f(r)}\frac{\partial K^\lm}{\partial r}
    -\frac{2}{rf(r)}\frac{\partial H_1^\lm}{\partial t}
    +\frac{1}r\frac{\partial H_0^\lm}{\partial r}+\frac{1}{r^2f(r)}\left(H_2^\lm-K^\lm\right)+
    \frac{\ell(\ell+1)}{2r^2f(r)}\left(\!\!\right.&\left.K^\lm-H_0^\lm\right)=\\
    &-8\pi T^{(3)\lm}~,
    \label{EE:3:e}
    \end{aligned}
    \eeq
    \\
    \beq
    \frac{\partial}{\partial r}\left[f(r)H_1^\lm\right]-\frac{\partial}{\partial t}\left(H_2^\lm+K^\lm\right)=\frac{4\pi i}{\mathfrak{m}(\ell)}rT^{(4)\lm}~,
    \label{EE:4:e}
    \eeq
    \\
    \beq
    -\frac{\partial H_1^\lm}{\partial t}+f(r)\frac{\partial}{\partial r}\left(H_0^\lm-K^\lm\right)+\frac{2M}{r^2}H_0^\lm+\frac{1}r\left(1-\frac{M}{r}\right)\left(H_2^\lm-H_0^\lm\right)=\frac{4\pi}{\mathfrak{m}(\ell)}rf(r)T^{(5)\lm}~,
    \label{EE:5:e}
    \eeq
    \\
    \beq
    \begin{aligned}
    -\frac{1}{f(r)}\frac{\partial^2K^\lm}{\partial t^2}+ & f(r)\frac{\partial^2K^\lm}{\partial r^2}+\frac{2}{r}\left(1-\frac{M}r\right)\frac{\partial K^\lm}{\partial r}-\frac{1}{f(r)}\frac{\partial^2H_2^\lm}{\partial t^2}+2\frac{\partial^2H_1^\lm}{\partial t\partial r} 
- f(r)\frac{\partial^2H_0^\lm}{\partial r^2}+\frac{2(r-M)}{r^2f(r)}\frac{\partial H_1^\lm}{\partial t}\\
    &-\frac{1}r\left(1-\frac{M}r\right)\frac{\partial H_2^\lm}{\partial r}-\frac{r+M}r\frac{\partial H_0^\lm}{\partial r}+
    \frac{\ell(\ell+1)}{2r^2}\left(H_0^\lm-H_2^\lm\right)=8\sqrt{2}\pi T^{(10)\lm}~,
    \label{EE:6:e}
    \end{aligned}
    \eeq
    \\
    \beq
    \frac{1}2\left(H_0^\lm-H_2^\lm\right)=\frac{8\pi}{\mathfrak{n}(\ell)}r^2T^{(9)\lm}~,
    \label{EE:7:e}
    \eeq
    \label{EE:even}
\end{subequations}
\end{widetext}
that is seven coupled equations for the four RW unknowns $H_0^\lm$, $H_1^\lm$, $H_2^\lm$ and $K^\lm$. For the odd parity, for which the RW gauge is $h_2^\lm=0$ for $\forall\ \ell\geq2$, we have

\begin{widetext}
\begin{subequations}
    \beq
    \frac{\partial^2h_0^\lm}{\partial r^2}-\frac{\partial^2h_1^\lm}{\partial t\partial r}-\frac{2}r\frac{\partial h_1^\lm}{\partial t}+\left[\frac{4M}{r^2}-\frac{\ell(\ell+1)}r\right]\frac{h_0^\lm}{rf(r)}=\frac{4\pi}{\mathfrak{m}(\ell)}r^{-1}f(r)^{-1}T^{(6)^\lm}~,
    \label{EE:1:o}
    \eeq
    \beq
    \frac{\partial^2h_1^\lm}{\partial t^2}-\frac{\partial^2h_0^\lm}{\partial t\partial r}+\frac{2}r\frac{\partial h_0^\lm}{\partial t}+\frac{(\ell-1)(\ell+2)f(r)}{r^2}h_1^\lm=-\frac{4\pi i}{\mathfrak{m}(\ell)}rf(r)T^{(7)\lm}~,
    \label{EE:2:o}
    \eeq
    \beq
    \frac{\partial}{\partial r}\Big[f(r)h_1^\lm\Big]-f(r)^{-1}\frac{\partial h_0^\lm}{\partial t}=-\frac{8\pi i}{\mathfrak{n}(\ell)}r^2T^{(8)^\lm}~,
    \label{EE:3:o}
    \eeq
    \label{EE:odd}
\end{subequations}
\end{widetext}
that is three coupled equations for the two RW unknowns functions $h_0^\lm$ and $h_1^\lm$.
\begin{widetext}
\begin{center}
\begin{table}[!htb]
  {\small
  \begin{tabular}{c*{5}{|c}}
  \hline
  \hline
  $i$&$1$&$2$&$3$&$4$&$5$\\ \hline
  $h^{(i)\lm}$ & $f(r)H_0^\lm$ & $-i\sqrt{2}H_1^\lm$ & $f(r)^{-1}H_2^\lm$ & $\frac{-i}{r}\mathfrak{m}(\ell)h^{(e)\lm}_0$ & $r^{-1}\mathfrak{m}(\ell)h^{(e)\lm}_1$\\
  $T^{(i)\lm}$ & $A^{(0)\lm}$ & $A^{(1)\lm}$ & $A^{\lm}$ & $B^{(0)\lm}$ & $B^{\lm}$\\
  $Y_\ab^{(i)\lm}$ & $a^{(0)\lm}$ & $a^{(1)\lm}$ & $a^{\lm}$ & $b^{(0)\lm}$ & $b^{\lm}$\\
  parity&even&even&even&even&even\\
  \hline
  \hline
  $i$&$6$&$7$&$8$&$9$&$10$\\ \hline
  $h^{(i)\lm}$ & $-r^{-1}\mathfrak{m}(\ell)h^{\lm}_0$ & $ir^{-1}\mathfrak{m}(\ell)h^{\lm}_1$ & $r^{-2}\mathfrak{m}(\ell)h^{\lm}_2$ & $\mathfrak{n}(\ell)G^\lm$ & $\sqrt{2}K^\lm-\frac{\mathfrak{m}(\ell)^2}{2\sqrt{2}}G^\lm$\\
  $T^{(i)\lm}$ & $Q^{(0)\lm}$ & $Q^{\lm}$ & $D^{\lm}$ & $F^{\lm}$ & $G^{(s)\lm}$\\
  $Y_\ab^{(i)\lm}$ & $c^{(0)\lm}$ & $c^{\lm}$ & $d^{\lm}$ & $f^{\lm}$ & $g^{(s)\lm}$\\
  parity&odd&odd&odd&even&even\\
  \hline
  \hline
  \end{tabular}
  \caption{This table provides the relation between Lousto and Price's \cite{lopr97b} and Zerilli's notation  \cite{ze70c}. We adopt the former convention that differs from the one used by Cunningham, Price and Moncrief (CPM) \cite{cuprmo78} for a normalisation factor. Index $i$ of the first row follows the order of the TSH as defined in \cite{ze70c}.}
  \label{table:Yi:to:YZ}
  }
\end{table}
\end{center}
\end{widetext}

\section{Notes on distributions}

We propose here to formally justify some computations used in the main body of the work. 
The concept of distribution is understood through the notion of test function. The set of test functions $\mathcal{D}(\Omega)$ belongs to the continuity class $\mathcal{C}^\infty$ on $\Omega\subset\mathbb{R}^n\to\mathbb{R}$ with a compact support (that is to say, identically zero outside of a certain range limits). We define any distribution $T$ in its integral form by
\beq
\left<T,\varphi\right>=\int_{\mathbb{R}^n}T(x)\varphi(x)dx~,
\eeq
where $\varphi$ is a test function of $\mathcal{D}(\Omega)$.

%%%%%%%%%%%%%%%%%%%%%%%%%%%%%%%%%%%%%%%%%%%%%%%%%%%%%%%%%%%%%%%%%%%%%%%%%%%%%%%%%%%%%%%%%%%%%%%
%%%%%%%%%%%%%%%%%%%%%%%%%%%%%%%%%%%%%%%%%%%%%%%%%%%%%%%%%%%%%%%%%%%%%%%%%%%%%%%%%%%%%%%%%%%%%%%
%%%%%%%%%%%%%%%%%%%%%%%%%%%%%%%%%%%%%%%%%%%%%%%%%%%%%%%%%%%%%%%%%%%%%%%%%%%%%%%%%%%%%%%%%%%%%%%
\subsection{Dirac Distribution $\delta$}
We define the Dirac distribution $\delta_{x_0}\defeq\delta(x-x_0)$ as the distribution { which associates $\varphi(x_0)$ to a function test $\varphi(x)$. }  It is given by its integral form
\beq
\left<\delta_{x_0},\varphi\right>=\int_{-\infty}^{+\infty}\delta(x-x_0)\varphi(x)dx=\varphi(x_0)~.
\eeq
Multiplying $\delta_{x_0}$ by an infinitely differentiable function is straightforward, since for any function $P\in\mathcal{C}^\infty$ and $\varphi\in\mathcal{D}(\mathbb{R})$ we have $P\varphi\in\mathcal{D}(\mathbb{R})$; therefore
\beq
\left<\delta_{x_0}P,\varphi\right>=\left<\delta_{x_0},P\varphi\right>=P(x_0)\left<\delta_{x_0},\varphi\right>~.
\eeq

Thus, in the most general case, if $P(t,r)$ is an infinitely differentiable function, and $\delta_{r_p}$ the Dirac distribution whose support is $r=r_p(t)$ we write
\beq
\left<\delta_{r_p} P,\varphi\right>=\int_{-\infty}^{+\infty}\delta(r-r_p(t))P(t,r)\varphi(r)dr
=P(t,r\!=\!r_p(t))\left<\delta_{r_p},\varphi\right>~,
\eeq
and
\beq
P(t,r)\delta(r-r_p(t))=\tilde{P}(t)\delta(r-r_p(t))~,
\label{prop:delta}
\eeq
where $\tilde{P}(t)\defeq P(t,r\!=\!r_p(t))$ is an infinitely one-dimensional differentiable function which results from the evaluation of $P(t,r)$ with the Dirac distribution $\delta$. The total derivative is 

\beq
\frac{d\tilde{P}(t)}{dt}=\Big[\partial_t P(t,r)+\dott{r}_p\partial_r P(t,r)\Big]_{r\!=\!r_p(t)}
\label{tot:derivative}
\eeq
where "$\rescale[0.5]{\circ}$" is the total derivative with respect to $t$.

%%%%%%%%%%%%%%%%%%%%%%%%%%%%%%%%%%%%%%%%%%%%%%%%%%%%%%%%%%%%%%%%%%%%%%%%%%%%%%%%%%%%%%%%%%%%%%%
%%%%%%%%%%%%%%%%%%%%%%%%%%%%%%%%%%%%%%%%%%%%%%%%%%%%%%%%%%%%%%%%%%%%%%%%%%%%%%%%%%%%%%%%%%%%%%%
%%%%%%%%%%%%%%%%%%%%%%%%%%%%%%%%%%%%%%%%%%%%%%%%%%%%%%%%%%%%%%%%%%%%%%%%%%%%%%%%%%%%%%%%%%%%%%%
\subsection{Derivative of Dirac distribution $\delta'$}
Using the property of derivation in distributional sense $\left<T',\varphi\right>=-\left<T,\varphi'\right>$ we directly get the derivative  $\delta'_{x_0}\defeq\partial_x\delta_{x_0}$ by replacing $T$ by $\delta_{x_0}$\beq
\left<\delta'_{x_0},\varphi\right>=-\left<\delta_{x_0},\varphi'\right>=-\varphi'(x_0)~.
\label{delta:derivative}
\eeq
Multiplying $\delta'_{x_0}$ by a function $P\in\mathcal{C}^\infty$, then

\beq
\begin{aligned}
\left<\delta'_{x_0}P,\varphi\right>&=\left<\delta'_{x_0},P\varphi\right>=-\left<\delta_{x_0},(P\varphi)'\right>
=-\left<\delta_{x_0},P\varphi'\right>-\left<\delta_{x_0},P'\varphi\right>
=P(x_0)\left<\delta'_{x_0},\varphi\right>-P'(x_0)\left<\delta_{x_0},\varphi\right>~,
\end{aligned}
\eeq
where, Eq. (\ref{delta:derivative}), the derivative of a test function is a test function. Thus, in the most general case, if $P(t,r)$ is an infinitely differentiable function and $\delta_{r_p}$ the Dirac distribution whose support is $r=r_p(t)$, we get
\beq
\begin{aligned}
\left<\delta_{r_p}' P,\varphi\right>&=\int_{-\infty}^{+\infty}\delta'(r-r_p(t))P(t,r)\varphi(r)dr =P(t,r\!=\!r_p(t))\left<\delta_{r_p}',\varphi\right>-\left.\frac{\partial P(t,r)}{\partial r}\right|_{r\!=\!r_p(t)}\left<\delta_{r_p},\varphi\right>~.
\end{aligned}
\eeq
We directly write
\beq
P(t,r)\delta'(r-r_p(t))=\tilde{P}(t)\delta'(r-r_p(t))-\tilde{Q}(t)\delta(r-r_p(t))~,
\label{prop:delta:prime}
\eeq
where $\tilde{Q}(t)\defeq\left.\partial_r P(t,r)\right|_{r\!=\!r_p(t)}$. We also evaluate the partial derivative of $\delta$ with respect to $t$, via a change of variable
\beq
\partial_t\delta(r-r_p(t))=-\dott{r}_p\partial_r\delta(r-r_p(t))
\label{t:derivative:delta}
\eeq

%%%%%%%%%%%%%%%%%%%%%%%%%%%%%%%%%%%%%%%%%%%%%%%%%%%%%%%%%%%%%%%%%%%%%%%%%%%%%%%%%%%%%%%%%%%%%%%
%%%%%%%%%%%%%%%%%%%%%%%%%%%%%%%%%%%%%%%%%%%%%%%%%%%%%%%%%%%%%%%%%%%%%%%%%%%%%%%%%%%%%%%%%%%%%%%
%%%%%%%%%%%%%%%%%%%%%%%%%%%%%%%%%%%%%%%%%%%%%%%%%%%%%%%%%%%%%%%%%%%%%%%%%%%%%%%%%%%%%%%%%%%%%%%
\subsection{Heaviside distribution $\mathcal{H}$}

The Heaviside distribution $\mathcal{H}_{x_0}\defeq\mathcal{H}(x-x_0)$ is defined as
\beq
	\mathcal{H}_{x_0} =
	\begin{cases}
	 0, & \text{{ if} }x<x_0\\
	 1, & \text{ if }x\geq x_0
	\end{cases}
\eeq
$\mathcal{H}_{x_0}$ is connected to the Dirac distribution via derivation
\beq
\left<\mathcal{H}'_{x_0},\varphi\right>=\left<\delta_{x_0},\varphi\right>~.
\label{heaviside:derivative}
\eeq

%\beq
%\left<\mathcal{H}',\varphi\right>=-\left<\mathcal{H},\varphi'\right>=-\int_{-\infty}^{+\infty}\mathcal{H}(x)\varphi'(x)dx=-\int_{0}^{+\infty}\varphi'(x)dx=\varphi(0)=\left<\delta,\varphi\right>~.
%\label{heaviside:derivative2}
%\eeq

In our case, being the support of $\mathcal{H}$ time dependent, it is necessary to clarify the partial derivatives $\partial_r\mathcal{H}(r-r_p(t))$ and $\partial_t\mathcal{H}(r-r_p(t))$ in terms of $\delta(r-r_p(t))$. According to Eq. (\ref{heaviside:derivative}), it is evinced  that
\beq
\partial_r\mathcal{H}(r-r_p(t))=\delta(r-r_p(t))~.
\label{r:derivative:heaviside}
\eeq
For the partial derivative with respect to $t$, we get by a change of variable $t\to r_p(t)$
\beq
\partial_t\mathcal{H}(r-r_p(t))=-\dott{r}_p\partial_r\mathcal{ H}(r-r_p(t))~.
\label{t:derivative:heaviside}
\eeq

%%%%%%%%%%%%%%%%%%%%%%%%%%%%%%%%%%%%%%%%%%%%%%%%%%%%%%%%%%%%%%%%%%%%%%%%%%%%%%%%%%%%%%%%%%%%%%%
%%%%%%%%%%%%%%%%%%%%%%%%%%%%%%%%%%%%%%%%% SUBSECTION %%%%%%%%%%%%%%%%%%%%%%%%%%%%%%%%%%%%%%%%%%
%%%%%%%%%%%%%%%%%%%%%%%%%%%%%%%%%%%%%%%%%%%%%%%%%%%%%%%%%%%%%%%%%%%%%%%%%%%%%%%%%%%%%%%%%%%%%%%
\section{Jump conditions: generic orbits}\label{ANNEXE_JC_GEN}

We list here the explicit forms of the jump { conditions} of the $\psi^\ell$ function and its derivatives until 2nd order. Jump conditions are computed on a geodesic given by $(R(t),\Phi(t),\Theta(t))$. The dot "$\cdot$" is the time derivative $d/dt$ such that $\dot{R} = dR/dt = f_R/\mE\sqrt{\mE^2-f_R}$ where we note $f_R\defeq f(R)$. We recall that $\lambda=(\ell-1)(\ell+2)/2$ and we introduce $\text{K} = (\pi \,m_0\,u^t)/(\lambda + 1 )$ { where} $u^t=dt/d\tau=\mE/f_R$, $\Lambda_1(R)=(\lambda R+3M)$ { and}  $\Lambda_2(R)=\lambda R^2(\lambda+1)$.  

\subsection*{0th order}

\beq
\jump{\psi^{\ell m}_e}=8\text{K}\frac{Rf_R}{\Lambda_1(R)}Y^{\lm\star}
\label{jump_e}~,
\eeq

\beq
\jump{\psi^{\ell m}_o}=-\frac{8\text{K}R}{\lambda}\mathcal{A}^{\lm\star}
\label{jump_o}~.
\eeq

\subsection*{1st order}
\beq
\begin{aligned}
\jump{\partial_r\psi^{\ell m}_e}=&\ 8\text{K}\frac{Rf_R(2\dot{R}\dot{u}^t+\ddot{R}{u}^t)}{{u}^t\Lambda_1\left(\dot{R}^2-f_R^2\right)}Y^{\lm\star}-8\text{K}\frac{\dot{R}^2\left[\lambda R^2\left(\lambda+1+{\ds\frac{2M}R}\right)+3M^2\right]}{R\Lambda_1^2\left(\dot{R}^2-f_R^2\right)}Y^{\lm\star} \\
&+ 8\text{K}\frac{f_R^2\left[\lambda R^2\left(\lambda+1+{\ds\frac{4M}R}\right)+9M^2\right]}{R\Lambda_1^2\left(\dot{R}^2-f_R^2\right)}Y^{\lm\star}-8\text{K}\frac{\dot{\Phi}^2R^2f_R^2}{\Lambda_1\left(\dot{R}^2-f_R^2\right)}Y^{\lm\star}
-4\text{K}\frac{\dot{\Phi}^2Rf_R}{\lambda\left(\dot{R}^2-f_R^2\right)}W^{\lm\star}~,
\end{aligned}
\label{jumpdr_e}
\eeq

\beq
\jump{\partial_r\psi^{\ell m}_o}=-\frac{8\text{K}}{\lambda\left(\dot{R}^2-f_R^2\right)}
\left[\frac{R\dot{R}}{{u}^t}\frac{d}{dt}(\mathcal{A}^{\lm\star}{u}^t)+\left(f_R^2+\dot{R}^2\right)\mathcal{A}^{\lm\star}\right]~,
\label{jumpdr_o}
\eeq

\beq
\begin{aligned}
\jump{\partial_t\psi^{\ell m}_e}=&\ 8\text{K}\frac{Rf_R}{\Lambda_1}\frac{d}{dt}Y^{\lm\star}
-8\text{K}\frac{Rf_R\left(\dot{R}^2+f_R^2\right)\dot{u}^t}{{u}^t\Lambda_1\left(\dot{R}^2-f_R^2\right)}Y^{\lm\star}
-8\text{K}\frac{Rf_R\dot{R}\ddot{R}}{\Lambda_1\left(\dot{R}^2-f_R^2\right)}Y^{\lm\star}+8\text{K}\frac{\dot{R}^3\left[(R(\lambda+1)+M\right]}{R\Lambda_1\left(\dot{R}^2-f_R^2\right)}Y^{\lm\star}\\
&- 8\text{K}\frac{\dot{R}f_R^2\left [R(\lambda+1)+3M\right]}{R\Lambda_1\left(\dot{R}^2-f_R^2\right)}Y^{\lm\star}+8\text{K}\frac{\dot{\Phi}^2\dot{R}R^2f_R^2}{\Lambda_1\left(\dot{R}^2-f_R^2\right)}Y^{\lm\star}
+4\text{K}\frac{\dot{\Phi}^2\dot{R}Rf_R}{\lambda\left(\dot{R}^2-f_R^2\right)}W^{\lm\star}~,
\end{aligned}
\label{jumpdt_e}
\eeq

\beq
\jump{\partial_t\psi^{\ell m}_o}=\frac{8\text{K}f_R^2}{\lambda\left(\dot{R}^2-f_R^2\right)}\left[\frac{R}{{u}^t}\frac{d}{dt}(\mathcal{A}^{\lm\star}{u}^t)+2\dot{R}\mathcal{A}^{\lm\star}\right]~.
\label{jumpdt_o}
\eeq

\subsection*{2nd order}

\beq
\begin{aligned}
\jump{\partial^2_r\psi^{\ell m}_e}=
&-\frac{4 m_0 \pi f_R^{3}\Lambda_1^{-1}}{ \lambda\left(\lambda +1\right)  {\left(\dot{R}^2 -f_R^2\right) }^{3} }
\Big[ 2f_R^{-2} \lambda  {R}^{4} {\dot{R} }^{4} {u}^t \ddot{Y}^{\lm\star}-4 \lambda  {R}^{4} {\dot{R} }^{2} {u}^t \ddot{Y}^{\lm\star}\\
&
+2 f_R^{2}  \lambda  {R}^{4} {u}^t \ddot{Y}^{\lm\star}-4f_R^{-2} \lambda  {R}^{4} {\dot{R} }^{4} \dot{u}^t  \dot{Y}^{\lm\star}+4 f_R^{2}  \lambda  {R}^{4} \dot{u}^t  \dot{Y}^{\lm\star}\\
&
 -4f_R^{-2} \lambda  {R}^{4} {\dot{R} }^{3} \ddot{R}  {u}^t \dot{Y}^{\lm\star}+4   \lambda  {R}^{4} \dot{R}  \ddot{R}  {u}^t \dot{Y}^{\lm\star}+4  \lambda  {R}^{2} \left(\lambda  R+R+M\right)  {\dot{R} }^{5} {u}^t \dot{Y}^{\lm\star}\\
&
 +4 f_R^{-1}  \dot{\Phi}^2 \lambda  {R}^{5} {\dot{R} }^{3} {u}^t \dot{Y}^{\lm\star} -8 f_R^{-1}  \lambda  {R}^{2} \left(\lambda  R+R+2 M\right)  {\dot{R} }^{3} {u}^t \dot{Y}^{\lm\star}\\
&
 -4 f_R  \dot{\Phi}^2 \lambda  {R}^{5} \dot{R}  {u}^t \dot{Y}^{\lm\star} +4 f_R  \lambda  {R}^{2} \left(\lambda  R+R+3 M\right)  \dot{R}  {u}^t \dot{Y}^{\lm\star}-6f_R^{-2} \lambda  {R}^{4} {\dot{R} }^{4} \ddot{u}^t  Y^{\lm\star}\\
&
+4   \lambda  {R}^{4} {\dot{R} }^{2} \ddot{u}^t  Y^{\lm\star} +2 f_R^{2}  \lambda  {R}^{4} \ddot{u}^t  Y^{\lm\star} +16   \lambda  {R}^{4} \dot{R}  \ddot{R}  \dot{u}^t  Y^{\lm\star}\\
&+4{\Lambda_1}^{-1} \lambda  {R}^{2} \left(\Lambda_2-3 M R+3 {M}^{2}\right)  {\dot{R} }^{5} \dot{u}^t  Y^{\lm\star}+4 f_R^{-1}  \dot{\Phi}^2 \lambda  {R}^{5} {\dot{R} }^{3} \dot{u}^t  Y^{\lm\star}\\
&-8 f_R^{-1}{\Lambda_1}^{-1} \lambda  {R}^{2} \left(\Lambda_2+8\lambda M R+21 {M}^{2}\right)  {\dot{R} }^{3} \dot{u}^t  Y^{\lm\star}-4 f_R  \dot{\Phi}^2 \lambda  {R}^{5} \dot{R}  \dot{u}^t  Y^{\lm\star}\\
&+4 f_R  \lambda  {R}^{2} \left(\lambda  R+R+5 M\right)  \dot{R}  \dot{u}^t  Y^{\lm\star}-4f_R^{-2} \lambda  {R}^{4} {\dot{R} }^{3} \dot{\ddot{R}}  {u}^t Y^{\lm\star}\\
&
+4   \lambda  {R}^{4} \dot{R}  \dot{\ddot{R}}  {u}^t Y^{\lm\star}+6f_R^{-2} \lambda  {R}^{4} {\dot{R} }^{2} {\ddot{R} }^{2} {u}^t Y^{\lm\star}+2   \lambda  {R}^{4} {\ddot{R} }^{2} {u}^t Y^{\lm\star}\\
&
+2{\Lambda_1}^{-1} \lambda  {R}^{2} \left(\Lambda_2-3 M R+3 {M}^{2}\right)  {\dot{R} }^{4} \ddot{R}  {u}^t Y^{\lm\star}-6 f_R^{-1}  \dot{\Phi}^2 \lambda  {R}^{5} {\dot{R} }^{2} \ddot{R}  {u}^t Y^{\lm\star}\\
&
-4 f_R^{-1}{\Lambda_1}^{-1} \lambda  {R}^{2} \left(\Lambda_2+4 \lambda  M R+9 {M}^{2}\right)  {\dot{R} }^{2} \ddot{R}  {u}^t Y^{\lm\star}-2 f_R  \dot{\Phi}^2 \lambda  {R}^{5} \ddot{R}  {u}^t Y^{\lm\star}\\
&
+2 f_R  \lambda  {R}^{2} \left(\lambda  R+R+5 M\right)  \ddot{R}  {u}^t Y^{\lm\star}\\
&-4 {\Lambda_1}^{-2}\lambda  R \left({\lambda }^{3} {R}^{3}+{\lambda }^{2} {R}^{3}+3 {\lambda }^{2} M {R}^{2}+9 \lambda  {M}^{2} R+9 {M}^{3}\right)  {\dot{R} }^{6} {u}^t Y^{\lm\star}\\
&
+4f_R^{-2}{\Lambda_1}^{-2}\lambda  \left(3 {\lambda } {R}^{2}\Lambda_2+ {\lambda }^{2} M {R}^{3}(8-5\lambda)+ {\lambda } {M}^{2} {R}^{2}(48-29\lambda)+{M}^{3} R(45-129\lambda)-153 {M}^{4}\right)  {\dot{R} }^{4} {u}^t Y^{\lm\star}\\
&
+4f_R^{-2}{\Lambda_1}^{-1}\dot{\Phi}^2 \lambda  {R}^{4} \left(\lambda  R+2 \lambda  M+6 M\right)  {\dot{R} }^{4} {u}^t Y^{\lm\star}+8 f_R^{-1}  \dot{\Phi}\ddot{\Phi}  \lambda  {R}^{5} {\dot{R} }^{3} {u}^t Y^{\lm\star}\\
&
-4  {\Lambda_1}^{-2}\lambda  \left(3 {\lambda } {R}^{2}\Lambda_2+ {\lambda }^{2} M {R}^{3}(9-4\lambda)+{\lambda } {M}^{2} {R}^{2}(51-14\lambda)+{M}^{3} R(45-60\lambda)-54 {M}^{4}\right)  {\dot{R} }^{2} {u}^t Y^{\lm\star}\\
&
-4 {\Lambda_1}^{-1} \dot{\Phi}^2 \lambda  {R}^{3} \left(\lambda  {R}^{2}-3 \lambda  M R+6 M R-15 {M}^{2}\right)  {\dot{R} }^{2} {u}^t Y^{\lm\star}-8 f_R  \dot{\Phi}\ddot{\Phi}  \lambda  {R}^{5} \dot{R}  {u}^t Y^{\lm\star}\\
&
+4 f_R^{2} {\Lambda_1}^{-2}\lambda  \left(3 {\lambda }^{2} {R}^{4}(\lambda+1)+ {\lambda }^{2} M {R}^{3}(2-\lambda)+{\lambda } {M}^{2} {R}^{2}(12-\lambda)+{M}^{3} R(9+3\lambda)-54 {M}^{4}\right)  {u}^t Y^{\lm\star}\\
&
-4 f_R^{2}  \dot{\Phi}^2 \lambda  M {R}^{3} {u}^t Y^{\lm\star}+2f_R^{-2}{\Lambda_1} \dot{\Phi}^2 {R}^{4} {\dot{R} }^{3} {u}^t \dot{W}^{\lm\star}-2  {\Lambda_1} \dot{\Phi}^2 {R}^{4} \dot{R}  {u}^t \dot{W}^{\lm\star}\\
&+2f_R^{-2}{\Lambda_1} \dot{\Phi}^2 {R}^{4} {\dot{R} }^{3} \dot{u}^t  W^{\lm\star}-2  {\Lambda_1} \dot{\Phi}^2 {R}^{4} \dot{R}  \dot{u}^t  W^{\lm\star}-3f_R^{-2}{\Lambda_1} \dot{\Phi}^2 {R}^{4} {\dot{R} }^{2} \ddot{R}  {u}^t W^{\lm\star}\\
&- {\Lambda_1} \dot{\Phi}^2 {R}^{4} \ddot{R}  {u}^t W^{\lm\star}+2 {\Lambda_1} \dot{\Phi}^2 {R}^{3} {\dot{R} }^{4} {u}^t W^{\lm\star}+4f_R^{-2}{\Lambda_1} \dot{\Phi}\ddot{\Phi}  {R}^{4} {\dot{R} }^{3} {u}^t W^{\lm\star}\\
&-2 f_R^{-1} {\Lambda_1} \dot{\Phi}^2 {R}^{2} \left(R-5 M\right)  {\dot{R} }^{2} {u}^t W^{\lm\star}-4  {\Lambda_1} \dot{\Phi}\ddot{\Phi}  {R}^{4} \dot{R}  {u}^t W^{\lm\star}-2 f_R {\Lambda_1} \dot{\Phi}^2 M {R}^{2} {u}^t W^{\lm\star}
\Big]~,
\end{aligned}
\label{jumpdr2_e}
\eeq

\beq
\begin{aligned}
\jump{\partial^2_r\psi^{\ell m}_o}=&-\frac{8 m_0 \pi}{\lambda  \left(\lambda +1\right)  {\left(\dot{R}^2 -f_R^2\right) }^{3}} \Big[
{R}^{4}\mathcal{A}^{\lm\star}\left( {\dot{R} }^{4} -{f_R}^{4}\right)\ddot{u}^t\\
&-{f_R}^{2} \mathcal{A}^{\lm\star} R \dot{R}  \left(3 {R}^{3} \ddot{R} +4 {R}^{2}-14 M R+12 {M}^{2}\right)  \dot{u}^t\\
& -\mathcal{A}^{\lm\star} R {\dot{R} }^{3} \left({R}^{3} \ddot{R} -2 {R}^{2}-2 M R+12 {M}^{2}\right)  \dot{u}^t
 +2 \mathcal{A}^{\lm\star} {R}^{3} {\dot{R} }^{5} \dot{u}^t+2 \dot{\cal{A}}^{\lm\star}  {R}^{4} {\dot{R} }^{4} \dot{u}^t\\
& -2 {f_R}^{4} \dot{\cal{A}}^{\lm\star}  {R}^{4} \dot{u}^t
 -2 {f_R}^{2} \mathcal{A}^{\lm\star} {\dot{R} }^{2} \left(3 {R}^{3} \ddot{R} -2 \lambda  {R}^{2}-2 {R}^{2}+4 \lambda  M R+10 M R-12 {M}^{2}\right)  {u}^t\\
& -{f_R}^{2} \dot{\cal{A}}^{\lm\star}  R \dot{R}  \left(3 {R}^{3} \ddot{R} +4 {R}^{2}-14 M R+12 {M}^{2}\right)  {u}^t\\
&-2 {f_R}^{4} \mathcal{A}^{\lm\star} \left({R}^{3} \ddot{R} + {R}^{2}(\lambda+1)-2M R(2+\lambda)+4 {M}^{2}\right)  {u}^t\\
& -\dot{\cal{A}}^{\lm\star}  R {\dot{R} }^{3} \left({R}^{3} \ddot{R} -2 {R}^{2}-2 M R+12 {M}^{2}\right)  {u}^t
+2 \dot{\cal{A}}^{\lm\star}  {R}^{3} {\dot{R} }^{5} {u}^t+\ddot{\cal{A}}^{\lm\star}  {R}^{4} {\dot{R} }^{4} {u}^t\\
& -2f_R\mathcal{A}^{\lm\star} R \left(\lambda  R+R-12 M\right)  {\dot{R} }^{4} {u}^t-{f_R}^{4} \ddot{\cal{A}}^{\lm\star}  {R}^{4} {u}^t\Big]~,
\end{aligned}
\label{jumpdr2_o}
\eeq
\beq
\begin{aligned}
\jump{\partial_r\partial_t\psi^{\ell m}_o}=&\frac{8f_Rm_0 \pi}{\lambda\left(\lambda +1\right)  {\left(\dot{R}^2 -f_R^2\right) }^{3}} \Big[
2f_R {R}^{4} {\dot{R} }\mathcal{A}^{\lm\star}\left(\dot{R}^2-f_R^2\right)\ddot{u}^t
-f_R \mathcal{A}^{\lm\star} R {\dot{R} }^{2} \left(3 {R}^{3} \ddot{R} +4 {R}^{2}-18 M R+20 {M}^{2}\right)  \dot{u}^t\\&
-{f_R}^{3} \mathcal{A}^{\lm\star} R \left({R}^{3} \ddot{R} +{R}^{2}-4 M R+4 {M}^{2}\right)  \dot{u}^t
+\mathcal{A}^{\lm\star} {R}^{2} \left(5 R-4 M\right)  {\dot{R} }^{4} \dot{u}^t
+4f_R\dot{\cal{A}}^{\lm\star}  {R}^{4} {\dot{R} }^{3} \dot{u}^t\\&
-4 {f_R}^{3} \dot{\cal{A}}^{\lm\star}  {R}^{4} \dot{R}  \dot{u}^t
-2 {f_R}^{3} \mathcal{A}^{\lm\star} \dot{R}  \left(3 {R}^{3} \ddot{R} +{R}^{2}(\lambda+1)-2M R(2+\lambda)+4 {M}^{2}\right)  {u}^t\\&
-f_R\dot{\cal{A}}^{\lm\star}  R {\dot{R} }^{2} \left(3 {R}^{3} \ddot{R} +4 {R}^{2}-18 M R+20 {M}^{2}\right)  {u}^t
-{f_R}^{3} \dot{\cal{A}}^{\lm\star}  R \left({R}^{3} \ddot{R} +{R}^{2}-4 M R+4 {M}^{2}\right)  {u}^t\\&
-2f_R\mathcal{A}^{\lm\star} {\dot{R} }^{3} \left({R}^{3} \ddot{R} -2 {R}^{2}(\lambda+1)+2M R(3+2\lambda)-4 {M}^{2}\right)  {u}^t
-2 {f_R}^{3} \ddot{\cal{A}}^{\lm\star}  {R}^{4} \dot{R}  {u}^t\\&
-2 \mathcal{A}^{\lm\star} R \left(\lambda  R+R-8 M\right)  {\dot{R} }^{5} {u}^t
+\dot{\cal{A}}^{\lm\star}  {R}^{2} \left(5 R-4 M\right)  {\dot{R} }^{4} {u}^t
+2f_R\ddot{\cal{A}}^{\lm\star}  {R}^{4} {\dot{R} }^{3} {u}^t\Big] ~,
\end{aligned}
\label{jumpdrdt_o}
\eeq

\beq
\begin{aligned}
\jump{\partial_r\partial_t\psi^{\ell m}_e}=&\frac{4 m_0 \pi \Lambda_1^{-2}}{\lambda\left(\lambda +1\right)\left(\dot{R}^2 -f_R^2\right)^3 }\Big[
2f_R\Lambda_1\lambda  {R}^{4} {\dot{R} } \left(\dot{R}^2-f_R^2\right)^2{u}^t \ddot{Y}^{\lm\star}\\&
+8f_R^3\Lambda_1\lambda  {R}^{4} {\dot{R} } \left(\dot{R}^2-f_R^2\right)\dot{u}^t \dot{Y}^{\lm\star}
-2f_R\Lambda_1\lambda  {R}^{4} {\ddot{R} } \left(\dot{R}^4-f_R^4\right){u}^t \dot{Y}^{\lm\star}\\&
+2 \lambda  {R}^{2} \left(\Lambda_2+6 \lambda  M R+6 M R+3 {M}^{2}\right)  {\dot{R} }^{6} {u}^t \dot{Y}^{\lm\star}
+2 {f_R}^{2}\Lambda_1 \dot{\Phi}^2 \lambda  {R}^{5} {\dot{R} }^{4} {u}^t \dot{Y}^{\lm\star}\\&
-2 {f_R}^{2} \lambda  {R}^{2} \left(\Lambda_2+12M R(\lambda+1)+9 {M}^{2}\right)  {\dot{R} }^{4} {u}^t \dot{Y}^{\lm\star}\\&
-2 {f_R}^{4} \lambda  {R}^{2} \left(\Lambda_2-2 M R(\lambda+3)+3 {M}^{2}\right)  {\dot{R} }^{2} {u}^t \dot{Y}^{\lm\star}
-2 {f_R}^{6}\Lambda_1 \dot{\Phi}^2 \lambda  {R}^{5} {u}^t \dot{Y}^{\lm\star}\\&
+2 {f_R}^{6} \lambda  {R}^{2} \left(\Lambda_2+4 \lambda  M R+9 {M}^{2}\right)  {u}^t \dot{Y}^{\lm\star}
-2f_R\Lambda_1\lambda  {R}^{4} {\dot{R} }^{5} \ddot{u}^t  Y^{\lm\star}\\&
-4 {f_R}^{3}\Lambda_1 \lambda  {R}^{4} {\dot{R} }^{3} \ddot{u}^t  Y^{\lm\star}
+6 {f_R}^{5}\Lambda_1 \lambda  {R}^{4} \dot{R}  \ddot{u}^t  Y^{\lm\star}
-2f_R\Lambda_1\lambda  {R}^{4} {\dot{R} }^{4} \ddot{R}  \dot{u}^t  Y^{\lm\star}\\&
+12 {f_R}^{3}\Lambda_1 \lambda  {R}^{4} {\dot{R} }^{2} \ddot{R}  \dot{u}^t  Y^{\lm\star}
+6 {f_R}^{5}\Lambda_1 \lambda  {R}^{4} \ddot{R}  \dot{u}^t  Y^{\lm\star}\\&
+2 \lambda  {R}^{2} \left(\Lambda_2+2 \lambda  M R+3 {M}^{2}\right)  {\dot{R} }^{6} \dot{u}^t  Y^{\lm\star}
+2 {f_R}^{2}\Lambda_1 \dot{\Phi}^2 \lambda  {R}^{5} {\dot{R} }^{4} \dot{u}^t  Y^{\lm\star}\\&
-2 {f_R}^{2} \lambda  {R}^{2} \left(\Lambda_2+24 \lambda  M R+12 M R+45 {M}^{2}\right)  {\dot{R} }^{4} \dot{u}^t  Y^{\lm\star}\\&
-2 {f_R}^{4} \lambda  {R}^{2} \left(\Lambda_2-2 \lambda  M R-12 M R+15 {M}^{2}\right)  {\dot{R} }^{2} \dot{u}^t  Y^{\lm\star}
-2 {f_R}^{6}\Lambda_1 \dot{\Phi}^2 \lambda  {R}^{5} \dot{u}^t  Y^{\lm\star}\\&
+2 {f_R}^{6} \lambda  {R}^{2} \left(\Lambda_2+4 \lambda  M R+9 {M}^{2}\right)  \dot{u}^t  Y^{\lm\star}
-2f_R\Lambda_1\lambda  {R}^{4} {\dot{R} }^{4} \dot{\ddot{R}}  {u}^t Y^{\lm\star}\\&
+2 {f_R}^{5}\Lambda_1 \lambda  {R}^{4} \dot{\ddot{R}}  {u}^t Y^{\lm\star}
+2f_R\Lambda_1\lambda  {R}^{4} {\dot{R} }^{3} {\ddot{R} }^{2} {u}^t Y^{\lm\star}
+6 {f_R}^{3}\Lambda_1 \lambda  {R}^{4} \dot{R}  {\ddot{R} }^{2} {u}^t Y^{\lm\star}\\&
+2 \lambda  {R}^{2} \left(\Lambda_2+2 \lambda  M R+3 {M}^{2}\right)  {\dot{R} }^{5} \ddot{R}  {u}^t Y^{\lm\star}
-2 {f_R}^{2}\Lambda_1 \dot{\Phi}^2 \lambda  {R}^{5} {\dot{R} }^{3} \ddot{R}  {u}^t Y^{\lm\star}\\&
-4 {f_R}^{2}\Lambda_1 \lambda  {R}^{2} \left(\lambda  R+R+3 M\right)  {\dot{R} }^{3} \ddot{R}  {u}^t Y^{\lm\star}
-6 {f_R}^{4}\Lambda_1 \dot{\Phi}^2 \lambda  {R}^{5} \dot{R}  \ddot{R}  {u}^t Y^{\lm\star}\\&
+2 {f_R}^{4} \lambda  {R}^{2} \left(\Lambda_2+10 \lambda  M R+6 M R+15 {M}^{2}\right)  \dot{R}  \ddot{R}  {u}^t Y^{\lm\star}\\&
-2 \lambda  R \left(\Lambda_2+2 \lambda  M R+3 {M}^{2}\right)  {\dot{R} }^{7} {u}^t Y^{\lm\star}\\&
+2f_R\lambda  \left(3 {\lambda }^{2} {R}^{3}+3 \lambda  {R}^{3}-4 {\lambda }^{2} M {R}^{2}+6 \lambda  M {R}^{2}-24 \lambda  {M}^{2} R+15 {M}^{2} R-48 {M}^{3}\right)  {\dot{R} }^{5} {u}^t Y^{\lm\star}\\&
+2f_R\dot{\Phi}^2 \lambda  {R}^{4} \left(\lambda  R+2 \lambda  M+6 M\right)  {\dot{R} }^{5} {u}^t Y^{\lm\star}
+4 {f_R}^{2}\Lambda_1 \dot{\Phi}\ddot{\Phi}  \lambda  {R}^{5} {\dot{R} }^{4} {u}^t Y^{\lm\star}\\&
-2 {f_R}^{3} \lambda  \left(3 {\lambda }^{2} {R}^{3}+3 \lambda  {R}^{3}-2 {\lambda }^{2} M {R}^{2}+4 \lambda  M {R}^{2}+8 \lambda  {M}^{2} R+9 {M}^{2} R+30 {M}^{3}\right)  {\dot{R} }^{3} {u}^t Y^{\lm\star}\\&
+12 {f_R}^{3}\Lambda_1 \dot{\Phi}^2 \lambda  M {R}^{3} {\dot{R} }^{3} {u}^t Y^{\lm\star}\\&
+2 {f_R}^{5} \lambda  \left({\lambda }^{2} {R}^{3}+\lambda  {R}^{3}-2 \lambda  M {R}^{2}+12 \lambda  {M}^{2} R-3 {M}^{2} R+24 {M}^{3}\right)  \dot{R}  {u}^t Y^{\lm\star}\\&
-2 {f_R}^{5} \dot{\Phi}^2 \lambda  {R}^{3} \left(\lambda  {R}^{2}+6 M R-6 {M}^{2}\right)  \dot{R}  {u}^t Y^{\lm\star}\\&
-4 {f_R}^{6}\Lambda_1 \dot{\Phi}\ddot{\Phi}  \lambda  {R}^{5} {u}^t Y^{\lm\star}
+f_R{\Lambda_1}^{2} \dot{\Phi}^2 {R}^{4} {\dot{R} }^{4} {u}^t \dot{W}^{\lm\star}
 -{f_R}^{5} {\Lambda_1}^{2} \dot{\Phi}^2 {R}^{4} {u}^t \dot{W}^{\lm\star}\\&
 +f_R{\Lambda_1}^{2} \dot{\Phi}^2 {R}^{4} {\dot{R} }^{4} \dot{u}^t  W^{\lm\star}
-{f_R}^{5} {\Lambda_1}^{2} \dot{\Phi}^2 {R}^{4} \dot{u}^t  W^{\lm\star}
-f_R{\Lambda_1}^{2} \dot{\Phi}^2 {R}^{4} {\dot{R} }^{3} \ddot{R}  {u}^t W^{\lm\star}\\&
-3 {f_R}^{3} {\Lambda_1}^{2} \dot{\Phi}^2 {R}^{4} \dot{R}  \ddot{R}  {u}^t W^{\lm\star}
+{\Lambda_1}^{2} \dot{\Phi}^2 {R}^{3} {\dot{R} }^{5} {u}^t W^{\lm\star}
+2f_R {\Lambda_1}^{2} \dot{\Phi}\ddot{\Phi}  {R}^{4} {\dot{R} }^{4} {u}^t W^{\lm\star}\\&
+6 {f_R}^{2} {\Lambda_1}^{2} \dot{\Phi}^2 M {R}^{2} {\dot{R} }^{3} {u}^t W^{\lm\star}
-{f_R}^{5} {\Lambda_1}^{2} \dot{\Phi}^2 {R}^{3} \dot{R}  {u}^t W^{\lm\star}
-2{f_R}^{5} {\Lambda_1}^{2} \dot{\Phi}\ddot{\Phi}  {R}^{4} {u}^t W^{\lm\star}\Big]~.
\end{aligned}
\label{jumpdrdt_e}
\eeq

\bibliography{references_spallicci_150913}
\end{document}